\documentclass[11pt]{article}
\pdfoutput=1

\usepackage[english]{babel}
\usepackage{fancyhdr}
\usepackage{amsmath}
\usepackage{amssymb}
\usepackage{amsfonts}
\usepackage{psfrag}
\usepackage[applemac]{inputenc}
\usepackage{graphicx,wrapfig}
\usepackage[bf,footnotesize]{caption2}
\usepackage{pstricks}
\usepackage{hyperref}

\newcommand{\Tr}{\mathop{\rm Tr}} 

\newcommand{\beq}{\begin{equation}}
\newcommand{\eeq}{\end{equation}}
\newcommand{\bea}{\begin{eqnarray}}
\newcommand{\eea}{\end{eqnarray}}
\newcommand{\bag}{\begin{align}}
\newcommand{\eag}{\end{align}}

\newcommand{\ie}{$\textnormal{i.e.}$ }

\newcommand{\GeV}{\,\mathrm{GeV}}
\newcommand{\TeV}{\,\mathrm{TeV}}

\newcommand{\eq}[1]{Eq.~(\ref{#1})}
\newcommand{\fig}[1]{Fig.~\ref{#1}}

\newcommand{\vev}[1]{\langle {#1} \rangle}

\newcommand{\Lslash}[1]{\slash{ \hspace{-2.5mm} #1}}
\newcommand{\Sslash}[1]{\slash{ \hspace{-2mm} #1}}

\newcommand{\SU}{\textrm{SU}}
\newcommand{\SO}{\textrm{SO}}
\newcommand{\U}{\textrm{U}}
\newcommand{\Sp}{\textrm{Sp}}

\newcommand{\higgs}{H}
\newcommand{\BL}{(B-L)_{\SO(10)}}
\newcommand{\triplet}{T}
\newcommand{\K}{\mathcal{K}}
\newcommand{\Gsym}{\mathcal{G}}
\newcommand{\qx}{q^c}
\newcommand{\lx}{l^c}
\newcommand{\tx}{t^c}
\newcommand{\laq}{\lambda_q}
\newcommand{\lax}{\lambda_x}

\begin{document}

\baselineskip=16pt
\setcounter{footnote}{0}
\setcounter{figure}{0}
\setcounter{table}{0}


\begin{titlepage}
\begin{flushright}
UAB-FT-689\\
\end{flushright}
\vspace{.3in}

\begin{center}
\vspace{1cm}

{\LARGE \bf
Composite GUTs: models and expectations at the LHC}

\vspace{.8cm}

\renewcommand{\thefootnote}{\fnsymbol{footnote}}

{\bf Michele Frigerio}$^{a,b,}$\footnote{E-mail: frigerio@ifae.es}, {\bf Javi Serra}$^{a,}$\footnote{E-mail: jserra@ifae.es}, {\bf Alvise Varagnolo}$^{a,}$\footnote{E-mail: alvise@ifae.es}

\vspace{.5cm}

\centerline{$^a$ \it  Institut de F\'isica d'Altes Energies, Universitat Aut\`onoma de Barcelona, E-08193 Bellaterra, SPAIN}
\centerline{$^b$ {\it CNRS, Laboratoire Charles Coulomb, UMR 5221, F-34095 Montpellier, FRANCE} }
\centerline{\it  Universit\'e Montpellier 2, Laboratoire Charles Coulomb, UMR 5221, F-34095 Montpellier, FRANCE}

\end{center}
\vspace{.8cm}

\begin{abstract}
\medskip
\noindent

We investigate grand unified theories (GUTs) in scenarios where electroweak (EW) symmetry breaking is triggered by a light composite Higgs, arising as a Nambu-Goldstone boson from a strongly interacting sector. The evolution of the standard model (SM) gauge couplings can be predicted at leading order, if the global symmetry of the composite sector is a simple group $\Gsym$ that contains the SM gauge group. It was noticed that, if the right-handed top quark is also composite, precision gauge unification can be achieved. We build minimal consistent models for a composite sector with these properties, thus demonstrating how composite GUTs may represent an alternative to supersymmetric GUTs. Taking into account the new contributions to the EW precision parameters, we compute the Higgs effective potential and prove that it realizes consistently EW symmetry breaking with little fine-tuning. The $\Gsym$ group structure and the requirement of proton stability determine the nature of the light composite states accompanying the Higgs and the top quark: a coloured triplet scalar and several vector-like fermions with exotic quantum numbers. We analyse the signatures of these composite partners at hadron colliders: distinctive ﬁnal states contain multiple top and bottom quarks, either alone or accompanied by a heavy stable charged particle, or by missing transverse energy.

\end{abstract}


\end{titlepage}

\tableofcontents

\renewcommand{\thefootnote}{\arabic{footnote}}
\addtocounter{footnote}{-3}

\newpage

\section{Introduction}

Two paradigms have been proposed to account for the stability of the electroweak (EW) scale against quantum corrections,
the so-called  gauge hierarchy problem.
One is a weakly coupled theory where the mass of  the elementary scalar responsible for electroweak symmetry breaking (EWSB), that is the Higgs boson,
is protected from ultraviolet scales by supersymmetry.
The other involves a new strongly coupled sector, which generates the EW scale
dynamically,
in analogy with the origin of the QCD scale.
The latter idea has been realized in several ways.
A particularly attractive possibility is to achieve EWSB thanks to a Higgs-like composite field, emerging as a Nambu-Goldstone boson (NGB)
 from the strongly interacting sector:
this is known as the composite-Higgs scenario \cite{oldcompoHiggs,newcompoHiggs,reviewContino}.
It is generically favoured by precision measurements with respect to the simplest form of technicolour  \cite{originalTC} or Higgsless models \cite{higgsless},
and it has a more economical particle content with respect to little-Higgs models \cite{littlehiggs}.

Whatever solution of the hierarchy problem is adopted, there are several robust indications for the unification of the gauge interactions at a scale $M_{GUT}$,
close but definitely smaller than the Planck scale $M_P$, where gravitational interactions become strong.
As a matter of fact, grand unified theories (GUTs) \cite{GUTs} elegantly account for the quantization of the electric charge,
the quantum numbers of quarks and leptons
and the basic relations between their Yukawa couplings, the cancellation of the gauge anomalies, the evidence for non-zero neutrino masses.
While GUTs have been intensively studied mostly in the supersymmetric framework \cite{susyGUTs}, the above-listed virtues of GUTs do not
rely on the existence of low-energy supersymmetry.
It is therefore sensible to try to realize precise gauge coupling unification in other extensions of the standard model (SM)
that provide a natural explanation of the EW-GUT hierarchy.

In comparison with supersymmetric scenarios, strongly coupled models
suffer from the obstruction to perturbative computations. Moreover, at first sight they do not share the striking prediction of precise gauge coupling unification,
which pertains to the minimal supersymmetric SM.
This gap has been significantly reduced over the years, thanks to the modern understanding of strongly coupled systems via the AdS/CFT correspondence \cite{adscft},
and in particular through the study of extra-dimensional scenarios of the Randall-Sundrum type
\cite{Randall:1999ee}, which permitted to extract both qualitative and quantitative information on EWSB, electroweak precision tests (EWPTs) and phenomenology.
Following such developments, the investigation of gauge coupling unification has been pursued
\cite{Pomarol:2000hp,warpedGUTs,AS}.
Actually, it turned out that some key features of strongly coupled models
can be studied with no need to specify  a dual extra-dimensional construction and independently from the details of the strong dynamics.
In particular, a proposal for precise gauge coupling unification in the composite-Higgs scenario was advanced \cite{Agashe:2005vg}.

In this paper we further investigate composite-Higgs scenarios which exhibit gauge coupling unification, from a purely four-dimensional perspective.
We will make use of two basic properties of the strongly interacting sector.
The first is an approximate conformal symmetry, spontaneously broken at low energies, around $\Lambda_c \sim $ few  TeV,
thus generating a mass gap in the spectrum of composite resonances.
For energies above
$\Lambda_c$ and up to the unification scale $M_{GUT}$, the composite sector alone is well described by a
strongly interacting  conformal field theory (CFT). The conformal symmetry fixes the behaviour
of the correlators of the composite sector, in particular those that affect the propagators of weakly coupled external fields. In the case of interest,
these are the SM gauge fields, with the corresponding gauge couplings.
It can be shown that the contribution of such composite sector to their running is logarithmic \cite{polyakov}.

The second crucial ingredient is the invariance of the composite sector under a set $\Gsym$ of global symmetry transformations,
analogous to the
approximate $\SU(3)_L \times \SU(3)_R$ chiral symmetry of QCD.
If the SM gauged group, $\Gsym_{SM} \equiv \SU(3)_C \times \SU(2)_L \times \U(1)_Y$, is embedded in a simple group $\Gsym$,
then gauge coupling unification becomes independent from the strong dynamics at leading order. This is because the one loop
contribution of the composite sector to the gauge coupling
beta functions is universal, that is to say $b_i^{comp}-b_j^{comp}=0$ for $i,j=1,2,3$. Moreover, the universal coefficient $b^{comp}$
is constant between $\Lambda_c$
and $M_{GUT}$ because of the conformal symmetry.\footnote{In principle,
the composite sector could exit the strong coupling regime before $M_{GUT}$, thus modifying $b^{comp}$.
This requires a breaking of the conformal symmetry at the intermediate scale where the transition between the two regimes occurs.
The universality of the beta function coefficients is maintained also in this case, as long as $\mathcal G$ is preserved. However,
this option would introduce an unnecessary model-dependence.
Therefore we will not consider it in this paper.}
At this point one is in the position to study quantitatively gauge unification,
and subsequently undertake the construction of composite GUTs.

It is remarkable that both the conformal symmetry and the $\mathcal G$-symmetry of the composite sector
are also instrumental to generate hierarchical Yukawa couplings, by an elegant mechanism
known as ``partial compositeness" \cite{Kaplan:1991dc,warpedcompo}.
The  light SM fermions are taken to be elementary particles external to the composite sector, weakly coupled to it by mixing
with composite fermionic operators.
The Yukawa couplings arise from this mixing
at low energies, and the hierarchies between them are due to the different scaling dimensions of the various operators
(at least one for each SM chiral fermion).
The required large anomalous dimensions  can be generated if the composite sector is strongly coupled over a large range of energies,
which naturally happens when it is close to an infrared-attractive fixed point, indicating that it is approximately conformal.
Since the composite sector is responsible for generating the EW scale, it must carry EW charges.
Now, in order to generate quark masses, it must also contain
operators  charged under colour.
Then, the composite operators transform non-trivially under the full $\mathcal G_{SM}$, which of course must be a subgroup of
the global symmetry $\mathcal G$ of the composite sector.
Therefore the only extra assumption to move from the ordinary composite-Higgs scenario  to the composite GUT scenario
is the requirement of $\mathcal G$ being simple.

Besides the breaking of the conformal symmetry at $\Lambda_c$, the strong dynamics does also lead to the spontaneous breaking of part of the global
symmetry, $\Gsym \rightarrow \K$. This is necessary in order to obtain the Higgs as a Nambu-Goldstone boson (NGB), in analogy
to the light pseudo-scalar mesons
of QCD, that are the approximate NGBs of  $\SU(3)_L \times \SU(3)_R \rightarrow \SU(3)_V$.
The unbroken subgroup $\K$ does not need to be simple, since this breaking is an infrared effect,
which does not modify the gauge coupling evolution at leading order over the large hierarchy.
Part of the global symmetries of the composite sector
will be eventually broken explicitly by the gauge and fermion couplings to the elementary fields.
Since these couplings are perturbatively small, the global symmetry $\mathcal G$ of the composite sector holds in good approximation
over the whole hierarchy between $\Lambda_c$ and $M_{GUT}$. Nonetheless, below $\Lambda_c$ the explicit breaking generates a non-trivial effective
potential for the NGBs of the composite sector, leading to EWSB.

In composite GUTs, the Higgs will be generically accompanied by other pseudo Nambu-Goldstone bosons (pNGBs) that fill with it a complete multiplet of the global symmetry
$\mathcal K$. These necessarily light extra scalar fields are a distinctive feature of composite GUTs, to be contrasted with
the usual weakly coupled GUTs, where the Higgs partners live at the GUT scale.
The set of light scalars typically includes a coloured triplet that can potentially mediate proton decay. In supersymmetric GUTs the proton decay
issue is usually cured imposing $R$-parity and making the triplet super-heavy ($\sim \! M_{GUT}$),
which requires to implement a doublet-triplet splitting mechanism.
In composite GUTs, instead,
the colour triplet (more in general, any operator generated by the strong dynamics)
that might mediate proton decay cannot be decoupled, since its mass scale is around or below $\Lambda_c \sim$ few TeV.
The remedy will be to forbid the triplet couplings to SM fermions,
and more in general to suppress baryon number violating operators, by imposing an appropriate symmetry.
Similarly, composite operators that mediate lepton number violation need to be suppressed, not to generate too large neutrino masses.

The last important feature of composite GUTs is the presence of extra vector-like fermions at the EW scale, that eventually are responsible for the
correction to the SM gauge coupling evolution, such that unification is achieved at $M_{GUT} \sim 10^{15} \GeV$, with a precision comparable to that of the MSSM. It is quite remarkable \cite{Agashe:2005vg} that  these extra fermions are automatically predicted, once one implements
in a straightforward way the attractive features of the composite-Higgs scenario described above.
Partial compositeness implies that the larger the mass of a SM fermion, the stronger its coupling is to the composite sector, and in turn the modification of its elementary properties. The degree of compositeness of the light SM fermions is thus small, but that of the top quark has to be large.
In fact, the well-motivated possibility exists that
 the right-handed top quark is an entirely composite chiral fermion.\footnote{
It will be clear in the following why the left-handed top should not be composite.
}
In this case, it must be accompanied by a set of composite partners, filling a complete multiplet of the global symmetry $\mathcal K$.
In order to make these chiral top partners massive and to cancel gauge anomalies, one is forced to introduce extra elementary fermions.
We will analyze in detail  the impact of these new exotic particles on precise unification,
EWPTs, and in EWSB,
as well as their manifestation at the large hadron collider (LHC). Their observation would constitute another crucial signature of composite GUTs.

In this paper we will not attempt to build an explicit ultraviolet completion for the composite GUT scenario, that is to say, we will not construct
a specific model at the scale $M_{GUT}$. This definitely  remains a very important task, in a territory that is presently largely unexplored.
At least one comment is in order
to settle the ground and avoid confusions.
The full GUT must possess a gauge symmetry $\Gsym_{GUT}$, which is a simple group containing $\Gsym_{SM}$.
One may conceive that $\Gsym_{GUT}$ is broken only in the elementary sector, which is promptly realized assuming that the GUT breaking fields do not
couple to the composite sector.\footnote{ To be concrete,
think of chiral fermions in a complex representation
of $\Gsym_{GUT}$, such as a $\textbf{5}$ or a $\textbf{10}$
of $\SU(5)$. One example is provided by the SM fermions, which form chiral $\SU(5)$ multiplets that only feel GUT breaking effects through Yukawa and gauge couplings. The required composite sector can be generated if there exist another set of such chiral fermions (\emph{i}) charged under an additional gauge interaction in the non-perturbative regime
and (\emph{ii}) sufficiently weakly coupled to the $\mathcal G_{GUT}$  breaking sector.}
Then, the composite sector would retain a global symmetry $\Gsym = \Gsym_{GUT}$.
However, the identification of the two groups may appear minimal but it is not necessary nor natural: the symmetry of the GUT theory may be larger,
and boil down to a low energy global symmetry $\Gsym \subset \Gsym_{GUT}$ or,  if the initial global symmetry of the strong sector was larger
than the gauged $\Gsym_{GUT}$, one could have $\Gsym\supset \Gsym_{GUT}$. The only requirement is
that $\Gsym_{SM}$ belongs to the intersection of $\Gsym$ and $\Gsym_{GUT}$.
We will see that the suppression of baryon and lepton number violation
implies further constraints on the interplay between these two groups.

The paper is organized as follows. In the next subsection we recall the general setup for the study of composite-Higgs scenarios,
for the non-practitioners.
In section \ref{GCE} we describe in detail the evolution of gauge couplings when the SM Higgs boson and the right-handed top quark are
composite states, and what the conditions for precision unification are. In section \ref{global} we discuss the global symmetries of the composite
sector that are required in order to respect the EWPTs, to implement gauge coupling unification and to avoid proton decay. The model
which emerges as the simplest viable possibility has a global symmetry $\mathcal G = \SO(11)$, with unbroken subgroup $\mathcal K = \SO(10)$.
The exotic fermion quantum numbers are then specified, and their contribution to EW precision parameters is computed.
In section \ref{ewsb} we compute the effective potential for the pNGBs in this model, and derive the constraints
for a satisfactory EWSB. In section \ref{pheno} we describe the collider phenomenology of the Higgs and top quark composite partners
in three different variants of the model.
We finally summarize the substantial features of our composite GUT models in section \ref{conclu}.

\subsection{The setup of composite-Higgs models} \label{setup}

The lagrangian for composite-Higgs models can be expressed, in the same spirit of Refs.~\cite{warpedcompo,Giudice:2007fh}, as
\beq
\mathcal{L} = \mathcal{L}_{elementary}^{\Gsym_{SM}} + \mathcal{L}_{composite}^{\Gsym \rightarrow \K} + \mathcal{L}_{mixing}^{\Gsym_{SM}
}~.\label{Lag}
\eeq
There is a sector of elementary weakly coupled fields, whose dynamics is described by $\mathcal{L}_{elementary}^{\Gsym_{SM}}$,
invariant under the SM gauge symmetries, $\Gsym_{SM}$. The field content of this sector is the one of the SM, without the Higgs.
In addition, there exists a new strongly interacting sector,
described by $\mathcal{L}_{composite}^{\Gsym \rightarrow \K}$,
made of composite bound states.
Such sector is characterized by a scale $m_{\rho}$, associated to the mass of the lightest massive resonances (massless composites are also present),
and by an inter-composite coupling $g_{\rho}$. The latter is larger than the elementary weak couplings (generically denoted by $g_{elem}$),
although it can be significantly smaller than the naive dimensional analysis (NDA) estimate in fully strongly interacting theories, that is $g_{\rho} \sim 4 \pi$.
The composite sector is invariant under a global symmetry  $\Gsym$, which contains $\Gsym_{SM}$ as a subgroup.
At a scale close to $m_{\rho}$, $\Gsym$ is spontaneously broken to $\K$, giving rise to a set of NGBs parametrizing the coset space $\Gsym/\K$; this
set includes the Higgs doublet $\higgs$.  The NGBs remain massless in the limit $g_{elem} \rightarrow 0$, and their dynamics is
described by a non-linear $\sigma$-model with characteristic scale $f$. This scale controls the interaction among the NGBs and it
is related to the composite sector parameters as
\beq
m_{\rho} = g_{\rho} f ~,
\label{mgf}
\eeq
in analogy with QCD, where \eq{mgf} relates the pion decay constant $f_\pi$ to the mass of the QCD resonances. Also, 
$\mathcal{L}_{composite}^{\Gsym \rightarrow \K}$ is approximately conformal
invariant at energies above $\Lambda_c \gtrsim m_{\rho}$, and it remains strongly coupled over the whole hierarchy between $\Lambda_c$ and $M_{GUT}$.

The interaction between the elementary and composite sectors is described by $\mathcal{L}_{mixing}^{\Gsym_{SM}}$, that in general respects
only the SM gauge symmetries.
We assume that $\mathcal{L}_{mixing}^{\Gsym_{SM}}$ has the form dictated by partial compositeness, which applies both to the coupling of the elementary gauge bosons as well as of the elementary fermions. Generically, the former can be expressed as
\beq
g_i A_{i\mu} \mathcal{J}_i^{\mu},
\label{gaugemix}
\eeq
where $A_i$ is the elementary gauge boson, coupled with strength $g_i$ to the corresponding composite sector current $\mathcal{J}_i$.
Analogously, there is a coupling for each elementary chiral fermion.
In order to describe the Yukawa couplings,
it is convenient to write these couplings as
\beq
\lambda_{\psi_L} \overline{\psi_L} \mathcal{O}_{\psi_L} + \lambda_{\psi_R} \overline{\psi_R} \mathcal{O}_{\psi_R} + h.c.~,
\label{fermionmix}
\eeq
where the chiral fermions $\psi_{L,R}$ are coupled to the composite operators $\mathcal{O}_{\psi_{L,R}}$ with strength $\lambda_{\psi_{L,R}}$.
The operators $\mathcal{O}_{\psi_{L,R}}$ transform under $\Gsym$ in such a way as to generate a coupling between $\psi_L$, $\psi_R$ and
$\higgs$ at low energies, below $m_{\rho}$. Therefore, the low energy values of $\lambda_{\psi_{L,R}}$
are constrained to reproduce the observed Yukawa couplings \cite{Kaplan:1991dc,warpedcompo}:\footnote{
Here we assume that each chiral elementary fermion couples dominantly to a unique composite operator, which is responsible
for inducing the corresponding Yukawa coupling. Nonetheless, extra couplings to other operators could be present,
if allowed by the gauge (and global) symmetries of the full lagrangian.
}
\beq
y_\psi ~\simeq~ \frac{\lambda_{\psi_L} \lambda_{\psi_R}}{g_{\rho}}~.
\label{Yuk}
\eeq

Of course both \eq{gaugemix} and \eq{fermionmix} must respect the SM gauge symmetries. In addition, extra global symmetries might be approximately preserved (in particular, consistency will require to impose baryon and lepton number conservation, as discussed later).
However, \eq{gaugemix} and \eq{fermionmix} do not respect the $\Gsym$ symmetry, therefore introducing a (weak) explicit breaking
of the global symmetries of the composite sector, in particular of the NGB symmetries.
As a consequence, an effective potential for the NGBs will be generated by loops of elementary fields. For instance, the mass of the Higgs field will receive corrections that scale like $g^2_{elem}m_{\rho}^2/(4\pi)^2 $,  with the scale $m_{\rho}$ acting as the cut-off for the  elementary loops.
This is in analogy to QCD, where the charged pion mass receives divergent loop corrections from the photon that are cut at the $\rho$ meson mass scale.

The effective potential, which is an expansion in the small explicit breaking couplings, $g_{elem}/g_{\rho}$, and in the number of elementary loops,
will induce a VEV for the Higgs, $v \simeq 246 \GeV$, breaking the EW symmetry.
This introduces the last parameter of our framework, $\xi \equiv v^2/f^2$, which describes the departure
from an elementary Higgs scenario, obtained in the limit $\xi \rightarrow 0$, or from
a so-called Higgsless scenario,
in the limit $\xi \rightarrow 1$
(in this case the longitudinal gauge boson scattering amplitudes are unitarized as in technicolour).
The deviations from the SM predictions introduced by the composite sector will then be proportional to $\xi$, and thus the EWPTs set an upper bound on this parameter, as reviewed in  section \ref{ewpt}. Besides, $\xi$ constitutes a rough measure of the degree of fine-tuning in the model.\footnote{Composite models \textit{naturally} tend to predict $v = f$ (or $v=0$), which is ruled out phenomenologically (see section \ref{ewpt}). Achieving a separation of scales requires a tuning of the parameters of the model, specifically of those responsible for the generation of the Higgs effective potential: this is the incarnation,
in composite-Higgs scenarios,
of what is customarily
dubbed as the little hierarchy problem.} We shall aim, for concreteness, at $\xi \simeq 0.1$, representing a 10\% fine-tuning, which would be
highly competitive with respect to alternative scenarios (e.g. supersymmetric constructions).
Note that $v \simeq 246 \GeV$ and $\xi \simeq 0.1$ imply $f \simeq 750 \GeV$, which we take as a reference value.

\section{Gauge coupling evolution \label{GCE}}

In the spirit
of GUTs, we assume that the SM gauge interactions have a common strength at high energies, which we observe at low energies as three different
gauge couplings due to the GUT spontaneous symmetry breaking, occurring at the scale $M_{GUT}$, and the consequent
differential running to lower energies.
Actually, an indication in favour of this assumption is provided by the SM particle content, since the evolution of  the SM gauge couplings from the EW scale to
high energies yields a rough convergence of their values, at the 20\% level.
The contribution of the SM fields to the renormalization of the gauge couplings
can be parametrized at one-loop by the $\beta$-function coefficients $b_i^{SM}$, where $i=1,2,3$ refer to the $\U(1)_Y$, $\SU(2)_L$, $\SU(3)_C$
groups, respectively.
While the contribution of the fermions is universal (\ie independent from $i$), since they fill complete $\SU(5)$ multiplets, the ones
of the gauge bosons and of the Higgs are not.

The usual test for gauge coupling unification at one-loop consists in the comparison of
the ratio of $\beta$-function coefficients,
$R\equiv(b_1-b_2)/(b_2-b_3)$,
with the value determined by the measurements of the gauge couplings at the scale $m_Z$, $R_{exp}=1.395\pm 0.015$.
The SM prediction is $R_{SM}\simeq 1.9$.
Although thresholds might arise at scales close to $M_{GUT}$  or at intermediate scales,
there are no observational nor theoretical reasons why they should be large enough to achieve precision unification.
On the other hand, the new physics associated with EWSB, which is required in particular to address the hierarchy problem,
may significantly contribute to the $\beta$-function coefficients and improve unification with respect to the SM.
We explain in the following how the needed states can arise in the context of composite-Higgs models \cite{Agashe:2005vg}.

\subsection{Composite sector contribution to the $\beta$-functions} \label{compo}

In general, a new strongly coupled sector may modify completely the gauge coupling evolution with respect to the SM.
If such a sector
is responsible for EWSB,
it necessarily affects at leading order the evolution of
the $\SU(2)_L$ and $\U(1)_Y$ couplings above the
compositeness scale $\Lambda_c$.
While in the SM the contribution of the Higgs doublet to the evolution is relatively small,
a larger number of degrees of freedom seems required to break the EW symmetry dynamically, so that no study of
gauge coupling unification is feasible if the contribution of the composite sector cannot be computed.

Besides the obstruction to perturbative computations around the scale $\Lambda_{c}$, that introduces a possibly large threshold,
we do not know the structure of the EWSB sector between $\Lambda_{c}$ and the unification scale $M_{GUT}$, making
the analysis of unification
highly model-dependent. Still, some deal of information may be extracted
on the basis of symmetry considerations and of the
consistency with
electroweak data, as we now describe.

As a first general consideration, recall that the composite sector is supposed to stabilize the hierarchy between the electroweak scale
and the unification (Planck) scale.
This is achieved thanks to its approximate conformal symmetry, broken only at low energies $\sim \Lambda_c$.
Then, if the composite sector is nearly conformal and weakly coupled to a set of external gauge fields,
its main contribution to the running of the external gauge couplings
will be logarithmic \cite{polyakov}.\footnote{
The fact that the sector is conformal does not mean that it cannot contribute to the scale dependence
of external fields coupled to it. What does not run is the intra-composite coupling.
Technically, the logarithmic running follows from the fact that the correction to the gauge boson propagators is given by $\vev{JJ}_{CFT}$ insertions,
where $J$ is the CFT current coupled to the gauge bosons, and conformal invariance implies
that $\vev{J(p) J(-p)} \propto p^2 \log p^2$ \cite{polyakov,holographicrunning}.}
This is precisely what happens in theories with no intermediate scales, like the SM between the electroweak and the Planck scale, or 
$\SU(3)_C \times \U(1)_{em}$
above $\Lambda_{QCD}$, where QCD is described by quarks and gluons.

\begin{figure}[!tp]
  \begin{center}
		\includegraphics[width=5cm]{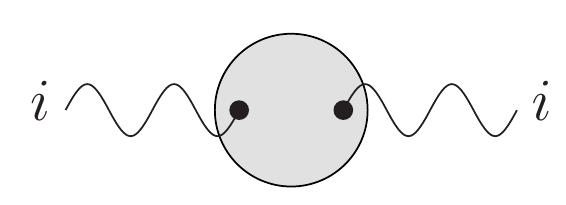}
  \caption{Diagrammatic representation of the leading-order contribution from the strong sector  to the SM gauge coupling running,
  parametrized in \eq{strongrun}.}
  \label{lead-run}
  \end{center}
  \end{figure}

Therefore, the contribution of the composite sector to the running of the gauge couplings $\alpha_i \equiv g_i^2/4 \pi$, as a function of the
renormalization scale $\mu$,
can be written as
\beq
\frac{d}{d \ln \mu} \left( \frac{1}{\alpha_i} \right) \supset \frac{b_i^{comp}}{2 \pi}~,
\label{strongrun}
\eeq
and can be visualized diagrammatically as in Fig.~\ref{lead-run}.
In general, the relative values of the coefficients $b_i^{comp}$ cannot be computed perturbatively, nor the absolute
size can be estimated in a model-independent way. Still, it was shown by Polyakov that $b_i^{comp} > 0$ \cite{polyakov}, and recent studies aim to put lower bounds on these coefficients,
 as a function of
the dimension of the scalar operators of a generic CFT \cite{cftbounds}.
These bounds could be of particular relevance for unification.  Here we will assume that $b_i^{comp}$ is small enough
for the SM gauge couplings not to hit a Landau pole before $M_{GUT}$.\footnote{
A warped extra-dimensional scenario yields
$b_i^{comp} = 2 \pi /(\alpha_i^{(5)} k) \sim N$,
where $k$ is the AdS curvature radius, $\alpha_i^{(5)}$ are the five-dimensional gauge couplings,
and $N$ is the number of colours of the dual conformal theory \cite{holographicrunning}.
However, the calculability in the warped extra-dimension requires a small ratio between the number of flavours and the number of colours, $F/N \ll 1$,
since this is the expansion parameter of the theory. Unfortunately, in the scenario discussed in this paper,
the number of flavours has to be large, due to the large global symmetry group $\Gsym$, while the
absence of a Landau pole requires $b_i^{comp} \alpha_i(M_{GUT})/2\pi\sim N \alpha_i(M_{GUT})/2\pi \ll 1$,
posing an upper bound on $N$. Therefore we will not rely on warped extra-dimension estimates nor on large-$N$ arguments in this work.}

The differential running, that is,
the dependence on the scale $\mu$ of the quantities $\delta_{ij}(\mu)\equiv 1/\alpha_i(\mu)-1/\alpha_j(\mu)$, is affected
at leading order by incomplete $\SU(5)$
representations, e.g., in the case of the SM, the gauge bosons and the Higgs doublet. One knows, therefore, the amount of ``$\SU(5)$ breaking"
that should be introduced with respect to the SM in order to achieve precision unification.
Then, the question is whether there are symmetries of the EWSB sector that allow to compute its contribution to the differential running,
independently from the strong dynamics.

A straightforward (perhaps, the only)
 possibility \cite{Agashe:2005vg} is to assume that the EWSB sector has a global symmetry $\Gsym$,
which is a simple group containing $\Gsym_{SM}$
(therefore $\Gsym$ can be $\SU(5)$ or a larger simple group).
In this case the EWSB sector does not contribute to $\delta_{ij}$ at the one-loop level,
because $b_i^{comp} = b^{comp}$ for $i=1,2,3$.
Besides, since the Higgs doublet $H$ arises as a light composite state from the $\Gsym$-symmetric sector, it does not contribute to the running.
At most, it gives a small contribution below $\Lambda_{c}$, the scale where $\Gsym$ is broken
spontaneously to $\K$, that may be non-simple. Similarly,
all low energy composite states may
contribute to the differential running only below $\Lambda_{c}$,
as a sub-leading threshold effect.

In particular, if some of the SM fields are composite, they do not contribute to the differential running above $\Lambda_c$, therefore
it is convenient to denote with $b_i^{elem}$
the $\beta$-function coefficients of the elementary SM fields only.
Specifically, when $H$ is part of the composite sector, the SM prediction
$R_{SM}\simeq 1.9$ is modified by the subtraction of $\higgs$, giving $R_{SM-\higgs} = 2$.
The extra required correction to achieve precise unification will be provided by the interactions between the elementary and the composite fermions,
as we now discuss.

\begin{figure}[!tp]
  \begin{center}
		\includegraphics[width=10cm]{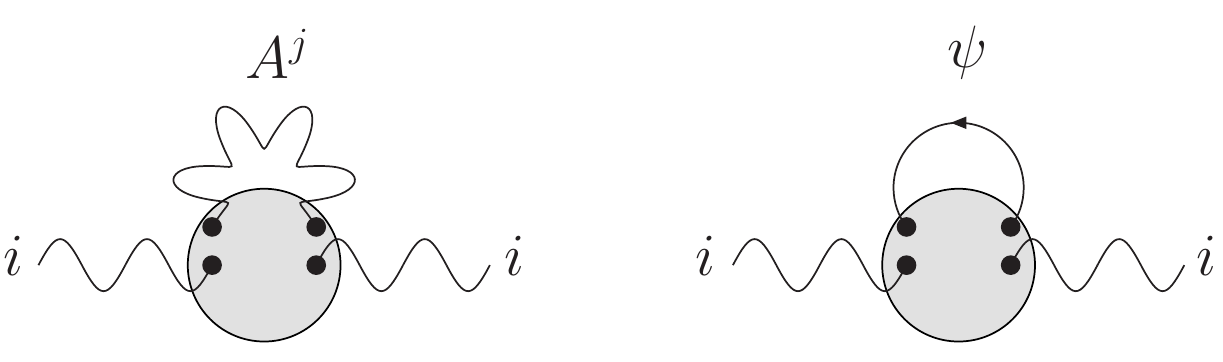}
  \caption{Example of sub-leading diagrams contributing to the differential running of the SM gauge couplings,
  on the left with a loop of elementary gauge bosons, and on the right with a loop of elementary fermions.}
  \label{elem-run}
  \end{center}
  \end{figure}

The interactions of the elementary fields with the composite sector break explicitly $\Gsym$ and thus their effect on the differential running must be quantified.
These are the SM gauge interactions of composite operators,
\eq{gaugemix},
as well as the fermion mixing terms,
\eq{fermionmix}.
The contribution of these interactions to the running can  be parametrized as \cite{Agashe:2005vg}
\beq
\frac{d}{d \ln \mu} \left( \frac{1}{\alpha_i} \right) \supset \frac{B^{comp}_{ij}}{2 \pi} \frac{\alpha_j}{4 \pi}
+ \frac{C^{comp}_{i\psi}}{2 \pi} \frac{\lambda_{\psi}^2}{16 \pi^2}~,
\label{subS}
\eeq
where $j$ is summed over SM gauge bosons, and $\psi$ over fermions. These are formally two-loop contributions, as shown in Fig.~\ref{elem-run},
but with unknown coefficients.
Since they are not universal, and not calculable a priori, they constitute an intrinsic theoretical uncertainty on unification in this scenario.\footnote{
Note that these
two-loop contributions can be interpreted as
threshold corrections
associated with the ultraviolet brane
in the warped extra-dimension picture.
They
can be explicitly computed by
integrating over the bulk, and they are enhanced by the logarithm of the ultraviolet-infrared hierarchy.}
These non-leading corrections can be as large as the leading ones if the mixing with the composite sector is large, as it is the case for the top
quark.

\subsection{Top compositeness and precision unification} \label{compotop}

Since the values of the SM gauge couplings $g_i$ are fixed by experiment,
the only couplings between the elementary and composite sectors that could
modify significantly the running
are the $\lambda_\psi$'s which,
in the framework of partial-compositeness, are
related to the Yukawa couplings
 as explained in section \ref{setup}.
Explicitly, below $\Lambda_c$ the couplings in \eq{fermionmix} generate, e.g. for a right-handed fermion $\psi_R$, the lagrangian
\beq
-\mathcal{L} \supset  (\lambda_{\psi_R}f) \overline{\psi_R} \Psi_L + M_{\psi_R} \overline{\Psi_R} \Psi_L
+ h.c. \, ,
\eeq
where $\Psi$ is a vector-like composite fermion (with the gauge quantum numbers of $\psi_R$)
that
arises as
 an excitation of the operator $\mathcal{O}_{\psi_{R}}$.
By diagonalizing the associated mass matrix, the massless SM fermion can be written as
$\psi_R^{SM} = \cos \theta_{\psi_R} \, \psi_R + \sin \theta_{\psi_R} \, \Psi_R$, with $\tan \theta_{\psi_R} = \lambda_{\psi_R}f / M_{\psi_R}$.
The fact that $M_{\psi_R} \simeq g_{\rho}f$ then leads to \eq{Yuk}.

The $\psi_R$ composite component becomes large when $\sin \theta_{\psi_R} \sim 1$, which requires a strongly coupled elementary field,
$\lambda_{\psi_R} \sim g_{\rho}$. Then,
the last term in \eq{subS} may become as large as a one-loop contribution:
\beq
\frac{d}{d \ln \mu} \left( \frac{1}{\alpha_i} \right) \supset \frac{C^{comp}_{i\psi_R}}{2 \pi} \frac{|\lambda_{\psi_R}|^2}{16 \pi^2}
\sim \frac{C^{comp}_{i\psi_R}}{2 \pi} \frac{g_\rho^2}{16\pi^2} \sim \frac{O(1)}{2 \pi}~,
\eeq
where in the last step we used
the rough strong coupling estimates $g_{\rho} \sim 4 \pi$
and
$C^{comp}_{i\psi_R} \sim O(1)$.\footnote{
The large-$N$ estimates would be $g_\rho \sim 4\pi/\sqrt{N}$
and $C^{comp}_{i\psi_R} \sim O(1) N$,
where $N$ is the number of colours of the strongly coupled theory (thought of as a QCD-like theory).}

Motivated by the large mass of the top quark, a natural possibility is to take the coupling of the right-handed top $t_R$ to be large,
thus making the SM top quark mostly composite.\footnote{
In general, the $t_L$ cannot be mostly composite, since gauge invariance would imply that $b_L$ is also mostly composite,
which is strongly disfavoured by measurements of the $Z b \bar{b}$ coupling.
However, this could be cured, as we will briefly review in section \ref{ewpt}, if the theory respects an extra parity symmetry \cite{Agashe:2006at}.
The alternative possibilities of mostly composite $b_R$ or
$\tau_R$ \cite{delAguila:2010vg} may also be interesting.}
In this case the distinction between composite and elementary fields becomes ambiguous:
the large $\Gsym$-violating coupling $\lambda_{t_R}$ introduces a large uncertainty in the prediction for unification.
The composite sector dynamics is significantly modified by $\lambda_{t_R}$, that cannot be treated as a small perturbation any longer.
To overcome this ambiguity, one is led to consider the possibility of full compositeness of $t_R$ as proposed in \cite{Agashe:2005vg},
that is a scenario with no elementary state
with the quantum numbers of $t_R$ in the low energy theory.
The role of the right-handed top is then played by a composite state,
denoted for simplicity $t_R$,
belonging to a chiral $\K$-multiplet
$T_R \equiv (t_R, x_R)$,
which is assumed to be massless before EWSB
(no partner $T_L$ exists).
Then, due to the unified $\Gsym$-symmetry of the composite sector above $\Lambda_c$, the contribution of $t_R$ must be subtracted from the
$\beta$-function coefficients $b_i^{elem}$.

If the low energy content of the elementary sector were just given by the SM without $\higgs$ and $t_R$, the ratio of $\beta$-function coefficients would be
$R_{SM-\higgs-t_R} \simeq 1.7$, which is closer to but still far from the experimental value.
However, a closer look at the composite-top scenario reveals that other chiral elementary fermions are needed to make the theory consistent.
We give below two independent arguments leading to this conclusion.
In particular, under well-motivated assumptions, we will show that a state $t'_L$ with SM charges $(\textbf{3},\textbf{1})_{\textbf{2/3}}$ should also be subtracted from the
differential running, so that the correct
expectation for unification turns out to be $R_{SM-\higgs-t_R-t'_L}\simeq 1.45$,
which is remarkably close to $R_{exp}$.\footnote{
The MSSM predicts $R_{MSSM}=1.4$, but this sharp agreement with experiment at one-loop level is deteriorated
when higher order corrections are included.}
In this scenario one may argue that precision unification is realized.

The first argument goes as follows. The elementary fermions of the complete GUT theory must belong to full $\Gsym_{GUT}$-multiplets.
The symmetry breaking at the GUT scale may well split the $\Gsym_{GUT}$-multiplets containing the elementary SM fermions,
giving a mass $\sim M_{GUT}$ to the elementary right-handed top, $t_R^{elem}$, but not to the other SM species. However,
the components which acquire a mass $\sim M_{GUT}$ must form vector-like pairs (or be Majorana fermions, e.g. sterile neutrinos).
Therefore,
the remaining massless states should form full $\Gsym_{GUT}$-multiplets
up to vector-like pairs of states.
The decoupling of each vector-like pair amounts to the subtraction of its contribution from the gauge coupling evolution.
In particular, $t_R^{elem}$ may decouple only if it pairs with an exotic fermion $t'_L$, and thus one achieves precision unification as described
in the previous paragraph.
Note that unification is enforced only by (\emph{i}) the Higgs and right-handed top compositeness, and (\emph{ii})
the $\Gsym$-symmetry of the composite sector.
In addition, exotic chiral fermions are predicted, in order to complete the $\Gsym_{GUT}$-multiplet of $t'_L$.

We remark that this first argument holds only when the GUT symmetry is fully realized in four dimensions. On the contrary, in extra-dimensional scenarios
where the GUT symmetry is realized in the bulk and it is broken explicitly on our brane by boundary conditions, the 4-dim chiral fermion zero-modes
do not need to fill $\Gsym_{GUT}$-multiplets (see e.g. Ref.~\cite{extradguts}).
In fact, in this case
there is no unified gauge symmetry on our 4-dim brane. In these scenarios it may still be sensible to study precision unification of the 4-dim gauge
couplings at $M_{GUT}$, since GUT scale thresholds can be kept small; then, one cannot appeal to the above argument to subtract $t'_L$ from
the running.

The second argument for the existence of exotic fermions \cite{Agashe:2005vg} applies when the $\K$-multiplet
containing the right-handed top also contains other states,
$T_R \equiv (t_R,x_R)$.
Then, the extra chiral fermions $x_R$ necessarily require conjugate partners,
in order to acquire a mass large enough to satisfy the experimental bounds. Also, such partners are needed to cancel the gauge anomalies, that were absent
with the SM fermion content, but would be generated by the chiral fermions $x_R$ alone.
Therefore, one must introduce exotic elementary fermions $x_L$, with the same charges of $x_R$
(see also Ref.~\cite{'tHooft:1979bh}).
It is equivalent, and perhaps more elegant,
to introduce a set of elementary exotic fermions $\{ t'_L,x_L \}$ that have the quantum numbers of a full $\K$-multiplet, with a lagrangian
\beq
-\mathcal{L} \supset  M_t \, \overline{t'_L} t_R^{elem}
+ \left(\lambda_{t'} \overline{t'_L}t_R+ \lambda_{x}\overline{x_L}x_R\right)f + h.c.  ~.
\label{trDec}
\eeq
The elementary $t'_L$ pairs with $t^{elem}_R$ making it super-heavy, with $M_t \sim M_{GUT}$,
the elementary $x_L$ pairs with the composite $x_R$ acquiring a
mass $m_{x} = \lambda_{x} f$, and the composite $t_R$ remains massless (neglecting the tiny mixing $\sim f/M_{GUT}$).

The exotic fermions contribute of course to the gauge coupling evolution, in a way that depends on  the choice of $\K$ and of the
$\K$-representation  containing $t_R$, \ie $T_R$.
However, when $\K$ is simple, the prediction for unification is univocal:
the composite fermions form full $\SU(5)$ representations, and so does the set $\{ t'_L,x_L \}$.
As a consequence, the addition of $x_L$ to the differential running  is equivalent to the subtraction of $t'_L$, realizing precise unification
as already discussed. Note that the case of simple $\K$ tallies with the first argument for exotic fermions.
When $\K$ is not simple, the prediction for unification becomes model-dependent. Precise unification can still be obtained, if the set of $x_L$
corrects appropriately the $\beta$-function coefficients, that is, if $R_{SM-\higgs-t_R+x_L}\simeq R_{exp}$.
We will come back briefly to this possibility at the end of section \ref{chain}.

A comment is in order on the field content of the elementary sector at the EW scale.
In general, composite-Higgs scenarios might have the potential to avoid the doublet-triplet splitting problem of
supersymmetric GUTs,
since $\higgs$ emerges from the composite sector at the EW scale, independently from the GUT symmetry breaking sector.
However, we have shown that to achieve unification one needs to introduce
light elementary fermions in split  $\SU(5)$ representations, contrary to the supersymmetric case.
We will see in section \ref{proton} that such splitting is needed also to prevent proton decay.

Finally, one may wonder if gauge anomalies constrain
the emergence of chiral fermions from the composite sector and/or the set of chiral fermions of the elementary sector.
In fact, if a $\Gsym$-symmetric
sector
that undergoes condensation
 were anomalous under the SM, one would predict that the composite spectrum
contains chiral fermions, because they must reproduce the anomaly  (see e.g. Ref.~\cite{'tHooft:1979bh}).
Of course such anomaly should be compensated by the elementary sector:
in particular, a composite $t_R$ must be compensated by the absence of $t_R^{elem}$, to recover the usual SM anomaly
cancellation. The composite sector may well be, instead, anomaly-free. In fact, this is automatically the case when its global symmetry $\Gsym$
contains a subgroup $\SO(10)\supset \Gsym_{SM}$.
Then, either the composite sector contains no chiral fermions at all, or it contains a set that is anomaly free (e.g. a full $\SO(10)$ representation).
In this case also the elementary sector should be anomaly free (e.g. the SM fermions plus a full $\SO(10)$ representation of
exotic fermions). In all cases, the composite-GUT scenario under consideration is consistent by construction,
since it has the SM chiral fermion content  plus a set of vector-like fermions  (both partially composite).

\section{Global symmetries of the composite sector \label{global}}

In this section we identify the global symmetries of the strongly interacting EWSB sector.
We begin with a review of the constraints coming from the electroweak precision tests (EWPTs), that will restrict the choice of the global symmetry group.
The reader already familiar with EWPTs can move directly to section \ref{chain}, where the minimal groups compatible with precision unification are classified.
We then confront with the constraints coming from baryon and lepton number conservation. Finally, having determined all the needed global symmetries,
we specify the quantum numbers of the exotic fermions associated with the composite top quark, and we estimate their contributions to EWPTs.

\subsection{Constraints from the electroweak precision tests \label{ewpt}}

In this section we shall briefly review the constraints on the EWSB sector
coming from EWPTs, specializing to the case of composite-Higgs models.
As customary, when evaluating extensions of the SM,
we shall resort to an effective field theory description,
following closely the analysis given in Ref.~\cite{Giudice:2007fh},
where the relevant higher dimensional operators have been studied
(we will use the same conventions).
These are generated by the strong dynamics, and in particular they generically arise through tree-level exchange of heavy resonances. Therefore, such operators will be suppressed by powers of $m_{\rho}$.
However, this naive estimate has to be refined with some further considerations:
(\emph{i}) The Higgs $H$ belongs entirely to the composite sector, and therefore it couples to it with strength $g_{\rho}$, so that
higher dimensional operators modifying the EW vacuum or the $\higgs$ properties shall be suppressed by powers of $g_{\rho}/m_{\rho} = 1/f$.
(\emph{ii}) The SM particles couple to the composite sector with strength dictated by partial-compositeness:
each gauge boson $A^i$ has coupling $g_i$, while each chiral fermion $\psi$ couples with strength $\lambda_{\psi}$.
This is particularly relevant for the composite right-handed top quark $t_R$, which couples with strength $\lambda_{t_R}=g_{\rho}$ as the Higgs;
the analysis of the viability and the consequences of a fully composite top quark has been presented in \cite{Pomarol:2008bh}.
(\emph{iii})
The low energy  particle content of the composite GUT models, below the scale $m_{\rho}$, is not that of the SM, since new light particles
arise as the $\K$-partners of both $\higgs$ and $t_R$.
The effects that these might have on precision observables are model dependent, and will be studied in section \ref{constraints},
after the relevant features of our scenario will have been settled.

We begin by recalling  that electroweak data strongly favour the presence of a light Higgs-like particle in the spectrum \cite{Barbieri:2004qk}.
The attempts to break the EW symmetry without a Higgs doublet, such as technicolour or Higgsless models, are generically difficult to reconcile with EWPTs, in particular due to large deviations in the Peskin-Takeuchi $S$ and $T$ parameters \cite{Peskin:1991sw}, with respect to the SM prediction.
This motivates a preference for composite-Higgs models, in particular those where $\higgs$ arises as a NGB from the strong dynamics,
since in this case a hierarchy between the EW scale and $m_{\rho}$ naturally arises \cite{Giudice:2007fh,Barbieri:2007bh}.

Next, let us motivate the requirement that the new physics should be
custodially symmetric, that is, it should (at least
approximately) respect the $\SU(2)_c$ custodial symmetry under which the three would-be
NGBs, eventually eaten by the EW gauge bosons, transform as a triplet \cite{Sikivie:1980hm}.
In our scenario, where EWSB is driven by the Higgs boson, this requirement translates into the requisite that the unbroken global symmetries of the composite sector, \ie the group $\K$, should contain a subgroup $\SO(4) \cong \SU(2)_L \times \SU(2)_R$, with $\higgs$ transforming in the representation
$\textbf{4} \cong (\textbf{2},\textbf{2})$. In this case, when the EW symmetry is broken by $\vev{\higgs} \propto {\rm diag}(v, v)$,
the $\SU(2)_c$ diagonal subgroup
of $\SU(2)_L \times \SU(2)_R$ remains unbroken.
The custodial symmetry is necessary to avoid large corrections to the tree level relation $m_W^2/ (m_Z \cos \theta_W)^2 = \rho = 1$,
that may arise from the operator
\beq
\frac{c_T}{2 f^2} | H^{\dag} \overleftrightarrow{D^{\mu}} H |^2~, \qquad   \widehat{T} \equiv \Delta \rho = - c_T \xi \, ,
\label{Tparam}
\eeq
with $c_T = O(1)$, as prescribed by NDA, and $\xi\equiv v^2/f^2$.
This operator can be generated purely by the strong dynamics, in the sense that $c_T$ does not vanish in the limit $g_{elem} \rightarrow 0$.
It is extremely constrained by present data: $-1.0 \lesssim 1000 \ \widehat{T}  \lesssim +2.1$,
where we projected on the $\widehat{T}$-axis the 95\% C.L. ellipse in the $\widehat{S}-\widehat{T}$ plane \cite{Nakamura:2010zzi}.
In order to minimize the fine-tuning in our scenario, \ie to allow for a higher value of $\xi$ compatible with the experimental constraints, the $\Delta \rho$
estimate given above implies that custodial symmetry must be imposed, because it automatically leads to $c_T = 0$.
Similar conclusions can be drawn by analyzing the corrections to $\rho$ due to $\SU(2)_c$-violating operators involving a composite
$t_R$, so that custodial symmetry emerges as a
generic requirement for the whole strong sector.\footnote{
Here we are neglecting model dependent contributions to $c_T$, due to couplings between the strong sector and the SM fields,
which depend on $g_{elem}$ and will be estimated later.}

The other major source of concern, especially in strong dynamics scenarios, comes from the $\widehat{S}$ parame\-ter.\footnote{
Other parameters associated to the EW gauge boson properties are higher order in the number of derivatives, and they typically
do not pose
strong constraints \cite{Barbieri:2004qk}.}
The composite sector will generically generate the operators
\beq
\frac{ic_W g}{2 m_{\rho}^2} (H^{\dag} \sigma_i \overleftrightarrow{D^{\mu}} H)(D^{\nu} W_{\mu \nu})^i + \frac{ic_B g'}{2 m_{\rho}^2}  (H^{\dag} \overleftrightarrow{D^{\mu}} H)(\partial^{\nu} B_{\mu \nu})~, \qquad \widehat{S} \equiv (c_W+c_B) \frac{m_W^2}{m_{\rho}^2} ~,
\label{Scont}
\eeq
with the NDA estimate $c_W,c_B = O(1)$.\footnote{
This estimate is actually confirmed in holographic composite-Higgs models \cite{newcompoHiggs,Contino:2006qr} or other strongly interacting EWSB models with hidden local symmetries,
see \cite{Piai:2010ma} for a review.}
The projection of the 95\% C.L. ellipse in the $\widehat{S}-\widehat{T}$ plane \cite{Nakamura:2010zzi} gives
$-1.7 \lesssim 1000 \ \widehat{S}  \lesssim +2.1$,
that leads to the constraint
$m_{\rho} \gtrsim 2.5 \TeV$.
For a benchmark value $f = 750 \GeV$, this
gives
$g_{\rho} \gtrsim 3.3$,
that lies within the window between $g_{elem}$ and $4 \pi$, the perturbativity limit for the coupling between resonances.
In other words, the bound on $\widehat{S}$ pushes to larger values of $g_{\rho}$ for a fixed $f$, which in turn
should not be much larger than $v$ to avoid fine-tuning.

In composite-Higgs models, both $\widehat{T}$ and $\widehat{S}$ receive an additional contribution, arising at one-loop level
because of the modified couplings of the Higgs to the gauge bosons. These couplings, which in the SM
are such that
$WW$ scattering is unitarized,
are suppressed due to the NGB nature of the Higgs boson. This leads to a mild sensitivity of the EW precision observables to the ultraviolet cut-off of
the effective lagrangian for the NGBs, $\Lambda \sim 4 \pi f / \sqrt{n_{NGB}} \sim m_{\rho}$ (as well as to the requirement of $WW$ scattering unitarization by massive vector resonances). The leading effect can be accounted for
by taking the SM expressions for $\widehat{T}$ and $\widehat{S}$ in the heavy Higgs approximation,
and replacing the Higgs mass $m_h$ with an effective mass \cite{Barbieri:2007bh}:
\beq
m_h^{eff} = m_h \left( \frac{\Lambda}{m_h} \right)^{\delta}.
\label{mheff}
\eeq
The exponent accounts for the modification of the  Higgs coupling to gauge bosons, that in the present scenario is given by $g_{hWW}=
\sqrt{1-\xi} \ g^{SM}_{hWW}$, so that we have $\delta=\xi$.
Then, one has
\beq
\Delta \widehat{T} \simeq - \frac{3 G_F m_W^2}{4 \sqrt{2} \pi^2} \tan^2 \theta_W \log \left[ \frac{m_h^{eff}}{m_h^{ref}} \right], \qquad
\Delta \widehat{S} \simeq \frac{G_F m_W^2}{12 \sqrt{2} \pi^2} \log \left[ \frac{m_h^{eff}}{m_h^{ref}} \right],
\label{TShiggs}
\eeq
where $G_F$ is the Fermi constant, $\theta_W$ the Weinberg angle, and $m_h^{ref} \simeq 117 \GeV$ is the reference Higgs mass
that we adopted to define the allowed experimental ranges for $\widehat{T}$ and $\widehat{S}$.

The last of the primary EWPTs that our scenario faces is the correction to the $Z b \bar{b}$ vertex.
Experimentally, the coupling of the $Z$ boson to the left-handed bottom quark current is known at the per mil level,
$-15 \lesssim 1000 \, \delta g_{b_L}/g_{b_L} \lesssim +2$
at the $2 \sigma$ level \cite{Kumar:2010vx}, if one allows for variations in $g_{b_R}$ (the coupling of the $Z$ to the right-handed bottom).\footnote{
Actually, the measured value of $g_{b_R}$ does not agree well with the SM prediction: the data on the forward-backward asymmetry
and the branching fraction of $Z$ into $b$'s suggest that $g_{b_R}$ should be larger than the SM value,
$g_{b_R}^{SM}
\simeq
 \sin^2\theta_W/3$,
by roughly $20\%$:
the best fit is given by $\delta g_{b_R}\simeq 0.016$,  with a $2\sigma$ range  $\delta g_{b_R} \in (0.000,0.030)$
\cite{Kumar:2010vx}.
This is the interval we adopted to determine the allowed range for $\delta g_{b_L}$.
If instead $\delta g_{b_R} = 0$ were enforced, the bound
would become $0 \lesssim 1000 \, \delta g_{b_L}/g_{b_L} \lesssim +2$ at the $2\sigma$ level.}
This suggests the introduction of a symmetry in the EWSB sector, and in its couplings to the elementary SM particles, in order to prevent large corrections to
$g_{b_L}$.
Such a symmetry has been identified in \cite{Agashe:2006at}:
the composite operator coupled to $b_L$ must transform under $\SU(2)_L \times \SU(2)_R$
as an eigenstate of the parity $P_{LR}$ exchanging $\SU(2)_L$ with $\SU(2)_R$, that is to say, it has $T_3^R=T_3^L$.
This possibility is realized in one of the models
we shall consider later. In this case the corrections to $g_{b_L}$ due to the composite sector are absent, more precisely $\delta g_{b_L} = 0$
at tree-level and for zero transferred momentum \cite{Agashe:2006at}.

If one cannot enforce such symmetry protection mechanism, strong constraints on the parameters of the model
 come from the limits on $\delta g_{b_L}/g_{b_L}$.
This effect can be parametrized by the higher dimensional operators
\beq
\frac{ic_L^{(1)}}{m_{\rho}^2} \left(H^{\dagger} D^{\mu} H\right)\left(\overline{q_L} \gamma^{\mu} q_L\right)
+ \frac{ic_L^{(3)}}{2 m_{\rho}^2} \left(H^{\dagger} \sigma_i D^{\mu} H \right)\left( \overline{q_L} \gamma^{\mu} \sigma^{i} q_L\right) + h.c. ~, \qquad
\frac{\delta g_{b_L}}{g_{b_L}} = \frac{ \left( c_L^{(1)} + c_L^{(3)} \right) \xi}{
2 \left(1-\frac{2}{3} \sin^2 \theta_W\right)
 g_\rho^2} ~,
\label{ZbbL}
\eeq
where $q_L$ is the top-bottom quark doublet.
The coefficients $c_L^{(1),(3)}$ depend on the coupling of $q_L$ to the composite sector, so that NDA gives $c_L^{(1),(3)} = \lambda_{q}^2 \times O(1)$.
Assuming no cancellations are present, the estimate $\delta g_{b_L}/g_{b_L} \sim (\lambda_{q}/g_{\rho})^2 \xi$
puts a strong bound on the degree of compositeness of $q_L$ and/or on $\xi$. Actually, since $t_R$ is fully composite in our scenario, $\laq \simeq y_t$ is determined by the requirement
to reproduce the observed top Yukawa. This implies a bound $m_{\rho} \gtrsim (1.4,3.9) \TeV$,
depending on the sign of the correction, which we cannot predict. This bound therefore can be even stronger than the one from $\widehat{S}$.
Nevertheless, the absence of a protection mechanism remains an open possibility.
As a matter of fact, the data on $g_{b_R}$ suggests that beyond the SM physics might affect significantly the SM fit for $g_{b_L}$
(see for instance Ref.~\cite{Kumar:2010vx}).

When the $P_{LR}$ symmetry introduced before is adopted,
the tree-level contribution to $\delta g_{b_L}$ vanishes ($c_L^{(1)} = -c_L^{(3)}$),
and we only expect loop corrections to give a new physics contribution of the order
$\delta g_{b_L}/g_{b_L} \sim (\delta g_{b_L}/g_{b_L})^{SM} (y_t/g_{\rho})^2 \xi$, where $(\delta g_{b_L}/g_{b_L})^{SM}$ is the SM top-loop contribution. This correction is safely below the experimental precision.
 Further loop corrections
 associated to $t_R$ compositeness are under control \cite{Pomarol:2008bh}.

In addition to the deviations from the SM predictions for $\widehat{T}$, $\widehat{S}$ and the $Z b \bar{b}$ coupling, also the couplings of $t_R$ will be significantly modified, due to its composite nature. However, there are no stronger constraints from present data.
Such deviations could be observed (and eventually top-compositeness could be discovered) in the near future \cite{Pomarol:2008bh}, by inspection of early LHC data.

\subsection{Minimal global symmetry breaking patterns} \label{chain}

The discussion of the previous sections leads to the following requirements on the global symmetry breaking pattern
$\Gsym \rightarrow \K$ of the composite sector:
\begin{description}
  \item[]\emph{i)} $\Gsym \supset \SU(5)$, in order to avoid leading order contributions to the differential running of the SM gauge couplings,
that would spoil the calculability of unification.
  \item[]\emph{ii)} $\K \supset \K_{min} \equiv \SU(3)_C \times \SU(2)_L \times \SU(2)_R \times \U(1)'$.
The $\SU(2)_R$ factor is needed to maintain a residual custodial symmetry after EWSB, while the extra abelian factor $\U(1)'$ is necessary
to properly embed
the hypercharge gauged group $\U(1)_Y$. In fact, the simplest embedding $Y=T^R_3$ turns out to be incompatible with the required hypercharges
of composite fermions (which mix with the SM elementary ones).
  \item[]\emph{iii)} The broken generators in $\Gsym/\K$ must include a $(\textbf{1},\textbf{2},\textbf{2})_\textbf{0}$ multiplet of $\K_{min}$,
  which corresponds to the NGBs with the quantum numbers of the Higgs doublet.
\end{description}

With these requirements, the rank of $\Gsym$ should be equal or larger than $5$. We find that there are only three possibilities with rank 5, listed in Table \ref{breakings}.
\begin{table}[t]
\begin{center}\begin{tabular}{|c|c|c|c|}
\hline &&&\\
&$\Gsym \rightarrow \K$ & $R_{NGB}$ & $\Sigma_i R^i_{\K_{min}}$ \\ &&&\\ \hline &&&\\
(\emph{a}) &$ \SO(11) \rightarrow \SO(10)$ & $\textbf{10}$ & $(\textbf{1},\textbf{2},\textbf{2})_{\textbf{0}} + (\textbf{3},\textbf{1},\textbf{1})_{-\textbf{2/3}} + (\bar{\textbf{3}},\textbf{1},\textbf{1})_{+\textbf{2/3}}$ \\ &&&\\
(\emph{b}) &$ \SO(11) \rightarrow \SO(7) \times \SU(2) \times \SU(2)$ & $(\textbf{7},\textbf{2},\textbf{2})$ & $(\textbf{1},\textbf{2},\textbf{2})_{\textbf{0}} + (\textbf{3},\textbf{2},\textbf{2})_{-\textbf{2/3}} + (\bar{\textbf{3}},\textbf{2},\textbf{2})_{+\textbf{2/3}}$ \\ &&&\\
(\emph{c}) &$\Sp(10) \rightarrow \Sp(8) \times \SU(2)$  & $(\textbf{8},\textbf{2})$ & $(\textbf{1},\textbf{2},\textbf{2})_{\textbf{0}} + (\textbf{3},\textbf{1},\textbf{2})_{-\textbf{2/3}} + (\bar{\textbf{3}},\textbf{1},\textbf{2})_{+\textbf{2/3}}$ \\
&&& or \\ &&& $(\textbf{1},\textbf{2},\textbf{2})_{\textbf{0}} + (\textbf{3},\textbf{2},\textbf{1})_{-\textbf{2/3}} + (\bar{\textbf{3}},\textbf{2},\textbf{1})_{+\textbf{2/3}}$
\\&&&\\ \hline
\end{tabular}
\end{center}
    \caption{Symmetry breaking patterns $\mathcal{G} \rightarrow \mathcal{K}$ satisfying the requirements \emph{i)}-\emph{iii)}, with the minimal restriction rank$(\Gsym)=5$. 
The NGB representations under $\mathcal{K}$ and $\mathcal{K}_{min}$ are reported in the third and fourth column, respectively.}
      \label{breakings}
\end{table}
We indicated with $R_{\rm NGB}$ the $\K$-representation of the broken generators in $\Gsym/\K$, \ie of the NGBs of the composite sector.
In the last column we provided the decomposition of $R_{NGB}$ under $\K_{min}$, with an arbitrary normalization of the $\U(1)'$ charges
(corresponding to the standard $B-L$ embedding, for the case $\K=\SO(10)$).
Note that in option $(\emph{c})$, $\SU(2)_L$ may or may not be identified
with the $\SU(2)$ factor external to $\Sp(8)$.

The group $\K$ indicated in Table \ref{breakings} is the maximal subgroup of $\Gsym$ satisfying the requirements \emph{i)}-\emph{iii)}, but in all the
three options the unbroken subgroup could actually be as small as $\K_{min}$.
Of course, if a non-maximal $\K$ is chosen, further NGBs appear, besides those listed in Table \ref{breakings}.
We expect the possibilities with rank$(\Gsym)>5$ to be  generalizations of these three cases with no qualitatively new features.\footnote{
Note that when the rank is larger than 5, the possibility appears of semi-simple groups of the kind $\SU(4)_A \times \SU(4)_B \times P_{AB}$ or $\SU(3)^3 \times Z_3$,
where unification is enforced by a permutation symmetry.}

The hypercharge of composite states depends on the embedding of $\U(1)_Y$ into $\SU(2)_R \times  \U(1)'$,
which is given in general by $Y = \alpha \, T^R_3 + \beta \, Y'$.
To enforce custodial symmetry in the simplest possible way, we asked for the Higgs doublet $\higgs$ to transform as $(\textbf{2},\textbf{2})_{\textbf{0}}$
under $\SU(2)_L \times \SU(2)_R \times \U(1)'$, so that one needs $\alpha = \pm 1$ to obtain $Y(\higgs) = \pm 1/2$,
while $\beta$ can be determined only by an additional requirement.
In particular, the hypercharge of the colour triplet NGBs, $\triplet$, that are present for all three breaking patterns  (\emph{a})-(\emph{c}),
is not fixed in general.

Consider for definiteness the case $\K = \SO(10)$ with the normalization $Y'=\BL$, that is, the customary $B-L$
generator within the global SO(10) symmetry, not to be confused with the $B-L$ symmetry of the SM.
If one requires the SM fermion quantum numbers to fit into a spinorial $\textbf{16}$ representation of $\SO(10)$, one finds two solutions:
$Y = T^R_{3} + Y'/2$ (standard $\SU(5)$ embedding into $\SO(10)$) or $Y=-T^R_3+Y'/2$ (flipped $\SU(5)$).
Then the colour triplet NGBs have $Y(T)=\pm1/3$.
In the following, we will also consider a different possibility, that is,
to embed $t_R$ into a $\textbf{10}$ representation of $\SO(10)$. In this case the two available solutions are
$Y = \pm T^R_{3} - Y'$ and the colour triplet NGBs have $Y(T)=\pm 2/3$.

Besides the light NGB scalar resonances, listed in Table \ref{breakings}, in the composite-$t_R$ limit one expects in general light fermionic resonances,
corresponding to the partners of $t_R$ filling a $\K$-multiplet, and to the exotic elementary fermions that pair with them
to form vector-like massive states. In principle, these fermion states can be absent all together, if the SM state $t_R\sim (\textbf{3},\textbf{1})_{\textbf{2/3}}$
forms a full $\K$-multiplet by itself. This requires $\K=\K_{min}$, $t_R \sim (\textbf{3},\textbf{1},\textbf{1})_{Y'(t_R)}$ and $Y(t_R) = \beta \, Y'(t_R)=2/3$.
In all other cases exotic fermions are needed, with quantum numbers determined by the choice of the symmetry breaking pattern
and of the $\K$-multiplet $T_R$ containing $t_R$.

As discussed in section \ref{compotop},
the set of exotic fermions determines the fate of gauge coupling unification. A sufficient condition to realize it accurately is to take $\K$ simple.
This is the case only for the symmetry breaking pattern $(\emph{a})$, with the largest possible unbroken subgroup, $\K=\SO(10)$.
This is the model whose phenomenology we will study in detail, motivated by unification.\footnote{
Besides, the coset $\SO(11)/\SO(10)$ has the advantage of being the smallest coset of our list,  with $n_{NGB}=10$.
This is a desirable feature, since the ultraviolet cut-off of our effective lagrangian for the NGBs is $\Lambda \sim 4 \pi f / \sqrt{n_{NGB}}$.}

When $\K$ is not simple, one can still hope to realize unification, if $R_{SM-\higgs-t_R+x_L}\simeq R_{exp}$,
where $x_L$ is the set of exotic fermions.
To realize the latter condition, one should carefully choose
$\K$ and the $\K$-multiplet $T_R$.
First, notice that in the SM $\alpha_1$ and $\alpha_2$ meet at $\sim 10^{13}$ GeV,
which is too early to unify with $\alpha_3$ as well as to prevent gauge-mediated proton-decay (see section \ref{proton}).
To delay unification one needs
a correction $\Delta(b_2-b_1)>0$. If $\K$ contains $\SU(2)_L$ as an isolated factor,
$T_R$ must be a singlet of $\SU(2)_L$
in order to contain $t_R$, and thus exotic fermions
do not contribute to $b_2$.
In order to increase $b_2$, one should resort to models where $\SU(2)_L$ is part of a larger simple group,
$\SU(2)_L \subset \K_L \subset \K$, and
$T_R$
contains $\SU(2)_L$-non-singlet components.
This is possible only for the symmetry breaking pattern $(\emph{c})$, with the maximal unbroken subgroup $\K=\Sp(8) \times \SU(2)$
and with $\SU(2)_L \subset \Sp(8)$.
However, inspecting small $\K$-representations, we did not find one that contains $t_R$ and leads to reasonably good unification:
the required contribution to $b_2$ is compensated by large contributions to $b_1$ and $b_3$, that go in the wrong direction.
We will not study further the case with $\K$ non-simple in the following.

\subsection{Constraints from proton stability \label{proton}}

As usual in GUTs, the stability of the proton is endangered by baryon number violating interactions. These manifest in the SM effective lagrangian as higher dimensional operators suppressed by the scale $\Lambda_B$ of baryon number violation.
In the elementary sector, this scale is the mass of GUT gauge bosons, $\Lambda_B \sim M_{GUT}$.
The interactions with the composite sector, however, may violate baryon number
(and lepton number) at much smaller scales $\sim m_{\rho}$, thus invalidating the whole program of composite GUTs.
Let us briefly comment, first, on the usual gauge-mediated proton decay at $M_{GUT}$,
and next move to the requirements to be imposed on the composite sector.

We explained how the differences of gauge coupling $\beta$-function coefficients, $b_i-b_j$,
which control the differential running at leading order, can be determined thanks to
the global symmetry $\Gsym$ of the composite sector.
This not only allows the prediction for $\alpha_3$ from the experimental values of $\alpha_{1,2}$,
but also fixes the value of the scale $M_{GUT}$, where $\alpha_1$ and $\alpha_2$ meet.
With our recipe for precision unification, that is, with an elementary field content given by $SM-\higgs-t_R-t'_L$,
one finds
$M_{GUT}  \simeq 6\cdot 10^{14}$ GeV,
which is a factor of $\sim 7$ smaller than the lower bound on GUT gauge boson masses in the minimal $\SU(5)$ model,
$M_V\gtrsim 4\cdot 10^{15}$ GeV.\footnote{
In the MSSM $M_{GUT}\simeq 2\cdot 10^{16}$ GeV,
however supersymmetry enhances Higgs-mediated p-decay (since it is induced by dimension 5 operators),
thus requiring an additional suppression.}
This gap shall be cured
either by two-loops corrections, that may be enhanced by the strong dynamics, or by GUT thresholds, or by
special structures of the Yukawa couplings leading to cancellations in the proton-decay operators (these
can relax the lower bound on $M_V$ by more than one order of magnitude \cite{Dorsner:2004xa}).
Also, if GUT breaking is realized in extra-dimensions by orbifolding, proton-decay operators can be forbidden or,
more in general, their structure can be significantly different \cite{extradguts}. We do not elaborate more on these issues, since we do not control the strong dynamics at the two-loop level
and we do not specify the theory at the GUT scale in this paper.

Let us now discuss possible low-energy sources of baryon and lepton number violation.
While in the SM $B$ and $L$ are accidental symmetries
of the renormalizable lagrangian
(due to the gauge symmetry and the SM field content),
in general the composite states may mediate $B$- and $L$-violating processes.
This is particularly worrisome in view of the unified $\Gsym$-symmetry of the composite sector, that will contain states with
the charges of the SM ones together with their $\Gsym$-partners.
Therefore, one a priori expects e.g. resonances at the scale $m_{\rho}$
with the same quantum numbers of the $\SU(5)$ gauge bosons $V$.
The couplings of these resonances to the SM elementary fermions
(in particular those contained in the proton), as given in \eq{fermionmix}, will induce $B$-violating operators of the form
\beq
\frac{\lambda_{i} \lambda_{j} \lambda_{k} \lambda_{l}}{m^2_{\rho}} \, \overline{\psi_i} \psi_j \overline{\psi_k} \psi_l \, \sim \frac{\sqrt{y_i y_j y_k y_l}}{f^2}\, \overline{\psi_i} \psi_j \overline{\psi_k} \psi_l ~,
\eeq
where we used the rough relation $\lambda_i \sim \sqrt{y_i  g_{\rho}}$, in order to illustrate that these operators cannot be
sufficiently suppressed, if one wants to properly reproduce the SM Yukawa couplings.\footnote{
Even if composite resonances coupling directly to the four elementary fermions are absent,
$B$-violating operators can  be generated through non-perturbative effects by the strong sector,
suppressed generically by the scale $\Lambda_c$.
 This suppression is again far too small.}
Similarly, the dimension-5 operator $\lambda_l^2 l_il_jHH/m_\rho$ could be induced, that would violate $L$,
generating Majorana neutrino masses far too large with respect to the observed ones.

One is then forced to postulate additional symmetries, to prevent too large $B$ and $L$ violations.
Note that additional symmetries are also required in supersymmetric models: in particular, with the particle
content of the MSSM, one needs to impose $R$-parity in order to forbid $B$ and $L$ violating dim-4 interactions;
this is sufficient in such weakly coupled theories, because one can assume that higher dimensional operators are generated only
at scales much larger than the EW scale. In the composite GUT scenario, the extra symmetry should forbid also, at the very least,
the dimension-5 and -6 operators.
This extra ``matter" symmetry,
denoted by $\Gsym_M$, shall be part of the global symmetries of the composite sector, either within the simple unified group,
$\Gsym_M \subset \Gsym$, or factored out, $\Gsym \times \Gsym_M$. Besides, it should be extended consistently to the elementary sector,
that is, it must be (to very good approximation) a symmetry
of the whole effective lagrangian in \eq{Lag}, and it should be preserved up to scales $\sim M_{GUT}$.
In other words, while $B$ and $L$ are accidental symmetries of the SM alone at low energy, the symmetry $\Gsym_M$
must be imposed by hand on the couplings of the SM elementary fields to the composite sector.

To see concretely how proton decay arises and how it can be forbidden, let us consider the simple case of $\K=\SU(5)$, left unbroken at $\Lambda_c$.
If the Higgs doublet $\higgs$ belongs to a composite multiplet $\textbf{5} \sim (\higgs,\triplet)$, the coloured triplet $\triplet$ can mediate
proton-decay, as it is well-known.
Clearly, there is no generator internal to $\K$ that can prevent the couplings of $\triplet$ to SM fermions,
without preventing at the same time the required couplings of $\higgs$.
Therefore, one is forced to introduce an extra
global symmetry external to $\K$.
The most obvious option is $\U(1)_B$
with (\emph{i}) the usual $B$-assignments of elementary fermions,
(\emph{ii}) $B(\higgs)=B(\triplet)=0$, as required since $\higgs$ and $\triplet$ belong to the same $\K$-multiplet;
(\emph{iii}) $B(T_R)=1/3$, in order to allow for the Yukawa coupling of
$t_R \in T_R$.\footnote{
Alternative choices could be e.g. $\U(1)_{3B+L}$,
which may be inspired by Pati-Salam unification
and allows for proton-decays into three leptons
(these require operators with dimension larger than 6 and may be sufficiently suppressed \cite{Gripaios:2009dq}).
In principle, even a discrete symmetry could be enough to guarantee the proton stability.}

The same reasoning can be applied to the realistic case $\K=\SO(10)$
(we saw that $\K=\SU(5)$ is too small to accommodate the custodial symmetry),
imposing a $\U(1)_B$ symmetry external to $\SO(10)$ (and therefore also to SO(11)).
However, in the $\SO(10)$ case there is a second, slightly more economical, possibility:
to identify the action of $B$ on the composite states with the action of one of the $\K$ generators. In fact, once the embedding of the SM subgroup
$\SU(3)_C \times \SU(2)_L \times \U(1)_Y$
into $\K$ is chosen,
one is left with one independent linear combination of the generators in the Cartan subalgebra
of $\SO(10)$, that is an independent $\U(1)_{X}$ symmetry,
 and one can take
$B(\phi_{comp})=X(\phi_{comp})$ for any composite field $\phi_{comp}$.
Since $B$ acts on elementary fermions as the usual baryon number, one needs $X(\higgs)=0$.
Recalling that the NGBs $(\higgs,\triplet)$ form a $\textbf{10}$ representation of SO(10), a viable choice is $X=Y'$.
In particular $B(T)=-2/3$.

As already mentioned, one also needs a symmetry to prevent neutrino masses larger than $\sim$ eV, as required by experiments,
that is, the composite sector should respect in good approximation lepton number. We do not repeat a detailed discussion
analogous to the one of baryon number, and we just adopt the most straightforward possibility: a $\U(1)_L$ global symmetry
with the usual $L$-assignments of elementary fermions and $L(\higgs)=L(\triplet)=L(T_R)=0$. Note that the whole composite sector
cannot be neutral under lepton number, since at least the charged leptons are acquiring their mass by mixing with
composite operators.

In summary, we assume that the full effective lagrangian in \eq{Lag},
respects a global symmetry $\Gsym_M = \U(1)_B \times \U(1)_L$.
In particular, the gauge and Yukawa interactions between the elementary and the composite sector generically do break  $\K$,
but must preserve both the $\Gsym_{SM}$ gauged subgroup, and the global $\Gsym_M$ symmetry.

Let us comment on the connection between the symmetry $\Gsym_M$ and the UV completion of our scenario at $M_{GUT}$.
It seems that the simplest possibility is to take $\Gsym_M$ external to the unified gauge group $\Gsym_{GUT}$,
since $\Gsym_{GUT}$ by itself always allows for $B$ (and $L$) violating operators  in the elementary sector.
This would imply, in particular, that elementary fermions with different $B$ shall belong to different
$\Gsym_{GUT}$ multiplets: a splitting of these multiplets is required at the GUT scale, with only the components with the correct $B$ remaining massless.
Interestingly, this is specular with respect to the case of supersymmetric GUTs, where SM fermions fill full $\SU(5)$ multiplets,
but a doublet-triplet splitting is necessary in the Higgs sector, to prevent proton decay.\footnote{Exceptions are possible:
a supersymmetric GUT model with no doublet-triplet was proposed in Ref.~\cite{Dvali:1995hp}.}
In composite GUTs, the doublet-triplet splitting issue appears to be transferred to the matter sector.
Still, there may be more subtle (and economical) ways to obtain the $\Gsym_M$ symmetry in the effective lagrangian, from the initial
$\Gsym_{GUT}$ symmetry of the full theory.
In a sense, this issue requires non-trivial model-building at the GUT scale, as it was the case historically
for doublet-triplet splitting in supersymmetric GUTs.

\subsection{Exotic fermions in the $\SO(11)/\SO(10)$ scenario} \label{exotics}

Let us recall that, in our scenario, the right-handed top $t_R$ is fully composite and it is part of the
$\K$-multiplet
$T_R = (t_R,x_R)$,
that includes in general other composite chiral fermions $x_R$.
The composite sector does not contain a left-handed partner of
$T_R$,
because $t_R$ should remain massless before EWSB.
Therefore, one needs to introduce exotic elementary fermions $x_L$, which form vector-like pairs with $x_R$ and cancel their anomalies.
In addition, when $\K$ is simple the contribution of $x_L$ to the gauge coupling evolution leads to precision unification,
as described in section~\ref{compotop}.

The exotic fermions $x$ should acquire a vector-like mass from the couplings $\lax$ between their composite and elementary chiral components.
These couplings (\emph{i}) should respect the symmetry $\Gsym_M$, to suppress baryon and lepton number violations;
(\emph{ii}) should be sufficiently large to avoid experimental lower bounds
on the masses of exotic fermions; (\emph{iii}) are constrained by EWPTs, analogously to the couplings $\lambda_\psi$ of the
SM fermions $\psi$ to the composite sector.
In this section we analyze the possible quantum numbers of exotic fermions
for the case that we are going to study phenomenologically: $\Gsym=\SO(11)\rightarrow \K=\SO(10)$.
In the various cases, we will also specify
the fermionic operators $\mathcal{O}_{\psi}$ coupling to the SM fermions.
These operators, besides containing a component with the $\Gsym_{SM}$ quantum numbers of $\psi$, must have a Yukawa-like
SO(10) invariant coupling to the $\textbf{10}$ containing the Higgs doublet.

The exotic charges depend on the choice of the $\SO(10)$ representation that contains the SM state $t_R \sim (\textbf{3},\textbf{1})_{\textbf{2/3}}$.
In order to illustrate the relevance of such choice for our results, we will compare the two simplest possibilities,
\bea
&(\emph{1})& T_R \sim \overline{\textbf{16}} = (\bar{\textbf{3}},\textbf{2},\textbf{1})_{-\textbf{1/3}} + (\textbf{1},\textbf{2},\textbf{1})_{\textbf{1}} + (\textbf{3},\textbf{1},\textbf{2})_{\textbf{1/3}}+(\textbf{1},\textbf{1},\textbf{2})_{-\textbf{1}} ~, \label{comp16}\\
&(\emph{2})& T_R  \sim \textbf{10} = (\textbf{1},\textbf{2},\textbf{2})_{\textbf{0}} + (\textbf{3},\textbf{1},\textbf{1})_{-\textbf{2/3}} + (\bar{\textbf{3}},\textbf{1},\textbf{1})_{\textbf{2/3}} ~, \label{comp10}
\eea
where the decomposition under the $\SO(10)$ subgroup $\SU(3)_C \times \SU(2)_L \times \SU(2)_R \times \U(1)'$ is provided (recall that
$Y' = (B-L)_{\SO(10)}$).\\

\noindent \textbf{Case (\emph{1})} corresponds to the usual $\SO(10)$ embedding of all SM fermions in a \textbf{16} multiplet, which is realized choosing for the hypercharge
the linear combination
$Y = \pm T^R_{3} + Y'/2$,
with $t_R$ corresponding to the component  of $(\textbf{3},\textbf{1},\textbf{2})_{\textbf{1/3}}$ with $T_{3}^R = \pm1/2$.
Then, the exotic elementary fermions $x_L$ have SM quantum numbers
\beq
(\emph{1})~~~
 \qx_L = (\bar{\textbf{3}},\textbf{2})_{-\textbf{1/6}}, \; \lx_L = (\textbf{1},\textbf{2})_{\textbf{1/2}}, \; b'_L = (\textbf{3},\textbf{1})_{-\textbf{1/3}},
\; \nu'_L= (\textbf{1},\textbf{1})_{\textbf{0}}, \; e'_L = (\textbf{1},\textbf{1})_{-\textbf{1}} ~,
\label{xL1}
\eeq
where $f'_L$ denotes a $\SU(2)_L$ singlet, not to be confused with the SM fermion $f_L$, which is a component of a $\SU(2)_L$ doublet.
Thus, these exotic fermions have the SM quantum numbers of a full vector-like fourth generation, except for the singlet top quark.

However, the symmetry $\Gsym_M$
differentiates them from an actual fourth generation, because it assigns to them exotic baryon and lepton numbers,
and thus it constrains the allowed couplings of the exotic fermions with the SM elementary fields and with the composites, in particular
$\higgs$ and $\triplet$.
As explained in section~\ref{proton}, we consider two possible assignments of baryon number,
either a group $\U(1)_{B_E}$ external to $\SO(11)$, or a group $\U(1)_{B_I}$ internal to $\SO(10) \subset \SO(11)$,
more precisely identified with the $Y' = (B-L)_{\SO(10)}$ generator
(the $\U(1)_L$ symmetry defined in section \ref{proton} is implicitly assumed everywhere, too).
\begin{itemize}
\item
In the case $\U(1)_{B_E}$, $B(x_R)=B(t_R)=1/3$ and,
in order to form vector-like states, the exotic fermions should all have $B(x_L)=1/3$.
All elementary fermions, $\psi = \{ x_L, q_L, b_R, l_L, e_R \}$ couple to the strong sector through operators transforming
under $\SO(11) \times \U(1)_{B_E}$ as $\textbf{32}_{B(\psi)}$ (which decomposes under SO(10) as $\textbf{32} = \textbf{16} + \overline{\textbf{16}}$).
In particular, $q_L$ mixes with a component with quantum numbers $(\textbf{3},\textbf{2},\textbf{1})_{\textbf{1/3}}$ under $\K_{min}$,
so that a Yukawa coupling with $\higgs \sim (\textbf{1},\textbf{2},\textbf{2})_{\textbf{0}}$ and $t_R \in (\textbf{3},\textbf{1},\textbf{2})_{\textbf{1/3}}$ is allowed.
Notice that both the exotics $x_L$ and $q_L$ could couple to the same composite operator, $\mathcal{O}_{(q,x)_L}$.
In this case the set of couplings $\lax$ and $\laq$ evolve to their low-energy values all with the anomalous dimension of $\mathcal{O}_{(q,x)_L}$. Therefore, if their ultraviolet values
do not present large hierarchies,
they are expected to be of the same order also at low energy.
Apart from the case of $x_L$ and $q_L$, each elementary fermion
mixes with a different composite operator, because of $B$ conservation. Also,
$\U(1)_B$ (along with the $\Gsym_{SM}$ gauge invariance) implies that the exotic fermions do not mix with the SM ones,
except for $b_R$ that can pair with $b'_L$ (they have the same $\Gsym_{SM} \times \U(1)_B$ charges).
The potential consequences of this mixing are discussed below.
\item
In the case $\U(1)_{B_I}$, $B(t_R)=1/3$ is different from the $t_R$ partners baryon number:
$B(\qx) = -1/3$, $B(\lx) = 1$, $B(b') = 1/3$, $B(\nu') = B(e')= -1$. The elementary fermions
$x_L$, $q_L$, and $b_R$ may all couple to the same operator in the \textbf{32} representation of SO(11).\footnote{
In this case, in order to reproduce the top-bottom mass difference, one is forced to take $\lambda_{b_R}\ll \laq$ at high energies.
In alternative, one may forbid  the coupling $b_R\mathcal{O}_{(q,x)_L}$ e.g. with an extra chiral $\U(1)$ symmetry,
and couple $b_R$ to an independent \textbf{32} operator; then the mass hierarchy can follow
from the different running of $\laq$ and $\lambda_{b_R}$ to low energies.}
The leptons cannot couple to \textbf{32}, because the states with their SM gauge charges carry baryon number.
We find that $l_L$ can couple, instead, to $\textbf{11} = \textbf{10} + \textbf{1}$, while $e_R$ couples to $\textbf{55} = \textbf{45} + \textbf{10}$,
allowing for the charged lepton Yukawa.
As for the case of $\U(1)_{B_E}$, there is a possible mixing between the elementary fermions $b_R$ and $b'_L$.
Also $q_L$ and $\qx_L$ can mix, since they have conjugate charges under $\Gsym_{SM} \times \Gsym_M$, but
this mixing can be forbidden by an extra chiral $\U(1)$ symmetry, as we assume in the following.
\end{itemize}

Note that, in the absence of $\U(1)_B$, the elementary exotics $x_L$ would pair, generically, not only with the composite states $x_R$,
but also with the elementary fermions with the same $\Gsym_{SM}$ quantum numbers, which we denote with $\psi_R$:
\beq
-\mathcal{L} \supset  m_x \overline{x_L} x_R + M_x \overline{x_L}  \psi_R  + h.c. \,.
\eeq
While $m_x = \lax f$, the elementary mass term $M_x$ can be as large as $M_{GUT}$ and this would
decouple the pair $(\psi_R,x_L)$, leaving as massless SM fermion the composite state $x_R$,
as it happens for $t_R$ (see \eq{trDec}).
The symmetry $\U(1)_B$, which was introduced to make the proton stable, comes to help for this independent issue:
consider for definiteness  the case with $\U(1)_{B_E}$, where $B(x_R)=B(x_L)=1/3$ and therefore
$M_x$ is forbidden for $B(\psi_R) \ne 1/3$.
Inspecting \eq{xL1},
only the elementary $b_R$ can
pair with the exotic $b'_L$ and may decouple with a large mass $M_b$,
leaving a mostly composite $b_R$.
This special property of $b_R$ is due to
its embedding with $t_R$ into a doublet of $\SU(2)_R$.
At this stage, case (\emph{1}) seems to predict that the composite $t_R$ is accompanied by
a composite $b_R$.
Unfortunately, this possibility is
not viable, since it would imply $y_b = y_t$.\footnote{
Even if this equality is avoided with some extra model-building, in the case of composite right-handed bottom the elementary pair $(b^{elem}_R,b'_L)$
should be subtracted from the running between the EW scale and $M_b$,
in the same way as $(t_R^{elem},t'_L)$ was subtracted between the EW scale and $M_t \sim M_{GUT}$ (see \eq{trDec}).
If also $M_b\sim M_{GUT}$, precision unification is spoiled because $R_{SM-\higgs-t_R-t'_L-b_R-b'_L}=1.2$:
one should choose $M_{b,t}$ somewhat below the GUT scale to fix unification.}
To avoid the problems with $b_R$-compositeness,
one needs a chiral symmetry forbidding the mass term $M_b \overline{b'_L}  b_R$,
e.g. a $Z_2$-parity with $Z_2(b_R) = -1$ and $Z_2(b'_L) = +1$.
However, any such symmetry distinguishing $b_R$ and $b'_L$ cannot be exact, because it would forbid at least one of 
the three couplings $y_t \overline{q_L} \higgs^c t_R$, $y_b \overline{q_L} \higgs b_R$ and $\lambda_{b'}f\, \overline{b'_L} b'_R$, that are necessary
to generate the top and bottom masses as well as $m_{b'}=\lambda_{b'}f$.\footnote{
To prove this, note that, under the chiral symmetry, the Higgs $H$ is neutral and $t_R$ has the same charge of $b_R'$, since they sit in the same
$\K$-multiplet.}
This signals that the chiral symmetry is only approximate and should be broken by the strong dynamics.
It also implies that in any of the limits $y_b \rightarrow 0$, $y_t \rightarrow 0$, $\lambda_{b'} \rightarrow 0$, one can recover the symmetry and thus have  $M_b \rightarrow 0$.
Then, one can estimate the minimum size of the breaking, $M_b \sim y_b y_t \lambda_{b'} / (4\pi g_{\rho})^2 f$ (through a loop with $q_L$),
that is negligible in comparison with $m_{b'}$ or even $m_{b} = y_b v/\sqrt{2}$.
In the case with $\U(1)_{B_I}$, the discussion of $b_R-b'_L$ mixing is identical.\\

\noindent \textbf{Case (\emph{2})} corresponds to a different choice for the hypercharge,
$Y = \pm T_3^R - Y'$,
such that $t_R$
can be identified with the $(\textbf{3},\textbf{1},\textbf{1})_{-\textbf{2/3}}$ component in \eq{comp10}.
This choice for $Y$ implies an unusual embedding of composite states with the quantum numbers of SM fermions in $\SO(10)$ multiplets:
$l_L$ and $t_R$ are contained in a $\textbf{10}$, $q_L$, $e_R$
in a $\textbf{45}$
 and $d_R$ in a $\textbf{120}$.
Then, one can couple $l_L$ to a composite operator transforming as a \textbf{11} of SO(11), $q_L$ to a \textbf{55}, et cetera.
This embedding is slightly less economical than in case (\emph{1}),
but note that the number and nature of the composite multiplets emerging from the strong dynamics is not restricted a priori.\footnote{
We remind that here we are dealing with the embedding of $Y$ into the global symmetry $\K$ of the composite sector,
while the way $Y$ is embedded in $\Gsym_{GUT}$ can be a standard one.}

Since $T_R$ transforms in a real representation  $\textbf{10}$, a mass term
$T_R T_R$ would be $\K$-invariant, but it is forbidden by a $\U(1)_B$ symmetry external to $\K$,
because we need
$B(t_R)=1/3$.
The option of taking $\U(1)_B$ internal to SO(10) is less appealing, since one would be forced anyway to introduce an external symmetry to keep
$t_R$ massless (before EWSB). Therefore, we will not consider this possibility for case (\emph{2}).
The exotic elementary fermions $x_L$ have the following SM quantum numbers:
\beq
(\emph{2})~~~
\lx_L = (\textbf{1},\textbf{2})_{\textbf{1/2}}, \; l'_L = (\textbf{1},\textbf{2})_{-\textbf{1/2}}, \, \tx_L = (\overline{\textbf{3}},\textbf{1})_{-\textbf{2/3}} ~,
\label{xL2}
\eeq
that is, there are two ``lepton" doublet resonances as well as a singlet ``top" one, with $B(x_L) = B(x_R) = B(t_R) = 1/3$
for all $x_L$.
Remarkably, $\U(1)_{B}$ is enough to forbid all the mass terms mixing the exotic fermions with the SM elementary fermions.
In particular, there is no $x_L$ with the quantum number of $b_R$.
As a consequence, $t_R$ remains the only composite SM fermion and precision unification works in the simplest way.
Also, notice that in this case $\laq$ and the exotic couplings, $\lambda_{\lx} \sim \lambda_{l'} \sim \lambda_{\tx}$, are not related.

\subsection{Contribution of the exotic fermions to the electroweak precision tests} \label{constraints}

In order to estimate the contribution of the exotic fermions to the EWPTs, we need to specify their couplings $\lax$ to the composite sector.
We will analyze this issue drawing a parallel with the analogous coupling of the top-bottom quark doublet $q_L$, since it also contributes significantly to the EW precision parameters:
\beq
-\mathcal{L}  \supset \overline{q_L} \laq \mathcal{O}_{q_L} + \sum_{x_L} \overline{x_L} \lax \mathcal{O}_{x_L},
\eeq
where for notational convenience we are writing a different composite operator $\mathcal{O}_{\psi}$ for each elementary fermion $\psi$,
knowing that all the $x_L$'s (and in case (\emph{1}) also $q_L$) actually couple to the same operator, but with different couplings.
The transformation properties of $\mathcal{O}_{q_L}$ and $\mathcal{O}_{x_L}$ under $\Gsym$, in particular under the subgroup $\SU(2)_L \times \SU(2)_R \times P_{LR}$, will determine the contributions to $\widehat{T}$ and $\delta g_{b_L}$ from the various $\lambda_\psi$'s.
The latter can be treated as spurion fields transforming under $\Gsym$ as the conjugate of $\mathcal{O}_{\psi_L}$.
As shown in section~\ref{exotics}, these transformation properties
are determined by those of the $T_R$ multiplet.

We will study, for each of our cases, the contributions to $\widehat{T}$ and $\delta g_{b_L}$ using the
effective lagrangian approach introduced in section~\ref{ewpt}, that is, we will estimate the contributions to the operators in \eq{Tparam} and \eq{ZbbL}, respectively. We first estimate the contributions arising after integrating out the strong sector resonances, which are suppressed by powers of $m_{\rho}$, and proportional to the degree of compositeness of the light fermions, that is to $\lambda_{q}$ and $\lambda_{x}$.
Next, after presenting the relevant low-energy operators, we compute the contributions generated by the mixing of the exotic fermions
with the SM ones, whose size is suppressed by powers of the exotic fermion mass, $m_x = \lambda_{x} f$.
We remind that, due to the full compositeness of $t_R$, $\lambda_{q} \simeq y_t$ is fixed.
One-loop contributions to $\widehat{S}$ from the exotics are much smaller than the tree-level correction from vector resonances discussed
in section~\ref{ewpt} (see \eq{Scont}), and therefore we do not consider them here. \\

\noindent \textbf{Case (\emph{1}).} Here $\mathcal{O}_{q_L}$ transforms under $\SU(2)_L \times \SU(2)_R$ as a (\textbf{2},\textbf{1}), which renders $\laq$ a singlet of custodial $\SU(2)_c$. Among the exotic couplings, $\mathcal{O}_{b'_L}$ transforms as (\textbf{1},\textbf{2}), thus $\lambda_{b'}
\sim \textbf{2}$ of $\SU(2)_c$, which then contributes to $\widehat{T}$ as $\lambda_{b'}^4$.\footnote{
This is because $\widehat{T}$ ``transforms'' under $\SU(2)_c$ as a \textbf{5} \cite{Kennedy:1991wa}.}
Also $\lambda_{e'},\lambda_{\nu'}  \sim \textbf{2}$ of $\SU(2)_c$, but if the two couplings are equal $SU(2)_R$-invariance is restored,
so the correction to $\widehat{T}$ goes as $(\lambda_{\nu'} - \lambda_{e'})^4$.
Regarding $\delta g_{b_L}$, as already pointed out after \eq{ZbbL}, there is a tree-level contribution through $\lambda_{q}$.
Summarizing,
\beq
\widehat{T}_{b'} \sim \frac{N_C}{16 \pi^2} \frac{\lambda_{b'}^4}{g_{\rho}^2} \xi ~, \qquad
\widehat{T}_{(\nu',e')} \sim \frac{1}{16 \pi^2} \frac{(\lambda_{\nu'} - \lambda_{e'})^4}{g_{\rho}^2} \xi ~, \qquad
\frac{\delta g_{b_L}}{g_{b_L}} \sim \frac{\lambda_{q}^2}{g_{\rho}^2} \xi ~.
\label{ttg}\eeq
These contributions to $\widehat{T}$, taken individually, set upper bounds on the mass of the exotic fermion $b'$,
$m_{b'}\lesssim 1.2 f (m_\rho/{\rm TeV})^{1/2}$,
and on the mass difference
$|m_{\nu'}-m_{e'}| \lesssim 1.5 f (m_\rho/{\rm TeV})^{1/2}$.
However, cancellations between the two terms could be present, thus
making the bounds milder;
also, for large $m_h$, the negative $\Delta \widehat{T}$ in \eq{TShiggs} can compensate these terms.
 The bound on $m_\rho$ from $\delta g_{b_L}$ was already discussed in section \ref{ewpt}.

In fact, a strong constraint from $\delta g_{b_L}$ is imposed on $\lambda_{b'}$, from the mixing of $b'$ and the bottom quark, which modifies
their couplings to the $Z$. Such mixing arises after EWSB from the Yukawa term $y_t \overline{q_L} \higgs b'_R$,
which is generated along with the top Yukawa. The leading contribution can be computed,
\beq
\frac{\delta g_{b_L}}{g_{b_L}} = - \frac{m_{t}^2}{m_{b'}^2} ~,
\label{gbl}
\eeq
and it yields the robust lower bound  $m_{b'} \gtrsim 1.4$ TeV.\footnote{
In the present scenario, the situation is somewhat better than with $t_R$ elementary \cite{Giudice:2007fh}.
In the latter case, in \eq{ttg} $\lambda_{b'}$ should be substituted by $\lambda_{t_R}$, with the  constraint $\lambda_{q} \lambda_{t_R} \simeq g_{\rho} y_t$,
which requires a very small $\xi$. Such a problem can be alleviated
by means of extra model building  \cite{Giudice:2007fh}.}
The associated correction to the $Wt_Lb_L$ coupling, $\delta g_{t_Lb_L} / g_{t_Lb_L} = -(m_t/m_{b'})^2/2$, is well below the experimental uncertainty.

As explained in section~\ref{ewpt}, the bound on $\delta g_{b_L}$ is derived allowing for $\delta g_{b_R}$ to vary in its $2\sigma$ range.
The $b_R$-$b'_R$ mixing
does not correct the $Z b_R \bar{b}_R$ vertex, because $b_R$ and $b'_R$ have the same EW charges.
However,
a contribution to $g_{b_R}$ comes from the composite resonances at $m_\rho$,
\beq
\delta g_{b_R} \sim
\frac{\lambda_{b_R}^2}{g_{\rho}^2} \xi
~.
\label{ZbRbR}
\eeq
In order for $\delta g_{b_R}$ to reach the experimental best value, one would need $\lambda_{b_R}/g_\rho \sim 0.4
(f/750\GeV)$,
assuming that the deviation has the proper sign, that is positive. In the minimal realization of our scenario,
this condition cannot be fulfilled, since $\lambda_{b_R}/g_{\rho} \simeq y_b/y_t \simeq 0.02.$ However, if the Yukawa coupling of the bottom
quark arises from the mixing of $q_L$ with a composite operator different from the top Yukawa one,
the required degree of compositeness of $b_R$ can be accommodated.
In that case $\lambda_{b_R}$ should be taken into account in the computation of the pNGB effective potential, but we will not pursue
this possibility in the following.

In addition, when baryon number is identified with the internal $(B-L)_{\SO(10)}$ symmetry, an extra mixing between $b_R$ and $b_R^c$ arises
through the coupling $\lambda_{b_R} b_R \qx_R H$.
This term modifies the coupling of the $Z$ to $b_R$ by
\beq
\delta g_{b_R} =
\frac{1}{2} \frac{\lambda_{b_R}^2}{\lambda_{\qx}^2} \xi ~.
\eeq
Such correction has the sign needed to improve the agreement with the data
and, taking $\lambda_{b_R} \simeq g_{\rho} (y_b/y_t)$,
it is as large as the best fit value for $\xi\sim 0.1$ and
$\lambda_{\qx}/g_{\rho} \sim 0.04$.
This indicates that, in order to improve the fit for $g_{b_R}$, the exotic fermion $b^c$ should have a mass close to the experimental
lower bound.

  \begin{figure}[!tp]
  \begin{center}
\includegraphics[width=8cm]{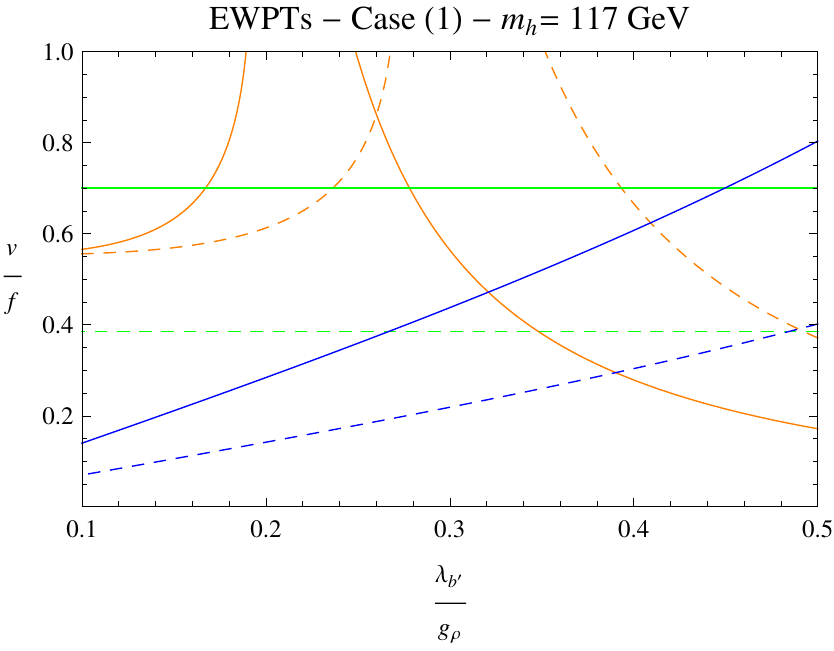}
\includegraphics[width=8cm]{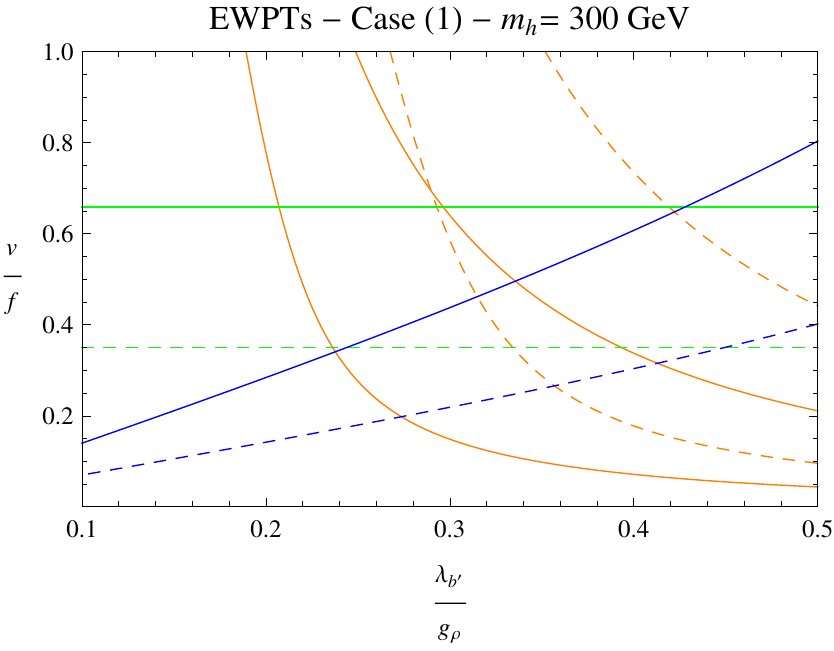}
\caption{
The bounds on $\sqrt{\xi}\equiv v/f$ coming from the EWPTs in case (\emph{1}), as a function of the
ratio $\lambda_{b'}/g_\rho$.
The solid (dashed) lines correspond to $g_\rho=8$ (4), while the left (right) panel
corresponds to a physical Higgs mass $m_h=117$ GeV ($300$ GeV).
The green horizontal lines give the 95\% C.L. upper bound from the $\widehat{S}$ parameter,
the region between the two orange curved lines corresponds to the 95\% C.L. allowed range for the $\widehat{T}$ parameter,
 and finally the blue
straight rising lines represent the $2\sigma$ upper bound from
$\delta g_{b_L}$.
}
  \label{ewpt1graf}
  \end{center}
  \end{figure}

The constraints in case (\emph{1}) are combined in \fig{ewpt1graf},
where we show the different
 bounds on $v/f$ as a function of $\lambda_{b'}/g_{\rho}$.
We present two plots, corresponding to two different Higgs masses, in order to show the preference of the EW data for a light Higgs.
Here we have assumed the first contribution to $\widehat{T}$ in \eq{ttg} to be positive (and neglected the second one), and the one to $\widehat{S}$
in \eq{Scont} to be also positive.
They are added to the contribution from the Higgs loops, given in \eq{TShiggs}.
The relevant contribution to $\delta g_{b_L}$ is the one in \eq{gbl}, that we constrained
allowing for $g_{b_R}$ to vary, since we have shown that sizable $\delta g_{b_R}$ can arise in this scenario.
We thus find that,
for low Higgs masses, close to the experimental bound,
 the deviations in $\widehat{T}$ and $g_{b_L}$ associated to $b'$ determine the maximum allowed value of $v/f$.
The upper bound on $v/f$ increases (decreases) with $g_{\rho}$ for small (large) values of $\lambda_{b'}/g_{\rho}$.
For higher Higgs masses, due to the associated negative contribution to $\widehat{T}$, a positive contribution to this parameter from $b'$ is required. This forces us to lie
to the right of the lower orange curved line (minimum allowed value of $\widehat{T}$), leading to a lower bound on $v/f$ and $\lambda_{b'}/g_{\rho}$.
The allowed region is further reduced by the constraint on $\delta g_{b_L}$ and, for smaller $g_{\rho}$, also the bound from $\widehat{S}$ can be relevant.\footnote{
In these plots we took $m_h$
to be independent from $\lambda_{b'}$.
Once the effective potential is computed  (see section~\ref{pot}), one will be able to study the
correlation between these two parameters.}
\\

\noindent \textbf{Case (\emph{2}).} The constraints on our scenario
from the exotic contributions to EWPTs are significantly milder when $T_R\sim ~$\textbf{10}.
This is because
$\mathcal{O}_{q_L}$ now transforms as $(\textbf{2},\textbf{2})_{-\textbf{2/3}}$ under $\SU(2)_L \times \SU(2)_R \times \U(1)'$
and consequently the bidoublet component coupling to $b_L$ has $T_3^R=T_3^L$.
Therefore, if the composite sector is $P_{LR}$ symmetric (which is the case when $\K=\SO(10)$), the $b_L$ coupling to the $Z$ is protected
at tree-level \cite{Agashe:2006at}.
However, there will be a contribution at one-loop, since the $\mathcal{O}_{q_L}$-component coupling to $t_L$ has $T_3^R\ne T_3^L$ and thus it
is not an eigenstate of $P_{LR}$.
The contributions to $\widehat{T}$ come from $\lambda_{q}$ and $(\lambda_{\lx}-\lambda_{l'})$, both transforming as  $\textbf{2}$
under $\SU(2)_c$.
One can estimate
\beq
\widehat{T}_{q} \sim \frac{N_C}{16 \pi^2} \frac{\laq^4}{g_{\rho}^2} \xi ~, \qquad
\widehat{T}_{(\lx,l')} \sim \frac{1}{16 \pi^2} \frac{(\lambda_{\lx} - \lambda_{l'})^4}{g_{\rho}^2} \xi~, \qquad
\frac{\delta g_{b_L}}{g_{b_L}} \sim \left( \frac{\delta g_b}{g_b} \right)_{SM} \frac{\lambda_{q}^2}{g_{\rho}^2} \xi~.
\label{Tgcase2}
\eeq
Given our reference value $\xi \sim 0.1$,
$\widehat{T}_{q}$ and $\delta g_{b_L}$ are well below the experimental constraints.

\begin{figure}[!tp]
  \begin{center}
\includegraphics[width=8cm]{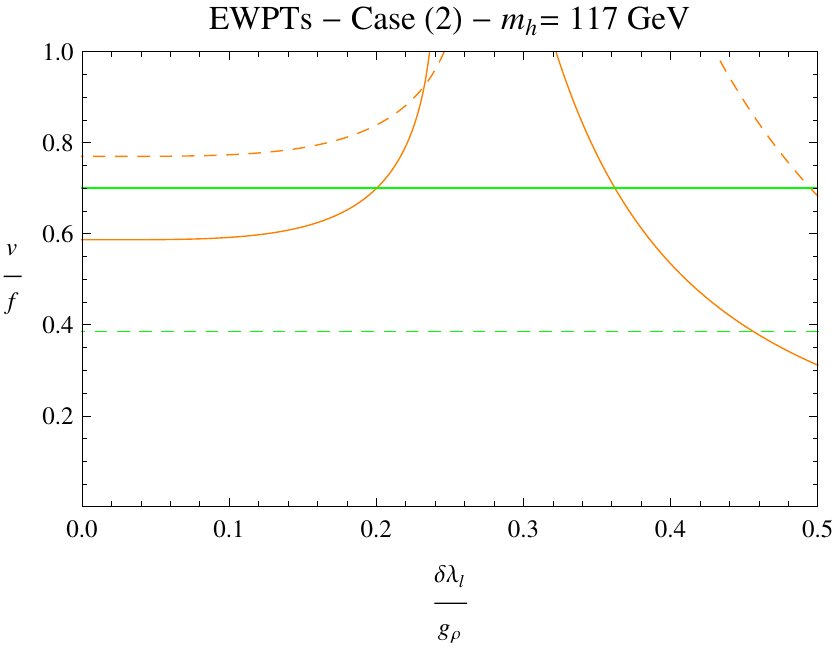}
\includegraphics[width=8cm]{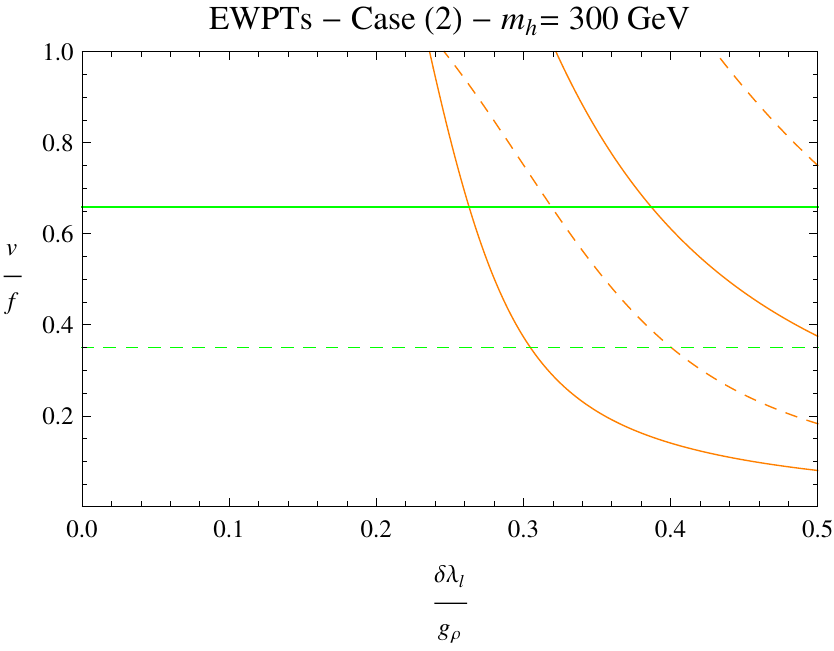}
\caption{ The bounds on $\sqrt{\xi}\equiv v/f$ coming from the EWPTs in case (\emph{2}), as a function of the coupling difference
$\delta\lambda_l/g_\rho\equiv |\lambda_{l^c}-\lambda_{l'}|/g_\rho$.
The solid (dashed) lines correspond to $g_\rho=8$ (4), while the left (right) panel
corresponds to a physical Higgs mass $m_h=117$ GeV ($300$ GeV). The green horizontal lines give the 95\% C.L. upper bound from the $\widehat{S}$
parameter,
while the two orange curved lines correspond to the 95\% C.L. lower and upper values of the $\widehat{T}$ parameter (the allowed region lies between the two lines).
There is no relevant constraint from
$\delta g_{b_L}$ in this case.}
  \label{ewpt2graf}
  \end{center}
  \end{figure}

The constraints on case (\emph{2}) are illustrated in \fig{ewpt2graf}, for two different values of $m_h$, as a function of $|\lambda_{\lx} - \lambda_{l'}|/g_\rho$.
The correction to $\delta g_{b_L}$ is not relevant in this case.
For small values of $m_h$ and $g_{\rho}$, the maximum allowed $v/f$ is determined by the bound on $\widehat{S}$.
As the value of $g_{\rho}$ increases, the bound becomes milder,
and the one from $\widehat{T}$ becomes important,
either when the custodial violating mass difference between $\lx$ and $l'$ is large,
or when it is too small to compensate the opposite sign contribution to $\widehat{T}$ from Higgs loops given in \eq{TShiggs}
(we are assuming a positive sign for the $\widehat{T}$ contributions shown in \eq{Tgcase2}).
For higher Higgs masses, the negative contribution to $\widehat{T}$ from Higgs loops increases to the point that an extra positive $\widehat{T}$ from the exotic fermions is demanded.
This puts a lower bound on the mass splitting $|m_{l^c}-m_{l'}|$.
Besides, the smaller $g_{\rho}$ is, the smaller the open parameter space, because of the combined constraints from $\widehat{T}$ and $\widehat{S}$.\\

All the aforementioned constraints on the parameters of our model shall be taken into account when
we will compute the conditions for EWSB, in section~\ref{pot}.
They shall provide guidance to identify the preferred value of the parameters $v/f$, $g_{\rho}$ and $m_x=\lambda_x f$.

\section{Electroweak symmetry breaking \label{ewsb}}

The NGBs of the composite sector are provided with a non-zero effective potential by the interactions with the elementary sector,
that explicitly break  the global symmetry $\Gsym$.
In this section,
by generalizing the formalism developed for minimal composite-Higgs models \cite{reviewContino},
 we will compute the effective potential for these pNGBs,
and show that the minimum can satisfy all requirements: electroweak symmetry is broken
while colour is not, \ie the Higgs doublet acquires a VEV $v$ while the other pNGBs do not, and at the same time
the induced ratio $v/f$ complies with the phenomenological constraints.
The mass spectrum of the pNGBs will be consequently estimated.

\subsection{The SO$(11)/$SO$(10)$ coset}

For the global symmetry breaking of the composite sector, $\Gsym \rightarrow \K$, we focus on option
(a) of section \ref{chain},  $\Gsym = \SO(11)$ and $\K = \SO(10)$, that generates ten NGBs,
transforming in the $\textbf{10}$ representation of $\SO(10)$,
which decomposes as $(\textbf{1},\textbf{2},\textbf{2})_{\textbf{0}} + (\textbf{3},\textbf{1},\textbf{1})_{-\textbf{2/3}} + (\bar{\textbf{3}},\textbf{1},\textbf{1})_{\textbf{2/3}}$ under the subgroup $\K_{min} \equiv \SU(3)_C \times \SU(2)_L \times \SU(2)_R \times \U(1)'$.
The $\SO(11)/\SO(10)$ coset can be parametrized by the NGB-matrix
\beq
U(\pi) = e^{i \sqrt{2} \pi^a(x) T_a / f}~,
\label{ngbmatrix}
\eeq
where $T_a$ are the broken $\SO(11)$ generators, given by $(T_a)_{IJ}=i(\delta_{I,11}\delta_{a,J}-\delta_{I,a}\delta_{11,J})/\sqrt{2}$,
and $\pi_a=(\tilde h_i,\tilde \phi_\alpha)$ are  the NGBs, where $i=1-4$ and $\alpha=1-6$ are $\SO(4)$
and $\SO(6)$
indices, respectively.

We find it convenient
to use an alternative parameterization, where the NGBs are associated to a dimensionless field $\Sigma$, which transforms linearly as
the \textbf{11} representation of SO(11) and acquires a VEV $\Sigma_0 = (0,\dots,0, 1)^T$:
\beq
\Sigma \equiv U(\pi) \Sigma_0 = \frac{\sin(\Pi/f)}{\Pi} \left[\tilde h,\tilde\phi,\Pi \cot (\Pi/f)\right]^T~,
\label{sigma}
\eeq
where $ \Pi \equiv \sqrt{\tilde h^{2} + \tilde\phi^{2}}$\,. One can redefine the NGB fields in terms of dimensionless variables,
$\sin (\Pi/f)/ \Pi \times \tilde h_i  \equiv  h_i$ and analogously  for $\tilde\phi_{\alpha}$, thus obtaining
\beq
\Sigma = \left( h,\phi,\pm\sqrt{1 - h^2 - \phi^2} \right)^T ~.
\eeq
With this coset parameterization it will be easier to decompose the $\Gsym$-invariant terms in the lagrangian.\footnote{
Moreover, one explicitly sees that the coset is just a parameterization of a sphere in 11 dimensions.}
Besides, the VEV of $h$ determines $v = \vev{h} f$, which is defined by $m_W^2 = g^2 v^2 /4$.
The four real scalars $h_i$, transforming as a $\textbf{4}$ of $\SO(4)$, are equivalent, up to an overall factor $f$ and a change of basis,
to a complex bidoublet $(i\sigma_2 H^*, H)\sim
(\textbf{2},\textbf{2})$ under $\SU(2)_L \times \SU(2)_R \simeq \SO(4)$,
while the six real scalars ${\phi}_{\alpha}$, transforming as a $\textbf{6}$ of SO(6),
correspond to a  complex triplet $T\sim \textbf{3}_{-\textbf{2/3}}$ under $\SU(3)_C \times \U(1)' \subset \SO(6)$.

The kinetic term for the pNGBs is given by
\beq
\frac{f^2}{2} | D_{\mu} \Sigma |^2
= |D_{\mu} H|^2 + |D_{\mu} T|^2
+ \frac{1}{4}
\frac{\left[\partial_{\mu}( H^{\dagger}  H) + \partial_{\mu}(T^{\dagger} T)\right]^2}
{f^2/2-|H|^2-|T|^2}~,
\eeq
where $D_{\mu} H$ and $D_{\mu} T$ are the covariant derivatives for the Higgs and the colour triplet.
The last term
represents the leading interactions among the pNGBs, that signal their composite nature.
These interactions lead both to corrections of $O(s/f^2)$ in processes with characteristic energy scale $\sqrt{s}$,
and to modifications of $O(v^2/f^2)$ (with respect to the renormalizable lagrangian) in the $H$ and $T$ couplings.
The latter arise because of the non-canonical kinetic terms after EWSB, which require a wave-function renormalization. As an example, the coupling of two $W$'s to the Higgs boson is modified 
as $g_{hWW} = g_{hWW}^{SM} \sqrt{1 - \xi}$, where $\xi\equiv v^2/f^2$. This deviation affects the EWPTs through \eq{mheff}.

\subsection{Explicit breaking of SO(11)} \label{eb11}

The composite sector is coupled to a set of elementary fields, that comprises the SM gauge bosons and fermions, except for the right-handed top quark,
as well as the exotic fermions.
Such interactions, introduced in \eq{gaugemix} for gauge bosons and \eq{fermionmix} for fermions, explicitly break
the global symmetry $\Gsym$, and thus induce a non-zero potential for the pNGBs at loop level.
In order to parametrize such interactions, it is useful to consider the $\Gsym$-breaking couplings as dimensionless spurion (non-propagating) fields
 $S$, that
transform in some definite representation of $\Gsym$. Then, the spurion VEVs $\vev{S}$
 will break $\Gsym$ (from now on we shall denote each spurion and the corresponding VEV with the same symbol).
There are two kinds of spurion fields:
\begin{itemize}
\item Gauge interactions:
  \beq
  A^{\mu}_{i_g} \Omega^{i_g}_{I_a} (T^{I_a})_{IJ} \mathcal{J}_{\mu}^{IJ}~,
  \label{gaugebreaking}
  \eeq
where $\mathcal{J}_{\mu}^{IJ}$ is the $\SO(11)$-current of the composite sector in a generic $\SO(11)$-representation carrying indices $I, J$, while
$T^{I_a}$ are the generators of $\SO(11)$ with $I_a=1,\dots,55$ varying over the adjoint representation \textbf{55},
$A^\mu_{i_g}$ are the SM gauge bosons with $i_g$ varying over the SM generators,
and the spurion $\Omega$ is a matrix of gauge couplings
parameterising the embedding of the gauge symmetries into the adjoint representation of $\SO(11)$.
Explicitly, one has
\beq
\Omega^{i_g}_{I_a} = \left( g_s \delta_{I_a}^{\alpha_C} + g \delta_{I_a}^{i_L} + g_1 \delta_{I_a}^{{Y}} \right)~,
\label{omega}\eeq
where $I_a = \alpha_C, \, i_L, \, Y$ identify the generators of $\SU(3)_C$, $\SU(2)_L$, $\U(1)_Y$ inside $\SO(11)$, and $g_1 \equiv \sqrt{5/3} \, g'$.
\item Fermion interactions:
\beq
\overline{\psi}_i (\lambda_{\psi})^{i}_{I} \mathcal{O}_\psi^{I} + h.c.~,
\label{fermionbreaking}
\eeq
where $\psi$ is a generic elementary fermion, $\mathcal{O}_\psi$ is the composite operator $\psi$ couples to,
$i$ and $I$ are respectively SM- and $\mathcal{G}$-indices, and the spurion $\lambda_{\psi}$ describes the embedding of $\psi$
into the $\mathcal{G}$-representation $\textbf{r}_\psi$ to which $\mathcal{O}_\psi$ belongs.
One can think of $\lambda_{\psi}$ as a projector that, acting on the operator $\mathcal{O}_\psi$, selects the component with the SM quantum numbers
of $\psi$, so that the interaction is gauge invariant.
Only the fermions with large couplings to the composite sector affect significantly the pNGB effective potential.
In our scenario, these are the left-handed top quark contained in $q_L$ and the exotic fermions $x_L$.
The SM quantum numbers of $x_L$ and the minimal choices for the $\mathcal{G}$-representations $\textbf{r}_{q}$ and $\textbf{r}_{x}$
have been discussed  in section~\ref{exotics}.
\end{itemize}
Technically, the spurions
transform separately under an elementary sector symmetry $\SU(3) \times \SU(2) \times \U(1)\equiv \Gsym_{SM}^{elem}$,
and a composite symmetry $\Gsym$, both global.\footnote{
The elementary global symmetry is in fact larger, since it includes extra U(1)'s in the form of global phase redefinitions of the elementary fields.}
When the spurions take a VEV, both $\Gsym_{SM}^{elem}$ and $\Gsym$ are broken, but the diagonal combination remains unbroken
and it is identified with the gauged SM symmetry $\Gsym_{SM}$.

\subsection{The effective lagrangian in the background of the NGBs \label{backL}}

The full low energy effective lagrangian $\mathcal{L}_{eff}$ can be obtained by integrating out the
heavy resonances of the
composite sector: $\mathcal{L}_{eff}$
includes the interactions among the SM fields, the NGBs and the exotic fermions, as well as other possible light resonances
\cite{Coleman:1969sm}.
In this section, however, we are interested only in the interactions that contribute to the NGB effective potential $V_{eff}(\pi)$,
which are encoded in the effective lagrangian for the elementary fields  in the background of $\Sigma$, \ie with no derivatives on $\Sigma$.
This is because only the elementary fields break the $\Gsym$-invariance explicitly, and because
$V_{eff}(\pi)$ is obtained from the effective action
in the zero-momentum limit (see e.g. Ref.~\cite{reviewContino}).
Readers not interested in the technical derivation of $\mathcal{L}_{eff}$ can move to section \ref{pot},
where the result for $V_{eff}(\pi)$ is presented and EWSB is discussed.

In order to identify the relevant terms in $\mathcal{L}_{eff}$, we will build $\Gsym$-invariants out of the spurions and $\Sigma$.
Since $\Sigma$ is dimensionless, terms with any power of $\Sigma$ should be included; the effective lagrangian
should not be expanded in the number of NGBs, but rather in the number of spurions.
It turns out that only a few terms, with at most two powers of $\Sigma$, are independent invariants, basically because $\Sigma^T\Sigma=1$.
A formal way to understand this is through the
Coleman-Callan-Wess-Zumino construction for lagrangians invariant under non-linearly realized symmetries \cite{Coleman:1969sm}.
In this formalism the non-derivative interactions of the NGBs are associated to the spurions only, redefined as
$\widetilde{S}(\pi) = U(\pi)[S]$.
These $\widetilde{S}$ transform generically
under a reducible representation of $\K$
and, at a given order in $\widetilde{S}$,
only a finite number of independent $\K$-invariants can be constructed.
We will do this counting in the following subsections before presenting the $\Gsym$-invariant effective lagrangian.
The same argument will apply for the terms in $V_{eff}(\pi)$, that will be computed in section \ref{pot}.

\subsubsection{Gauge bosons}

The gauge spurion $\Omega$ transforms in the adjoint of SO(11), which decomposes under SO(10) as $\textbf{55} = \textbf{45} + \textbf{10}$.
In order to identify the $\K$-invariants quadratic in the gauge couplings
(\ie the leading order ones),
we shall inspect the singlets contained in $\Omega^2$. Since $\textbf{55} \times \textbf{55}
= (\textbf{45} + \textbf{10}) \times (\textbf{45} + \textbf{10}) = \textbf{1}_{45\times 45} + \textbf{1}_{10\times 10} + \cdots$, there are two $\K$-invariants. One combination of them is independent from the NGBs, since it is also invariant under $\Gsym$; to see this, note that
$\textbf{55} \times \textbf{55} = \textbf{1}_\Gsym + \cdots$ contains one $\Gsym$-singlet, denoted as $\textbf{1}_\Gsym$.

The effective lagrangian
for the elementary gauge bosons in the background of $\Sigma$ can be written in momentum space as
\beq
\mathcal{L}_{eff}^A (p) = \frac{1}{2} P^T_{\mu \nu}(p) A_{i_g}^{\mu}(p) A_{j_g}^{\nu}(p) \Omega^{i_g}_{I_a} \Omega^{j_g}_{J_a}
\left[ \Pi_0^A(p^2) (T^{I_a})_{IJ} (T^{J_a})^{JI} +
\Pi_1^A(p^2) \Sigma^I (T^{I_a})_{IJ} (T^{J_a})^{JK} \Sigma_K \right]~,
\label{leffA}\eeq
where $P^T_{\mu\nu}(p)=g_{\mu\nu}-p_\mu p_\nu/p^2$ is the transverse projector,
and $\Pi_{0,1}^A(p^2)$ are form factors that encode the effects of the strong dynamics.
Actually $\Pi_0^A$, or more precisely $d \, \Pi_0^A/dp^2 |_{p^2=0}$, represents the contribution of the composite sector to the $\beta$-function
of the SM gauge couplings.
Here $I, J, K$ are SO(11) indices in the fundamental representation.
The normalization of the $\SO(11)$ generators and their contraction with the $\Sigma$-components read
\bea
  (T^{I_a})_{IJ} (T^{J_a})^{JI} &=&  \Tr[T^{I_a} T^{J_a}] = \delta^{I_a J_a} ~,\\
\Omega^{i_g}_{I_a} \Omega^{j_g}_{J_a} \Sigma^I (T^{I_a})_{IJ} (T^{J_a})^{JK} \Sigma_K &=& g^2 h_i (T^{i_L})_{ij} (T^{j_L})_{jk} h_k
+ 2 g g_1 h_i (T^{i_L})_{ij} (T^{Y})_{jk} h_k \nonumber \\
&+& g_1^2 h_i (T^{Y})_{ij} (T^{Y})_{jk} h_k + g_1^2 \phi_{\alpha} (T^{Y})_{\alpha \beta} (T^{Y})_{\beta \gamma} \phi_\gamma \nonumber \\
&+& g_s^2 \phi_{\alpha} (T^{\alpha_C})_{\alpha \beta} (T^{\beta_C})_{\beta \gamma} \phi_\gamma
+ 2 g_1 g_s \phi_{\alpha} (T^{\alpha_C})_{\alpha \beta} (T^{Y})_{\beta \gamma} \phi_\gamma ~,
\eea
where $i, j, k$ are SO(4) indices, $\alpha, \beta, \gamma$ are SO(6) indices, and $T^{\alpha_C}, T^{i_L}, T^{Y}$ are respectively the generators of $\SU(3)_C$, $\SU(2)_L$, $\U(1)_Y$.

\subsubsection{Fermions in case (\emph{1}): $\textbf{r}_q = \textbf{r}_x = \textbf{32}$}

Let us consider case $(\emph{1})$ of section \ref{exotics}, with both $q_L$ and $x_L$ coupling to a composite operator
in the \textbf{32} spinor representation of SO(11), in particular $q_L$ couples to the \textbf{16} of SO(10) and $x_L$ to the $\overline{\textbf{16}}$.
Therefore both $\lambda_{q}$ and $\lambda_{x}$ transform in the \textbf{32}.
The $\K$-invariants quadratic in
these couplings
are easily counted by inspecting the tensor product $\textbf{32} \times \textbf{32}' = (\textbf{16} + \overline{\textbf{16}}) \times (\textbf{16}' + {\overline{\textbf{16}}}') =
\textbf{1}_{16\times {\overline{16}}'} + \textbf{1}_{\overline{16} \times 16'} +\cdots = \textbf{1}_\Gsym + \textbf{1} + \cdots$, that is one $\Gsym$-invariant
independent from the NGBs, plus one invariant that depends on them.

The effective lagrangian for these elementary fermions in the background for $\Sigma$ can be written as
\beq
\mathcal{L}_{eff}^\psi(p) = \sum_{\psi,\psi'} \overline{\psi}_{i_\psi}(p) \left[
\Pi^{\psi\psi'}_0(p^2) (\lambda_{\psi})^{i_\psi}_{I_s} (\lambda^*_{\psi'})_{j_{\psi'}}^{I_{s}}
+ \Pi^{\psi\psi'}_1(p^2) (\lambda_{\psi})^{i_\psi}_{I_s} \Sigma_I (\Gamma^{I})_{J_s}^{I_s}(\lambda^*_{\psi'})_{j_{\psi'}}^{J_{s}}
 \right] \Sslash{p} \, \psi'^{j_{\psi'}}(p)~,
\label{lpsi}\eeq
where the sum runs over $(\psi,\psi')=(q_L,q_L),(q_L,x_L),(x_L,q_L),(x_L,x_L)$, with $x_L$ spanning the exotic fermions
listed in \eq{xL1}.
Here $i_\psi,j_{\psi'}$ are the SM
elementary
indices of $\psi,\psi'$ respectively, while $I$ and $I_s,J_s$ are
$\SO(11)$ indices in the fundamental and spinor representation, respectively, contracted by
the $\Gamma$ matrices of SO(11).
Notice that we are not including in \eq{lpsi} terms of the kind $\psi^T C \psi'$ (where $C$ is the charge-conjugation matrix), since they are forbidden by the baryon number symmetry $\U(1)_{B_E}$, that is assumed to hold in this section.
On the contrary, these terms are allowed for the case of $\U(1)_{B_I}$.
The modifications introduced by the addition of these extra terms will be discussed in section~\ref{441}.

As explained in section~\ref{exotics}, the exotic fermions couple to all the components of
$T_R \sim \overline{\textbf{16}}$, other than $t_R$.
Therefore, only these
${\overline{\bf16}}$-components
of the spurion $\lambda_{x}$
are non-zero.
In particular any contraction involving the $\textbf{16}$-components vanishes.
Similarly, in $\lambda_{q}$ only the $\bf 16$-component with the quantum numbers of $q_L$ is non-zero.
Keeping this in mind, the $\SO(11)$ contractions in \eq{lpsi} read explicitly
\bea
(\lambda_{q})^{i_q}_{I_s} (\lambda^*_{q})_{j_{q}}^{I_s} &=& \Tr[(\lambda_{q})^{i_q} (\lambda^*_{q})_{j_{q}}] =
\delta^{i_q}_{j_{q}}
\lambda_q^2
~,\\
(\lambda_{x})^{i_x}_{I_s} (\lambda^*_{x})_{j_{x}}^{I_s} &=&
\delta^{i_x}_{j_{x}}
\lambda_x^2
~,\\
(\lambda_{q})^{i_q}_{I_s} (\lambda^*_{x})_{j_{x}}^{I_s} &=&
 \delta^{i_q}_{j_x} \lambda_q
 \lambda_x
  = 0 ~, \\
(\lambda_{q})^{i_q}_{I_s} \Sigma_I (\Gamma^{I})_{J_s}^{I_s} (\lambda^*_{q})_{j_{q}}^{J_s} &=&
\delta^{i_q}_{j_{q}}
\lambda_q^2
\Sigma_{11}  = \pm
\delta^{i_q}_{j_{q}}
\lambda_q^2
\sqrt{1 - h^2 - \phi^2} ~,
 \label{qq}\\
(\lambda_{x})^{i_x}_{I_s} \Sigma_I (\Gamma^{I})_{J_s}^{I_s} (\lambda^*_{x})_{j_{x}}^{J_s} &=&
\mp \delta^{i_x}_{j_{x}}
\lambda_x^2
\sqrt{1 - h^2 - \phi^2} ~,
\label{xx}\\
\left[
 (\overline{q_L})_{i_q} \,
\Sslash{p} \, (x_L)^{j_x}  \right]
(\lambda_{q})^{i_q}_{I_s} \Sigma_I (\Gamma^{I})_{J_s}^{I_s} (\lambda^*_{x})_{j_{x}}^{J_s} &=&
\lambda_q
(\overline{q_L})_{\alpha i}
\, \Sslash{p} \bigl[
\lambda_{b'}
 (b'_L)^{\alpha} \higgs^{i}  \nonumber \\
&&   + \lambda_{\lx} (\lx_L)^{i} \triplet^{\alpha}
+ \lambda_{\qx} (\qx_L)_\beta^i \triplet^*_\gamma \epsilon^{\alpha\beta\gamma}
\bigr] \label{three}~, \quad \quad
\eea
where, with a slight abuse of notation, we dubbed $\lambda_{q,x}$
the values of the non-zero components of the spurions $(\lambda_{q,x})^{i_{q,x}}_{I_s}$
(these values can be taken real with no loss of generality).
The sign difference between \eq{qq} and \eq{xx} comes from the antisymmetry of the $\SO(11)$ invariant $\bf{32} \times \bf{32} \times \bf{11}$,
and in \eq{three} we employed the decomposition of the $\SO(10)$ invariant $\bf{16} \times \bf{16} \times \bf{10}$ into SM components.

\subsubsection{Fermions in case (\emph{2}): $\textbf{r}_q = \textbf{55}$, $\textbf{r}_x = \textbf{11}$}

Let us consider now case $(\emph{2})$ of section \ref{exotics}, with $q_L$ coupled to a ${\bf 45}$ of SO(10), and $x_L$ to a ${\bf 10}$,
so that we take $\lambda_{q}$ to transform as ${\bf 55}={\bf 45}+{\bf 10}$ and $\lambda_{x}$ as ${\bf 11}={\bf 1}+{\bf 10}$ under $\SO(11)$.
The $\K$-invariants quadratic in the elementary-composite couplings correspond to the $\K$-singlets in the tensor products
$\textbf{55} \times \textbf{55} = \textbf{1}_\Gsym + \textbf{1} + \cdots$, $\textbf{11} \times \textbf{11}
 = \textbf{1}_\Gsym + \textbf{1} + \cdots$, and $\textbf{55} \times \textbf{11}
 = \textbf{1} + \cdots$.

In this case the effective lagrangian for the elementary fermions in the background of the $\Sigma$ field is given by
\bea
\mathcal{L}_{eff}^\psi &=&
(\overline{q_L})_{i_q}
 \left[ \Pi^{qq}_0 (\lambda_{q})^{i_q}_{I_a} (\lambda^*_{q})_{j_{q}}^{I_a} + \Pi^{qq}_1 (\lambda_{q})^{i_q}_{I_a} \Sigma^I (T^{I_a})_{IJ} (T_{J_a})^{JK} \Sigma_K (\lambda^*_{q})_{j_{q}}^{J_a} \right] \Sslash{p} \, (q_L)^{j_q} \nonumber \\
&+&
(\overline{x_L})_{i_x}
\left[ \Pi^{xx}_0 (\lambda_{x})^{i_x}_{I} (\lambda^*_{x})_{j_{x}}^{I}
+ \Pi^{xx}_1 (\lambda_{x})^{i_x}_{I} \Sigma^I \Sigma_J (\lambda^*_{x})_{j_{x}}^{J} \right] \Sslash{p} \, (x_L)^{j_x} \nonumber \\
&+&
\left\{ (\overline{q_L})_{i_q}
 \left[ \Pi^{qx}_1 (\lambda_{q})^{i_q}_{I_a} \Sigma^I  (T^{I_a})_{IJ} (\lambda^*_{x})_{j_{x}}^{J} \right] \Sslash{p} \, (x_L)^{j_x} + h.c. \right\}~,
\label{lpsi2}\eea
where $x_L$ represents the set of exotic fermions listed in \eq{xL2} and only the corresponding components of $\lambda_{x}$
are non-zero.
The $\SO(11)$ contractions in \eq{lpsi2} read explicitly
\bea
(\lambda_{q})^{i_q}_{I_a} (\lambda^*_{q})_{j_{q}}^{I_a} &=&
\delta^{i_q}_{j_{q}}
\lambda_q^2 ~,\\
(\lambda_{x})^{i_x}_{I_a} (\lambda^*_{x})_{j_{x}}^{I_a} &=&
\delta^{i_x}_{j_{x}}
\lambda_x^2~,\\
\left[
(\overline{q_L})_{i_q} \,
 \Sslash{p} \, (q_L)^{j_q} \right] \times &&\nonumber\\
\times (\lambda_{q})^{i_q}_{I_a} \Sigma^I (T^{I_a})_{IJ} (T_{J_a})^{JK} \Sigma_K (\lambda^*_{q})^{J_a}_{j_q} &=&
-\lambda_q^2 \,
(\overline{q_L})_{i \alpha} \,
 \Sslash{p} \, (q_L)^{j \beta} \,
 \left(
\delta^i_j T^*_\beta T^\alpha + H^i H^*_j \delta^\alpha_\beta \right) /
2 ~, \quad \quad \\
\left[
(\overline{x_L})_{i_x} \,
 \Sslash{p} \, (x_L)^{j_x} \right]
(\lambda_{x})^{i_x}_{I} \Sigma^I \Sigma_J (\lambda^*_{x})_{j_{x}}^{J}&=&
\left[\lambda_{\lx} (\overline{\lx_L})_i H^i + \lambda_{l'} (\overline{l'_L})^i H^*_i+\lambda_{\tx} (\overline{\tx_L})^\alpha \triplet^*_\alpha \right]
\times  \nonumber \\
&\times& \Sslash{p}
\left[\lambda_{\lx} H^*_j (\lx_L)^j + \lambda_{l'} H^j (l'_L)_j
+\lambda_{\tx} \triplet^\beta (\tx_L)_\beta  \right]
~,\\
\left[
(\overline{q_L})_{i_q} \,
 \Sslash{p} \, (x_L)^{j_x} \right]
(\lambda_{q})^{i_q}_{I_a} \Sigma^I  (T^{I_a})_{IJ} (\lambda^*_{x})_{j_{x}}^{J}&=&
\lambda_q\lambda_{l'} (\overline{q_L})_{\alpha i} \, \Sslash{p} \, (l'_L)^i \triplet^\alpha / \sqrt{2}
~.
\eea

\subsection{The NGB effective potential \label{pot}}

The effective potential $V_{eff}(\pi)$ for the NGBs can be obtained rigorously from the effective lagrangian
$\mathcal{L}_{eff}$ presented in section~\ref{backL},
by integrating out the elementary fields.
In fact, up to some convenient specifications, $V_{eff}(\pi)$ may be identified with
the most general $\Gsym$-invariant potential $V(\Sigma,S)$ constructed with the NGB field $\Sigma$ and the spurions $S$.
The coefficient in front of each invariant
can be evaluated, in principle, integrating over
the form factors introduced in $\mathcal{L}_{eff}$.
Similar integrals have been computed only in specific models with an extra-dimensional dual \cite{newcompoHiggs}.\footnote{
In principle, they could also be determined from (not yet measured) experimental data on the form factors,
as it has been done in QCD for the meson chiral lagrangian.}
In the present scenario, we can only provide NDA estimates for such integrals, and the uncertainty will be parametrized by one dimensionless
coefficient for each invariant, expected to be $O(1)$.
Note that, depending on the specific form of $\mathcal{L}_{eff}$,
some $\Gsym$-invariants may not be generated by integrating out the elementary fields at one-loop level,
rather they appear at higher orders only: this occurrence can be easily verified by inspecting  $\mathcal{L}_{eff}$.

\begin{figure}[!tp]
  \begin{center}
	  \includegraphics{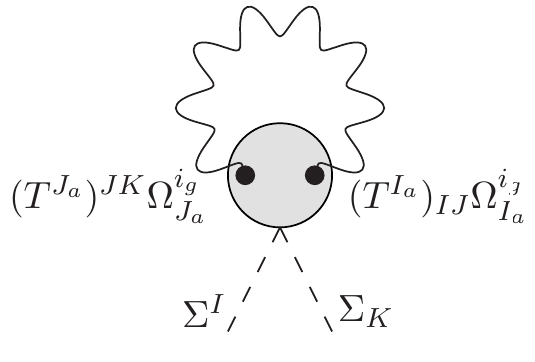}
  \caption{Leading order contribution in the number of gauge spurions and loops to the effective potential for the NGBs.}
  \label{gauge-loop}
  \end{center}
  \end{figure}

In order to build the invariants,
the elementary indices of the spurions have to be contracted, because the potential $V_{eff}$ should respect the SM
elementary sector symmetries. In other words, only terms with no elementary external lines contribute to it.
For this reason, we will construct $V_{eff}$ contracting $\Sigma$ with
\bea
&&\Omega^2_{I_a J_a} \equiv \Omega^{i_g}_{I_a} \Omega^{i_g}_{J_a} ~, \\
&& (\lambda_\psi^2)_{I}^{J} \equiv (\lambda_{\psi})^{i}_{I} (\lambda^*_{\psi})_{i_{}}^{J} ~.
\eea
As far as the  gauge interactions are concerned, there is only one invariant that depends on the NGBs, which corresponds to the
second term in \eq{leffA} and is given by
\bea
\label{a1}
A_1 =  \Omega^2_{I_a J_a} \Sigma^I (T^{I_a})_{IJ} (T^{J_a})^{JK} \Sigma_K =  \frac{3}{2}
g^2
h^2 + \frac{8}{3}
g_s^2
\phi^2 ~.
\eea
The corresponding Feynman diagram is depicted in \fig{gauge-loop}.
The factors $3/2$ and $8/3$
are given by twice the quadratic Casimir of $\SU(N)$, $C_2(N) = (N^2 -1) / (2N)$, for $\SU(2)_L$ and $\SU(3)_C$ respectively
(we neglect the hypercharge contributions).

In the potential, $A_1$ is multiplied by the integral over the form factors (see e.g. \cite{reviewContino}), given by
\beq
3 \int \frac{d^4 p}{(2\pi)^4} \frac{\Pi_1^A(p^2)}{p^2} =
a_1 \frac{3}{16 \pi^2} \frac{m_{\rho}^4}{g_{\rho}^2} ~.
\eeq
The uncertainty in this last expression is accounted for by the $O(1)$ coefficient $a_1$.
The factor of 3 counts the number of Lorentz polarizations of the gauge boson.
Here we estimated the integral taking a momentum cut-off $m_{\rho}$, a factor $1/16 \pi^2$ for the elementary gauge loop,
and another ``loop factor''
$1/g_{\rho}^2$ from the strong sector  (described pictorially by the shaded blob in \fig{gauge-loop}). This is needed in order to match the expectation
$V_{eff} \sim \Lambda^4 /16 \pi^2$, that must hold when all the interactions become strong at the cut-off scale $\Lambda \sim m_\rho$,
that is when $g_{\rho}, g_i \sim 4 \pi$.\footnote{
This factor also matches the NDA estimate for a loop of $N$ constituent ``techni-quarks'', given by $N/(16\pi^2)$, when one takes the large-$N$
estimate for the inter-composite coupling, $g_{\rho} = 4 \pi / \sqrt{N}$.}
The sign of $a_1$ is not fixed by the low-energy theory we are working with, although in calculable examples
\cite{newcompoHiggs,Witten:1983ut} it turns out to be positive (see also \cite{Preskill:1980mz}).

The other invariants entering $V_{eff}$ depend on interactions of the elementary fermions with  the composite sector.

\subsubsection{Case $(\emph{1})$} \label{441}

\begin{figure}[!tp]
  \begin{center}
	  \includegraphics{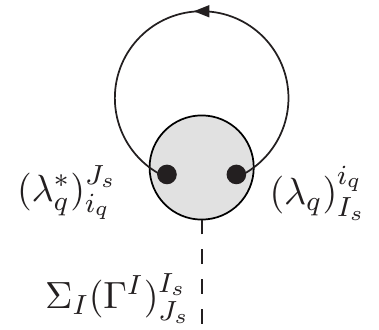}
  \caption{Leading order contribution in the number of $q_L$-spurions and loops to the effective potential for the NGBs.}
  \label{qL-loop}
  \end{center}
  \end{figure}

Let us consider first case $(\emph{1})$, where the composite operators coupling to $q_L$ and $x_L$
transform in the spinor representation, $\textbf{r}_q = \textbf{r}_x = \textbf{32}$, and let us assume
a baryon number symmetry $\U(1)_{B_E}$ external to $\SO(11)$.
The set of leading invariants (two spurions, one loop) is given by
\bea
&& B^{(1)}_1 = (\lambda_{q}^2)_{I_s}^{J_s} \Sigma_I (\Gamma^{I})_{J_s}^{I_s} = 6 \,
\lambda_q^2
 \sqrt{1 - h^2 - \phi^2} ~,
\label{b11} \\
&& C^{(1)}_1 = (\lambda_{x}^2)_{I_s}^{J_s} \Sigma_I (\Gamma^{I})_{J_s}^{I_s} = - 13 \,
\lambda_x^2
 \sqrt{1-h^2 - \phi^2} ~.
\label{c11}
\eea
For notational convenience, we take the same value of $\lambda_x$ for all the exotic fermions.
Thus, the factors $6$ and $13$ account for the number of components of $q_L$ and $x_L$, respectively, running in the loop.
The diagram for the invariant $B^{(1)}_1$ is shown in \fig{qL-loop}. In the potential, it will be multiplied by
\beq
2 \int \frac{d^4 p}{(2\pi)^4} \Pi_1^{qq}(p^2) =
b_1^{(1)} \frac{2}{16 \pi^2} \frac{m_{\rho}^4}{g_{\rho}^2} ~.
\eeq
Again, the sign of the order one coefficient $b_1^{(1)}$ is not determined.
The factor of 2 accounts for the elementary fermion polarisations. In the case of $x_L$, an analog integral $2\int d^4 p/ (2\pi)^4 \Pi_1^{xx}(p^2)$
multiplies the invariant $C_1^{(1)}$.\footnote{When integrating out the exotic fermions, one may worry about the contribution from the composite resonances $x_R$ to the integral over the form-factors. Since the $x_R$ are massless in the limit $\lambda_x \rightarrow 0$, they may introduce infrared divergences that result in a
logarithmic enhancement, $\log (m^2_{\rho}/m^2_x)$. Since this logarithm is not very large anyway, we consider it as part of the uncertainty parametrized by the order one coefficients.}

\begin{figure}[!tp]
  \begin{center}
  	  \includegraphics{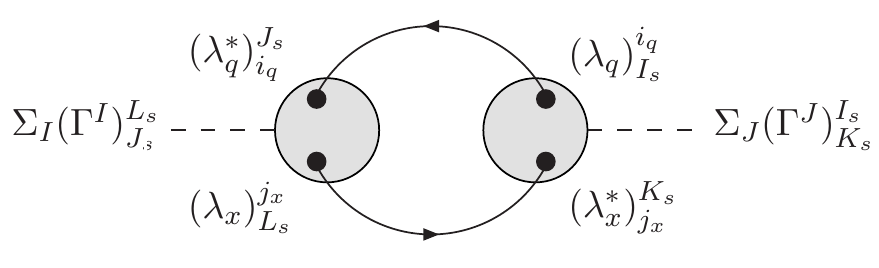}
  \caption{One of the next to leading order contribution ($D_2^{(2)}$ in \eq{d22}) to $V_{eff}$,
  quartic in the number of ($q_L$ and $x_L$) spurions and with
  one loop of the elementary sector.}
  \label{qLxL-loop}
  \end{center}
  \end{figure}

At the next order, with four insertions of the spurions and one elementary fermion loop,
the invariants are\footnote{
We neglect one-loop $O(\lambda^4)$ contributions
obtained by adding to diagrams like the one in Fig.~\ref{qL-loop} an extra composite sector blob in the elementary propagator, since they
have the same functional dependence on the NGBs as $B_1^{(1)}$ and $C_1^{(1)}$,
with a relative suppression by an extra factor $\lambda^2/g_\rho^2$.}
\bea
\label{b22}
&& B^{(2)}_2 = (\lambda_{q}^2)_{I_s}^{J_s}  (\lambda_{q}^2)_{L_s}^{K_s} \Sigma_I (\Gamma^{I})_{K_s}^{I_s} \Sigma_J (\Gamma^{J})_{J_s}^{L_s} = 6
\lambda_q^4
\left( 1 - h^2 - \phi^2 \right)  ~,\\
&& C^{(2)}_2 = (\lambda_{x}^2)_{I_s}^{J_s}  (\lambda_{x}^2)_{L_s}^{K_s} \Sigma_I (\Gamma^{I})_{K_s}^{I_s} \Sigma_J (\Gamma^{J})_{J_s}^{L_s} = 13
\lambda_x^4
\left( 1 - h^2 - \phi^2 \right) ~,\\
&& D^{(2)}_2 = (\lambda_{q}^2)_{I_s}^{J_s}  (\lambda_{x}^2)_{L_s}^{K_s} \Sigma_I (\Gamma^{I})_{K_s}^{I_s} \Sigma_J (\Gamma^{J})_{J_s}^{L_s} =
\frac 12 \lambda_x^2 \lambda_q^2
(3 \, h^2 + 6 \, \phi^2) \label{d22}~.
\eea
The factors 3 and 6 in front of $h^2$ and $\phi^2$ in \eq{d22}
come from the sum over the SM degrees of freedom running in the loop, in the case $x_L=b'_L$  and $x_L=\lx_L,\qx_L$ respectively
(see \eq{three}).
The diagram corresponding to $D^{(2)}_2$ is shown in \fig{qLxL-loop}, and in the effective potential it carries a factor\footnote{
We are neglecting the invariants $B^{(1)}_1B^{(1)}_1$, $B^{(1)}_1C^{(1)}_1$ and  $C^{(1)}_1C^{(1)}_1$, that are also of $O(\lambda^4)$,
but they correspond to two loops of elementary fermions and only one composite blob. Therefore,
their contribution is suppressed by an extra loop factor $1/(16\pi^2)$ and enhanced by a factor $g_\rho^2$, with respect to \eq{int2}.
Also, they are enhanced by an extra power of the number of elementary fermion components in the loop. Anyway these invariants have the same
functional dependence on the NGBs as $B_2^{(2)}$ (or $C_2^{(2)}$), so we can absorb the contribution of the former in the order one coefficient of the latter.}
\beq
2 \int \frac{d^4 p}{(2\pi)^4} \left[ \Pi_1^{qx}(p) \right]^2 =
d_2^{(2)} \frac{2}{16 \pi^2} \frac{m_{\rho}^4}{g_{\rho}^4} ~.
\label{int2}
\eeq

Let us notice that, contrary to previous models in the literature \cite{newcompoHiggs,Gripaios:2009pe}, all the large fermionic contributions to $V_{eff}$
arise from kinetic-type couplings, not from Yukawa-type couplings. This is because the only
large Yukawa, the one of the top quark, is not a coupling of $\Sigma$ to the elementary fermions only, since $t_R$ is fully composite.
As a consequence such coupling does not contribute to $V_{eff}$.

Armed with the list of the relevant invariants we can now discern whether the radiatively generated potential $V_{eff}$
allows for $\SU(2)_L \times \U(1)_{Y}$ breaking to $\U(1)_Q$ while preserving $\SU(3)_C$, and we can study
the conditions to obtain a phenomenologically viable $v/f$ ratio at the minimum of $V_{eff}$.
The potential is a function of $h^2\equiv \sum_ih_i^2$ and $\phi^2\equiv \sum_\alpha \phi_\alpha^2$ only.
Close to the origin, the EW symmetry will be broken if the quadratic term $\mu_h^2 h^2$ in the potential is negative,
while  colour will be preserved if
at the same time the term $\mu_\phi^2 \phi^2$ is positive.
This happens when the fermionic contributions of the top quark and of the exotic fermions to $\mu_{h,\phi}^2$ are negative and large enough
to overcome
the gauge contribution to $\mu_h^2$ generated by the $W$'s, but not the larger contribution to $\mu_\phi^2$ generated by the gluons.
In this case $V_{eff}$ is minimized by
$\vev{h} \ne 0$
 and $\vev{\phi}=0$.

In order to present in a compact form the minimization conditions, we will set from the beginning $\vev{\phi} = 0$ and verify a posteriori the
consistency of this assumption, by requiring the physical mass of the coloured triplet to be positive: we checked that
this condition is necessary and sufficient to guarantee that the global minimum of the potential does not break colour.
Combining the contributions of eqs.~(\ref{a1}), (\ref{b11})-(\ref{c11}) and (\ref{b22})-(\ref{d22}), one obtains
\beq
V_{eff}(h) = \pm \alpha \sqrt{1-h^2} - \beta h^2 = \alpha \cos(\tilde h/f) - \beta \sin^2(\tilde h/f) ~,
\label{veh}
\eeq
where the expressions for $\alpha$ and $\beta$ are
\bea
\alpha &\simeq& \frac{m_{\rho}^4}{16 \pi^2} \left( 12 b_1^{(1)} \frac{\lambda_{q}^2}{g_{\rho}^2} - 26 c_1^{(1)} \frac{\lambda_{x}^2}{g_{\rho}^2} \right)
~,\label{alpha}\\
\beta &\simeq& \frac{m_{\rho}^4}{16 \pi^2} \left( - \frac{9}{2} a_1 \frac{g^2}{g_{\rho}^2} + 12 b_2^{(2)} \frac{\lambda_{q}^4}{g_{\rho}^4} + 26 c_2^{(2)}
\frac{\lambda_{x}^4}{g_{\rho}^4} - 3 d_2^{(2)} \frac{\lambda_{q}^2 \lambda_{x}^2}{g_{\rho}^4} \right)~.\label{beta}
\eea
The minimum of the potential is displaced from $h=0$ provided that $\beta>|\alpha|/2$; in this case the minimum sits at
$\vev{h}=\sqrt{1-\alpha^2/(2\beta)^2}$, or equivalently $\cos(\vev{\tilde h}/f) = - \alpha/(2 \beta)$.
The ratio between the electroweak scale and the NGB decay constant is given by $\sqrt{\xi} \equiv v/f = \vev{h} = \sin(\vev{\tilde{h}}/f)$,
and it is constrained by EWPTs, as described in sections \ref{ewpt} and \ref{constraints}.
In order to achieve sufficiently small values
of $\xi$ one has to require $|\alpha|/2$ to be close to $\beta$.

\begin{figure}[!t]
\begin{center}
\includegraphics[width=7.3cm]{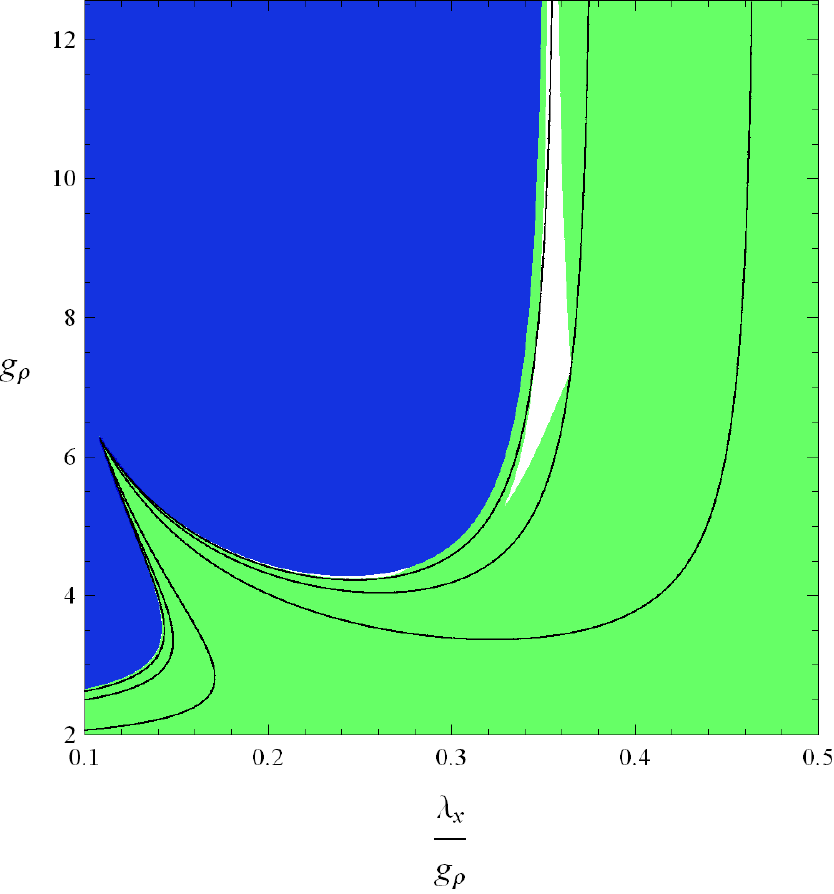}\qquad\qquad
\includegraphics[width=7.3cm]{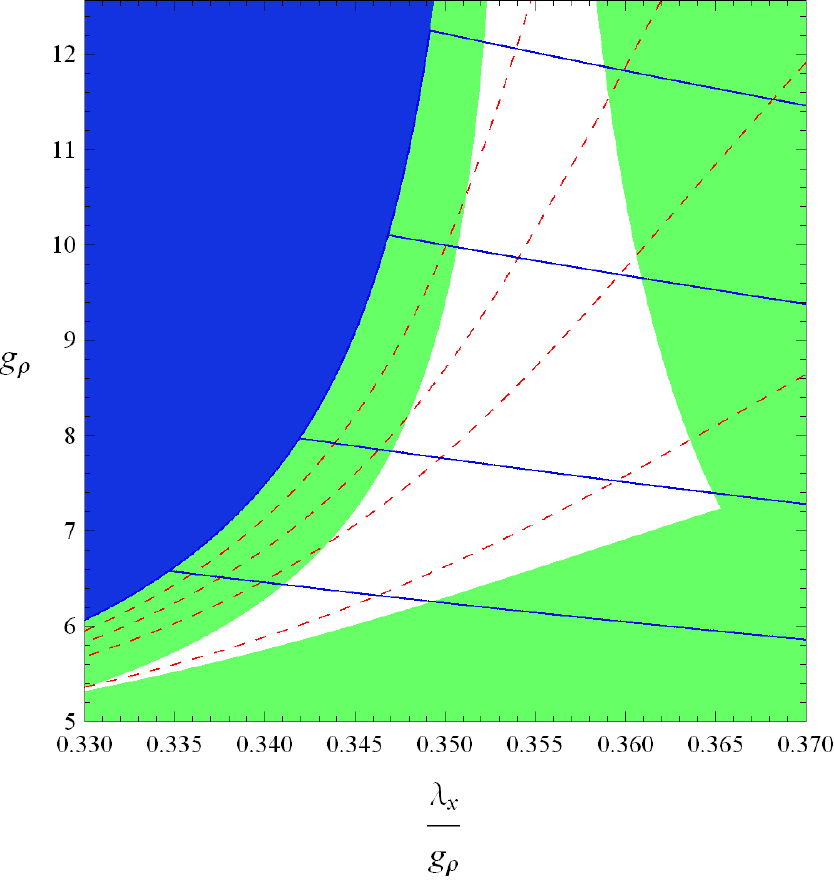}
\caption{
The allowed range for the inter-composite coupling, $g_\rho$, as a function of the composite-elementary coupling of the exotic fermions,
$\lambda_x/g_\rho$, for the case $(\emph{1})$ with an illustrative choice of the order one coefficients in the effective potential:
$a_1= b_1^{(1)}=c_1^{(1)}=1/2$, $b_2^{(2)}=c_2^{(2)}=-d_2^{(2)}=2$.
The right panel is a zoom of the left panel for an interesting range of the parameters.
We fixed $\lambda_q=1$ and $v=246$ GeV.
The requirement of EWSB (defined by $\beta>|\alpha|/2$) excludes the blue (dark) shaded region.
The requirement of unbroken colour (defined by $m^2_T>0$) is always satisfied in the displayed range of parameters.
EWPTs exclude the green (light) shaded region.
The parameter $\xi$ increases from zero, at the boundary of the region with no EWSB, to larger values as one moves away from this boundary.
This is explicitly shown by the isocontours of constant $f$ (black solid lines) in the left panel: $f = 1000, 500, 300 \GeV$, from left to right.
In the right panel we also show curves of constant $m_T=1,1.5,2,3$ TeV (red dashed lines, from right to left) and of constant
$m_h=450,550,700,850$ GeV (blue solid lines,
from bottom to top).
}\label{lxgr}
\end{center}
\end{figure}

Since $t_R$-compositeness fixes $\lambda_q \simeq y_t\simeq 1$, the only free parameters of the model are $\lambda_{x}$ and $g_\rho$
(here we take the same value $\lambda_x$ for all exotic fermions), as well as a set of unknown order one coefficients,
that are expected to lie in a narrow range around $\pm 1$.
One also needs
$\lambda_x/g_\rho$ to be somewhat smaller than
one, in order for our perturbative computation of the effective potential to be valid.\footnote{
Besides, $\lambda_x$ cannot be arbitrarily small, since this parameter fixes the masses of the exotic fermions, which are constrained by direct experimental
searches. We will see in section~\ref{pheno} how these lower bounds depend on the quantum numbers of each exotic fermion.}
In \fig{lxgr} we display the allowed parameter space in the $\lambda_x/g_\rho - g_\rho$ plane, for fixed values of the order one coefficients.
The region allowed by EWSB and EWPTs is
sensitive to the latter choice of coefficients, therefore this figure is intended only to illustrate the
correlations between the parameters of the model;
the specific values of the parameters displayed should not be taken as a univocal prediction of the model.

One can check that the leading order fermion contribution to $\mu^2_{h,\phi}$
is given by $|\alpha|/2$ and thus has the same sign of the gauge contribution.
Therefore, the next-to-leading fermion terms in $\beta$
are the actual responsible for EWSB.
They shall overcome the weak gauge term in $\beta$ as well as the leading fermion terms in $\alpha$, both positive,
in order to fulfill the condition $\beta > |\alpha|/2$.
This can be achieved
when (\emph{i}) $\lambda_x/g_\rho$ is not very small and (\emph{ii}) the coefficients of the next-to-leading fermion terms add up to
a significantly larger value than those of the leading ones.
In particular, there can be an accidental cancellation between the first and second term in \eq{alpha}, e.g. for
$b_1^{(1)}\simeq c_1^{(1)}$ and $\lambda_x \simeq 0.7$.
These requirements are illustrated in \fig{lxgr} (left panel), where we took next-to-leading coefficients equal to $2$ and leading ones equal to $1/2$,
in order to enlarge the region with EWSB.
One notices that, for large values of $g_{\rho}$, $\lambda_x/g_{\rho}$ has to be larger than a
certain value, determined by the $W$ and $q_L$ loop contributions to $V_{eff}$.
For small $g_{\rho}$,
smaller values of $\lambda_x/g_{\rho}$ are allowed, especially when the cancellation in $\alpha$ occurs.
All in all, the need to enhance the $O(\lambda^4)$ terms over the $O(\lambda^2)$ ones, in order to achieve EWSB,
requires some moderate tuning of the parameters of the model.

In the region where EWSB is achieved,
the constraints from the EWPTs put an upper and a lower bound on the allowed value of $v/f$,
depending on $\lambda_x/g_{\rho}$,
as we showed in \fig{ewpt1graf}.
These bounds translate
 into the green shaded region in \fig{lxgr}, left panel.
Consider first large values of $\lambda_x/g_{\rho}$  ($\gtrsim 0.3$):
the region where $\xi$ is small (close to the region with unbroken EW symmetry) is excluded by
the lower bound on $\widehat{T}$. This is because the Higgs mass is large
(see \eq{mh1} below) and needs to be compensated by a positive contribution to
$\widehat{T}$ from the exotic fermion $b'$ (recall we are taking $\lambda_{b'} = \lambda_x$ for all $x$'s).
The upper bound on $\widehat{T}$ excludes large values of $\xi$, where this same contribution from $b'$ becomes too large. Thus, the allowed range for $\widehat{T}$ translates into an allowed window for $\lambda_x/g_{\rho}$,
mildly dependent on $g_{\rho}$.
Finally,
$\delta g_{b_L}$ puts a lower bound on $\lambda_x f$, which cuts the allowed region at small values of $g_\rho$.
For small $\lambda_x/g_\rho$ ($\lesssim 0.3$),
this same constraint from $\delta g_{b_L}$ leaves
only a very narrow allowed strip, which is still interesting because it corresponds to lower Higgs masses.
One can check that $f$ can be as small as
$\sim 500 \GeV$, in agreement with Fig.~\ref{ewpt1graf}.
The masses of the exotic fermions are expected to be
\beq
m_x \simeq \lambda_x f \simeq 1.9 \TeV \left( \frac{\lambda_x}{2.5} \right) \left( \frac{f}{750 \GeV} \right),
\eeq
but we recall that a significant difference between the various $\lambda_x$'s is possible.

The physical mass of the Higgs in the EWSB minimum is given by
\beq
m_h^2 \simeq
 \frac{2 \beta}{f^4} v^2
 \sim
{N_x} \frac{\lambda_x^4}{16 \pi^2} v^2 \simeq (440 \GeV)^2 \left( \frac{\lambda_x}{2.5} \right)^4 ~,
\label{mh1}
\eeq
where the estimate assumes that the largest contribution comes from the $O(\lambda_x^4)$ loop, generated by the $N_x=13$ exotic fermions,
which is a good approximation for intermediate and large values of $g_{\rho}$.
As expected, the Higgs mass scale is controlled by the Higgs VEV $v$, rather than by $f$.
The dominant contribution from exotic fermion loops raises the value of $m_h$ with respect to minimal composite-Higgs models
without gauge unification.
Regarding the mass of the colour triplet, this can be written as
\beq
m_T^2 \simeq  \left[ a_1 g_{\rho}^2 \left(16 g_s^2 - 9 g^2 \right) + d_2^{(2)} 6 \lambda_q^2 \lambda_x^2 \right] \frac{f^2}{16\pi^2}
\sim
N_g \frac{g_s^2}{16 \pi^2} m_{\rho}^2 \simeq \left( 1.2 \TeV \right)^2 \left( \frac{m_{\rho}}{4.5 \TeV} \right)^2
~,
\eeq
where $f^2$ is determined, in turn, by the minimization condition,
$f^2=v^2/(1-(\alpha/2 \beta)^2)$.  In the estimate we assumed that the loop with the $N_g=8$ gluons dominates.
The triplet mass scale is controlled by $m_{\rho}=g_\rho f$, therefore $m_T^2$ is  enhanced with respect to $m_h^2$ by a factor $1/\xi$.
The gluon loops as well as the exotic fermion loops (at large $\lambda_x$) push the mass of the triplet to the $\TeV$ range.
The isocurves of constant $m_h$ and $m_T$ are also shown in the right panel of Fig.~\ref{lxgr}.\\

Let us comment briefly on the structure of $V_{eff}$ when the baryon number symmetry $\U(1)_{B_I}$ is adopted instead of $\U(1)_{B_E}$.
In this case the effective lagrangian contains also terms of the type $\psi^T C \psi'\Phi$ ($\psi,\psi'=q_L,x_L$, $\Phi=H,H^*,T,T^*$), that
were forbidden by $\U(1)_{B_E}$.
 They do not generate leading contributions (two spurions, one loop) to $V_{eff}$, but they do generate
one loop contributions with four spurions, analog to the one depicted in Fig.~\ref{qLxL-loop}, with the arrow of one elementary fermion
reversed. The corresponding terms in $V_{eff}(h)$ are proportional to $\lambda_{x}^4h^2$,
completely analog to the third term in \eq{beta}. Therefore, they do not modify qualitatively the analysis of EWSB
(there are also new terms depending on the colour triplet, $\sim \lambda_q^4\phi^2$ or $\sim \lambda_x^4\phi^2$, that contribute to $m_T^2$).

\subsubsection{Case $(\emph{2})$}

In case $(\emph{2})$ the spurions $\lambda_{q}$ and $\lambda_{x}$ transform in the representations
$\textbf{r}_q = \textbf{55}$ and $\textbf{r}_x = \textbf{11}$ respectively.
One can again perform the computation of the leading invariants generated by the fermion interactions, which take the following form:
\bea
B^{(1)}_1 &=& (\lambda_{q}^2)_{I_a}^{J_a} \Sigma^I (T^{I_a})_{IJ} (T_{J_a})^{JK} \Sigma_K
= - \frac{\lambda_{q}^2}{4}
\left(3 h^2 + 2 \phi^2\right) ~,
\label{b11ii}\\
C^{(1)}_1 &=& (\lambda_{x}^2)_{I}^{J} \Sigma^I \Sigma_J =
\frac{1}{2}\left[
(\lambda_{l^c}^2+\lambda_{l'}^2) h^2 +\lambda_{t^c}^2 \phi^2 \right]
~,
\label{c11ii}\\
B^{(2)}_2 &=& (\lambda_{q}^2)_{I_a}^{J_a}  (\lambda_{q}^2)_{L_a}^{K_a} \left[\Sigma^I (T^{I_a})_{IJ} (T_{K_a})^{JK} \Sigma_K \right]
\left[\Sigma^L (T^{L_a})_{LM} (T_{J_a})^{MN} \Sigma_N \right] \nonumber \\
&=& \frac{\lambda_{q}^4}{16} \left(3 h^4 + 2 h^2 \phi^2 + 2 \phi^4 \right)
~,\\
C^{(2)}_2 &=& (\lambda_{x}^2)_{I}^{J}  (\lambda_{x}^2)_{L}^{K} \Sigma^I  \Sigma_K \Sigma_J \Sigma^L =
\left[C^{(1)}_1\right]^2
~,\\
D^{(2)}_2 &=& (\lambda_{q}^2)_{I_a}^{J_a}  (\lambda_{x}^2)_{I}^{J} \Sigma^K (T^{I_a})_{KJ} \Sigma_L (T_{J_a})^{LI}
= \frac{\lambda_{q}^2 \lambda_{l'}^2}{4}  (2 \phi^2)
\label{d22ii}~,
\eea
where the couplings $\lambda_x$ of the three exotic fermions, $x=l^c,l',t^c$, can be different in general.

As in case $(\emph{1})$, it is consistent to assume that colour remains unbroken (that is, $\vev{\phi}=0$),
as long as the mass of the coloured triplet is verified to be positive in the desired minimum.
In this case, combining the contributions of Eqs.~(\ref{a1}) and (\ref{b11ii})-(\ref{d22ii}), one finds that
 the effective potential is
\beq
V_{eff}(h) = \alpha h^4 - \beta h^2 = \alpha \sin^4(\tilde h/f) - \beta \sin^2(\tilde h/f) ~,
\label{vehii}\eeq
which has a non-trivial minimum ($\vev{h}\ne 0,1$) for $0 < \beta < 2 \alpha$:
in this interval $\vev{h}\equiv \sin(\tilde h/f)\equiv v/f=\sqrt{\beta/(2\alpha)}$.
The coefficients of the potential are given by
\bea
\alpha &\simeq& \frac{m_{\rho}^4}{16 \pi^2} \left( \frac{3}{8}  b_2^{(2)} \frac{\lambda_{q}^4}{g_{\rho}^4}
+  \frac{1}{2} c_2^{(2)} \frac{(\lambda_{l^c}^2+\lambda_{l'}^2)^2}{g_{\rho}^4} \right) ~,\\
\beta &\simeq& \frac{m_{\rho}^4}{16 \pi^2} \left( - \frac{9}{2} a_1 \frac{g^2}{g_{\rho}^2} +  \frac{3}{2}  b_1^{(1)} \frac{\lambda_{q}^2}{g_{\rho}^2}
- c_1^{(1)} \frac{\lambda_{l^c}^2+\lambda_{l'}^2}{g_{\rho}^2} \right)~.
\eea
With respect to case $(\emph{1})$, it is easier  to turn $\beta$ positive and thus drive EWSB, since the negative weak gauge contribution
can be compensated by leading order fermion
contributions. Still, a sufficiently small $\xi$ requires to tune $\beta$ to be small compared with the next-to-leading order terms
that determine the size of $\alpha$.
A favourable choice is to take negative values for $b^{(1)}_1$ and $c^{(1)}_1$, since reasonable values of $\lambda_{\lx,l'}$
can cancel the $W$ and $q_L$ contributions making $\beta$ small. Notice that $\lambda_{t^c}$ does not contribute to the Higgs potential,
thus the mass of $t^c$ is not related to EWSB.

\begin{figure}[!t]
\begin{center}
\includegraphics[width=7.3cm]{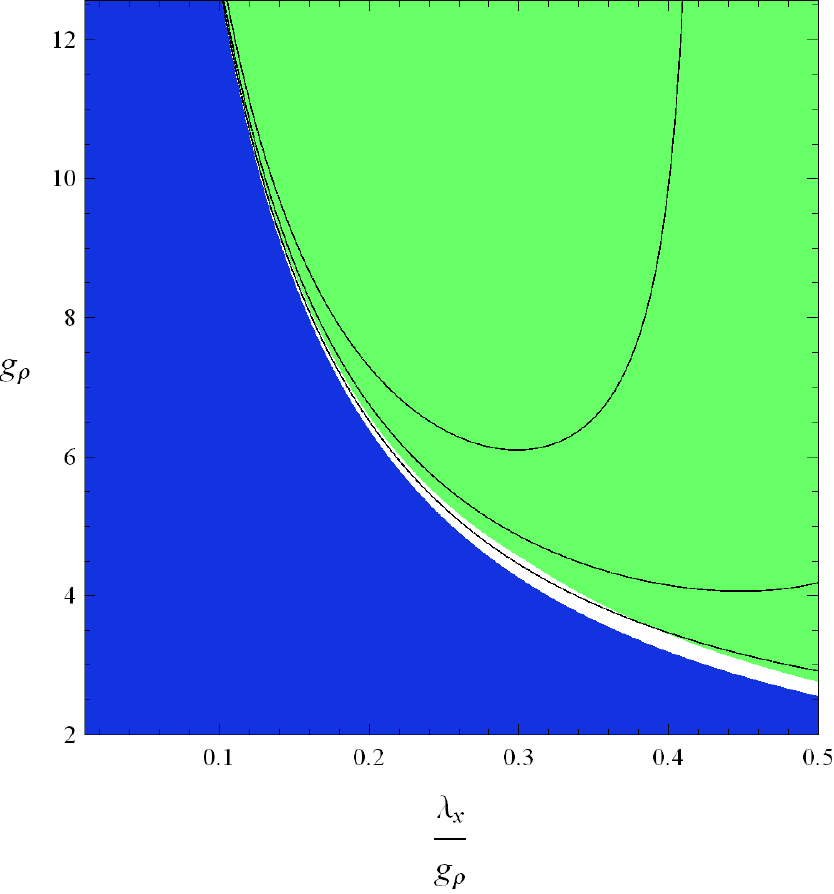}\qquad\qquad
\includegraphics[width=7.3cm]{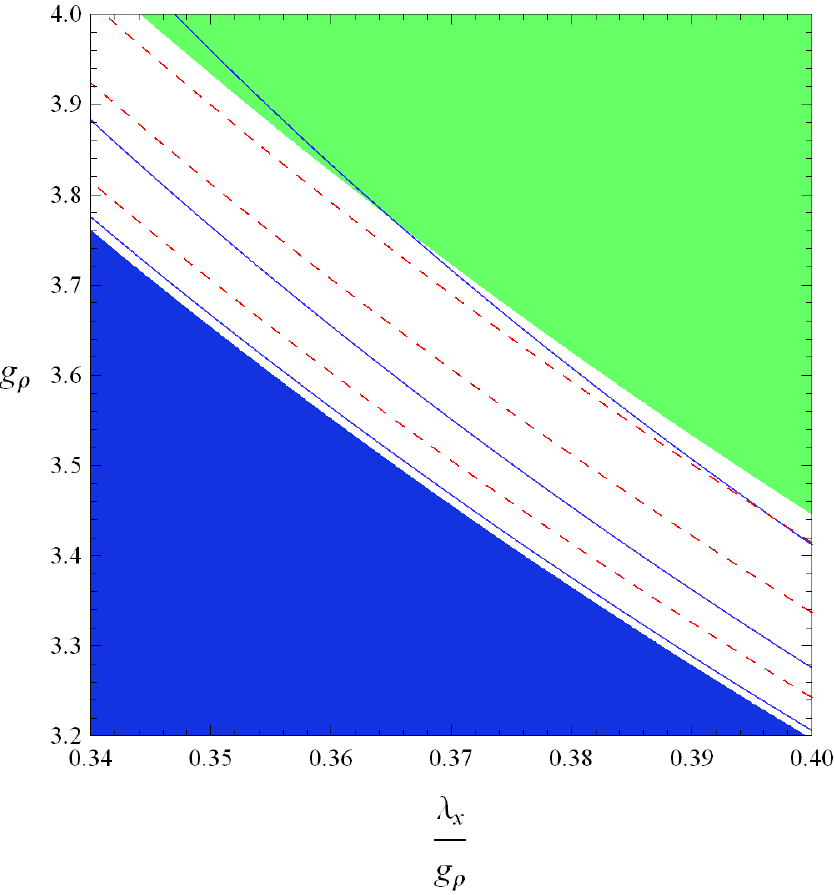}
\caption{
The allowed range for
$g_\rho$ as a function of
$\lambda_x/g_\rho$ for the case $(\emph{2})$, with
effective potential coefficients
$a_1= -b_1^{(1)}=-c_1^{(1)}=1/2$, $b_2^{(2)}=c_2^{(2)}=2$.
The right panel is a zoom of the left panel.
The requirement of EWSB (defined by $0<\beta<2\alpha$) excludes the blue (dark) shaded region.
The requirement of unbroken colour (defined by $m^2_T>0$) is always satisfied in the displayed range of parameters.
The EWPTs
exclude the green (light) shaded region.
In the allowed unshaded region,
$\xi$ increases
from zero, at the EWSB boundary, to its maximal allowed value, close to the EWPTs boundary.
This is explicitly shown by the isocontours of constant $f$ (black solid lines) in the left panel: $f = 750, 500, 300 \GeV$, from bottom to top.
In the right panel we also show curves of constant $m_T=750, 900, 1500 \GeV$ (red dashed lines, from top to bottom) and of constant $m_h= 195, 200, 210 \GeV$ (blue solid lines, from bottom to top).
}\label{lxgr2}
\end{center}
\end{figure}

In Fig.~\ref{lxgr2} we display the allowed parameter space in the $\lambda_x/g_\rho - g_\rho$ plane, taking
$\lambda_x=(\lambda_{l'}+\lambda_{l^c})/2=\lambda_{t^c}$, $\lambda_{l^c}-\lambda_{l'} = 2 \lambda_x / 5$ and a fixed choice of the order one coefficients.
Once again, we warn the reader that the region allowed by EWSB and EWPTs is quite sensitive to this choice
and the figure is intended only to illustrate one possible range of variation of the physical parameters: e.g.,
for fixed values of $\lambda_x$ and $g_\rho$,
$\xi$ scales as the ratio of leading over next-to-leading
coefficients (in Fig.~\ref{lxgr2} we took this ratio to be $1/4$).
The green shaded region is excluded by EWPTs, with the stronger constraint, for small $g_{\rho}$, coming
from $\widehat{S}$.
However, for larger values of $g_{\rho}$ the bound from $\widehat{S}$ becomes milder, allowing for a larger $\xi$, and the bound from $\widehat{T}$ becomes relevant.
We considered a non-zero difference between $\lambda_{l^c}$ and $\lambda_{l'}$
in order to generate a positive contribution to $\widehat{T}$ from the exotic fermions, that partially cancels
the negative one associated with $m_h$.
The lower bound on $f$ is attained for large $g_\rho$,
consistently with \fig{ewpt2graf}:
we find a minimum $f \sim 600 \GeV$, that is comparable with case (\emph{1}).
Notice that $\lambda_x$ is approximately constant in the allowed region of  Fig.~\ref{lxgr2} (left panel),
because the latter is bounded by the two curves $\beta=0$ and
$2.1 \times 10^{-3}=\widehat{S}\propto \xi/g_\rho^2$, that in this case are both independent from $g_\rho$
(the minimization of $V_{eff}$ implies $\xi \propto g_{\rho}^2$).
Taking the coefficients of the various terms to be close to one,
a value  $\lambda_x \sim 1$ is obtained,
allowing for the masses $m_x\simeq \lambda_x f$ to be relatively lighter with respect to case (\emph{1}).

The mass of the Higgs in the EWSB minimum is given by
\beq
m_h^2 \simeq \frac{8 \alpha}{f^4} v^2 \left( 1 - \frac{v^2}{f^2} \right) \sim ~ N_x \frac{\lambda_{x}^4}{4 \pi^2} v^2
\simeq (80 \GeV)^2 \left( \frac{\lambda_{x}}{1} \right)^4 ~,
\eeq
where the  largest contribution comes from the $N_x=4$ exotic fermions (only $l^c$ and $l'$ contribute in this case),
so that the Higgs mass is correlated with their mass.
This order of magnitude estimate implies a much lighter Higgs than in case (\emph{1}), with some range of parameters that is even incompatible
with the LEP lower bound.
The message is that
in case (\emph{2}) the Higgs is expected to be light, while in case (\emph{1}) it prefers to be heavier.
Concerning the mass of the coloured triplet, it is given by
\beq
m_T^2 \simeq \frac{f^2}{8\pi^2} \left[ g_{\rho}^2 (8 a_1 g_s^2 -  b_1^{(1)} \lambda_q^2
+  c_1^{(1)} \lambda_{t^c}^2) +  d_2^{(2)} \lambda_q^2 \lambda_{l'}^2 \right]
+ \frac{v^2}{32\pi^2}  \left[ b_2^{(2)} \lambda_q^4 + 4 c_2^{(2)} (\lambda_{l^c}^2+\lambda_{l'}^2)\lambda_{t^c}^2
\right]~,
\eeq
where the minimization of $V_{eff}$ fixes $f^2=v^2 (2\alpha/\beta)$.
 In this case the triplet mass receives two large contributions, from gluons and from $t^c$, that can be estimated as
\bea
m_T^2 &\sim& N_g \frac{g_s^2}{16 \pi^2}m_\rho^2
\simeq \left( 1.2 \TeV \right)^2 \left( \frac{m_{\rho}}{4.5 \TeV} \right)^2 ~,
\nonumber \\
&\sim& \frac{\lambda_{t^c}^2}{16 \pi^2} m_{\rho}^2
\simeq \left( 0.4 \TeV \right)^2 \left( \frac{\lambda_{t^c}}{1} \right)^2 \left( \frac{m_{\rho}}{4.5 \TeV} \right)^2
~.
\eea
If these two terms add up, the triplet is heavy, but they may have instead opposite sign and partially compensate, allowing for a lighter $T$.
For illustration, the isocurves of constant $m_h$ and $m_T$ are shown in the right panel of Fig.~\ref{lxgr2} for a fixed set of $O(1)$ parameters,
that leads to a partial cancellation in the triplet mass, whose minimal allowed value ends up to be lighter than in case (\emph{1}).

\section{Collider phenomenology of Higgs and top quark partners \label{pheno}}

The deviations from the elementary SM Higgs properties due to compositeness
have been studied in detail in e.g. Ref.~\cite{Giudice:2007fh}:
the most promising processes are
longitudinal vector boson scattering and strong double-Higgs production, as well as the measurement
of anomalous Higgs couplings.
The signatures associated to top-compositeness have been also analyzed, in a similar fashion, in Ref.~\cite{Pomarol:2008bh}.
Evidence can be searched in modified top couplings to gauge bosons, or through the enhancement of processes
like $pp \rightarrow t\bar{t}t\bar{t}$ at large partonic center of mass energy \cite{Pomarol:2008bh,4tops}.

Here we focus on the new physics more specifically related to the scenario of composite unification, that is to say, on
the composite partners of $H$ and $t_R$, that fill with them complete $\K$-multiplets, for the case $\K=\SO(10)$
(recall that the states accompanying $t_R$ are also complemented with elementary chiral partners to form a set of vector-like fermions).
These composite partners (in short, ``comparts")
are the counterpart of supersymmetric partners (``sparticles") in the case of supersymmetric unification.

The Higgs $H$ is accompanied by a colour triplet $T$, sharing its character of pNGB. As such it acquires a mass only through radiative corrections, roughly a loop factor smaller than the compositeness scale. Therefore $T$ is necessarily light, if we demand a natural EW scale.
From the analyses of EWSB in section \ref{pot},
we expect the $\triplet$ mass to lie in the range $\sim (0.5, 2.5) \TeV$, for values of $\xi$
between the maximum allowed by EWPTs and $\sim 0.1$.
 This is in contrast to the vast majority of GUT models, where a colour triplet scalar is also present, although with a mass of the order of $M_{GUT}$.

The fermionic partners of the $t_R$, the exotic fermions $x$'s, are responsible for the precise unification of gauge couplings at $M_{GUT}$.
Therefore, their presence at low energies would constitute an evidence for the unified character of the SM forces. They are intimately related to the composite nature of the top, and as a consequence their observation at colliders will be associated to top (and bottom)
physics, either through production or decay.
Notice that all the $x$'s are lighter than the other resonances of the composite sector, which makes them a perfect target for colliders such as the LHC, even in the unfortunate case that the scale of compositeness is just too high to be accessible.
While the mass of $b'$ in case (\emph{1}) is tightly constrained by EWPTs, $m_{b'} \gtrsim 1.4 \TeV$, the masses of the rest of the fermionic comparts are only required to lie in the region allowed by EWSB.
We recall that all the $x$'s couple, each with a different $\lambda_x$, to the same composite sector operator.
If all couplings have a similar value,
the $x$'s acquire a common mass $m_x \simeq \lambda_x f$.
Then, from the analysis of section \ref{pot}, $m_x$ should be
in the range $\sim(1, 3) \TeV$  to trigger EWSB in case (\emph{1}),
while lighter exotic fermions in the range $\sim(0.5, 1.5) \TeV$ are expected in case (\emph{2}).
Mild hierarchies between the couplings $\lambda_x$ are  well possible,
allowing for fermionic comparts as light as their current experimental bound, discussed below.

The Higgs is expected to be the lightest composite state (besides the top quark), and its mass grows with the number $N_x$ and the
coupling $\lambda_x$ of the exotic fermions.
In the case (\emph{1}) $m_h$ can be as heavy as $\sim 800 \GeV$, while values smaller than $400$ GeV
are possible but require extra tuning of the parameters.
In the case (\emph{2}) $m_h$ can lie between the current experimental bound, 115 GeV, and $\sim 250 \GeV$.

Before considering specific realizations of our scenario, let us make some general considerations on production and decay of the comparts.
We expect them to be mostly pair-produced via gauge interactions. This will certainly be the case
for the coloured comparts. The production cross section of a pair of heavy quarks at the LHC with $\sqrt{s} = 14 \TeV$ is around $ 2 - 0.05$ pb for masses $0.5 - 1$ TeV \cite{Contino:2008hi}. For a scalar colour triplet, the pair production cross section is of the order of $0.5 - 0.01$ pb for masses $0.5 - 1$ TeV \cite{Chen:2008hh}. Double production will also be substantial for the case of weakly interacting comparts.
Cross sections for pair production of heavy leptons are around 20 fb for 500 GeV masses at the LHC with $\sqrt{s} = 14 \TeV$ \cite{delAguila:2010es}.

Once produced, the open decay channels for the comparts crucially depend on their baryon number assignment. We will see that, in some models,
only a few of them can decay in SM states only, predominantly involving top quarks and (composite) longitudinal gauge bosons. Therefore, among the rest of the comparts, the lightest is stable, with interesting consequences for detection at LHC as well as for dark matter.
However, we will also present a model with no stable comparts, with interesting third generation diquark as well as triquark signatures.
Of course, the decay chains
will depend on the details of the mass spectrum in a model-dependent way.

\subsection{The model with $t_R \in \bf{\overline{16}}$ and $\U(1)_{B_E}$} \label{phenomodel1}

We focus first on the embedding of the composite $t_R$ in the spinor representation of $\K=\SO(10)$.
It is convenient to display the quantum numbers of the SM states (Table \ref{tabSM}) and the ones of the comparts
(Table \ref{tabCP}), to promptly identify the allowed interactions between them.
However, recall that not all the interactions allowed by $\Gsym_{SM} \times \U(1)_B$ will be necessarily generated
by the elementary-composite couplings introduced in section \ref{eb11},
because these may preserve some extra symmetries, associated with additional generators of SO(10);
still, we expect these
 symmetries to be broken by extra (smaller) couplings
of the elementary fields to different composite operators.

We discuss
in this section
the model where baryon number is identified with a
group $\U(1)_{B_E}$ external to $\SO(10)$,
so that the partners of $\higgs$ ($t_R$) have the same baryon number as $\higgs$ ($t_R$).
In the next section we will consider the other possibility, a
group $\U(1)_{B_I}$ internal to $\SO(10)$,
more precisely identified with the $(B-L)_{\SO(10)}$ generator.
It will be shown that the phenomenology of the comparts at colliders is substantially  different in the two cases.

\begin{table}[t]\begin{center}
\begin{tabular}{|c|ccccc|c|}
\hline
& $q_L$ & $t_R$ & $b_R$ & $l_L$  & $e_R$ & $\higgs$ \\ \hline &&&&&& \\ [-2ex]
$\SU(3)_C$ & 3 & 3 & 3 & 1 & 1 & 1\\  [1ex]
$\SU(2)_L$ & 2 & 1 & 1 & 2 & 1 & 2\\  [1ex]
$\U(1)_Y$ & $\frac 1 6$ & $\frac 2 3$ & $-\frac 1 3$ & $-\frac 1 2$ & $-1$ & $\frac 1 2$ \\ [1ex] \hline &&&&&& \\ [-2ex]
$\U(1)_B$ & $\frac 1 3$ & $\frac 1 3$ & $\frac 1 3$ & 0 & 0 & 0\\ [1ex]
\hline
\end{tabular}
\caption{The gauge and baryon number assignments of the SM fields. \label{tabSM}}
\end{center}\end{table}

\begin{table}[t]\begin{center}
\begin{tabular}{|c|ccccc|c|}
\hline &&&&&& \\ [-2.5ex]
& $q^c$ & $b'$ & $l^c$ & $\nu'$ & $e'$ & $\triplet$ \\ \hline &&&&&& \\ [-2ex]
$\SU(3)_C$ & $\bar{3}$ & 3 & 1 & 1 & 1 & 3 \\  [1ex]
$\SU(2)_L$ & 2 & 1 &  2 & 1 & 1 & 1\\  [1ex]
$\U(1)_Y$ & $-\frac 1 6$ & $-\frac 1 3$ & $\frac 1 2$ & 0 & $-1$ & $-\frac 1 3$ \\  [1ex]\hline &&&&&& \\ [-2ex]
$\U(1)_{B_E}$ & $\frac 1 3$ &  $\frac 1 3$ & $\frac 1 3$ & $\frac 1 3$ & $\frac 1 3$ & 0\\ [1ex]
$\U(1)_{B_I}$ & $-\frac 1 3$ &  $\frac 1 3$ & 1 & -1 & -1 & $-\frac 2 3$ \\ [1ex]
\hline
\end{tabular}
\caption{The gauge and baryon number assignments of the composite partner fields in the case (\emph{1}), that is with $t_R \in {\bf\overline{16}}$.
Two choices for the baryon number are displayed, depending on if $\U(1)_B$ is taken to be external or internal to $\SO(10)$. \label{tabCP}}
\end{center}\end{table}

The effective lagrangian of the comparts contains, besides the obvious compart mass terms and gauge interactions,
the Yukawa couplings as well as higher dimensional interactions.
The allowed Yukawa interactions with the SM fields
can be classified on the basis of the number of comparts involved:
\bea
{\rm 2~comparts:} && \laq \overline{q_L} l^c_R T~,~~~ \laq \overline{q_L}  q^c_R T^* ~; \nonumber \\
&& \widetilde \lambda_{\lx} \overline{l^c_L} \nu'_R H ~,~~~ \widetilde \lambda_{\nu'} \overline{\nu'_L} l^c_R H^* ~,~~~
\widetilde \lambda_{b_R,\nu'} \overline{b_R} \nu'_L T ~; \label{2co}\\
{\rm 1~compart:} && \laq \overline{q_L} b'_R H~. \label{1co}
\eea
We do not present here a
 list of higher dimensional
operators; we will rather specify those controlling relevant
processes when needed, for example in the case of the compart $e'$, that has no allowed Yukawa couplings.
The second line of \eq{2co} contains terms that are not generated by the couplings $\lambda_\psi$, which induce the SM and exotic
fermion masses.
The reason is that such couplings accidentally respect $\U(1)_{(B-L)_{\SO(10)}}$, while the terms in the second line of \eq{2co} do not.
However,
this $\U(1)$ can be broken in general by extra elementary-composite couplings, generating such additional terms.
Their coefficients $\widetilde\lambda_{\psi}$
should be smaller than $\lambda_q$ or $\lambda_x$, not to spoil the analysis of the potential carried out in section~\ref{pot}.
However, they are important for the decay of some of the comparts.

The only compart that can be singly produced or decay into a SM final state is $b'$.
This can be understood in terms of the $Z_3$ symmetry known as baryon triality, that can be defined
as $B_3 \equiv (3B -n_C)_{\rm mod~3}$,
where $n_C$ is the number of colour indices ($n_C=1$ for the $\bf 3$ representation, $-1$ for the $\bf \bar{3}$).
All SM particles  as well as $b'_L$ have $B_3=0$.
On the contrary $B_3(q^c_L)=B_3(T)=2$ and $B_3(l^c_L)=B_3(\nu'_L)=B_3(e'_L)=1$.
This implies that the lightest of these comparts will be stable.

The compart $b'$, being a vector-like $\SU(2)_L$-singlet fermion, is not strictly a fourth family down-type quark, but still some
of the bounds on the latter apply.
The coupling in \eq{1co} opens the decay channels $b' \rightarrow b h,~tW_L,~bZ_L$
(where the subscript $L$ denotes longitudinal polarizations), all with similar strength.
CDF searches exclude at 95\% CL bottom-like quarks with masses $m_{b'} \lesssim 340(270)$ GeV decaying to $tW$ ($bZ$) \cite{b4PRL}(\cite{b4PRD}). Also, the CMS collaboration has already analyzed the double production of bottom-like quarks decaying to $tW$, excluding masses below 360
GeV at 95\% CL \cite{Collaboration:2011em}.
Both constraints are weaker than the indirect bound $m_{b'}\gtrsim 1.4 \TeV$ from the $Z b\bar{b}$ coupling, discussed in section \ref{constraints}.
The latter limit strongly reduces the discovery prospects of $b'$ at the LHC.
For such high masses, single production, associated with a bottom or top quark and a spectator jet, becomes the dominant production mechanism, with $\sigma(m_{b'}=1-2 \, \TeV) \sim 0.2 - 0.005$ pb for $\sqrt{s}=14$ TeV \cite{Mrazek:2009yu}.\\

Concerning the other comparts,
let us discuss a couple of interesting possible scenarios.
Suppose first that the pNGB $T$ is lighter than all fermion comparts.
In fact, as discussed in section \ref{441},
we have roughly
$m_T^2/m_x^2 \sim N_g (g_s^2/16\pi^2) (g_\rho^2/\lax^2)$,
which is typically smaller than one.
We checked
that, in the parameter region compatible with EWSB and EWPTs, $T$ can be the lightest compart.

Then, the fermion comparts decay into SM$+T$ final states as
\beq
t^c\rightarrow b T~,~b^c\rightarrow t T~,~~~e^c\rightarrow tT^*~,~\nu^c \rightarrow bT^*~,~~~\nu' \rightarrow bT^*~,~~~e'\rightarrow bb\bar{t} T^* ~.
\eeq
The decays of $\qx = (b^c, t^c)$ and $\lx = (e^c, \nu^c)$ go through the first two couplings in \eq{2co}.
Besides, the heaviest of the $\SU(2)_L$ components of $\qx$ and $\lx$ can decay to the lightest one and a $W$ boson, either real or virtual.
The decay of $\nu'$ proceeds through the last term in Eq.~\ref{2co}.\footnote{
This term involves a coupling of $b_R$ to the composite sector. We notice that such coupling
could be responsible for an important deviation of the $Z b_R \overline{b_R}$ coupling from the SM prediction,
as explained below \eq{ZbRbR}.}
The decay of $e'$ proceeds through a dim-7 operator involving four fermions and $T$.
We find that the leading one is $\overline{b_R} e'_R  \overline{b_R}  t_R T$ (various contractions understood),
that carries the NDA suppression factor $(\tilde \lambda_{b_R} / g_{\rho})^2 \times 1/f^3$. Then the decay rate is small,
$\Gamma(e' \rightarrow bb\bar{t} T^*) \sim (\tilde \lambda_{b_R}/g_{\rho})^4 (m_{e'}^7/ f^6) /(4\pi)^5  \sim (0.5 \, \mu \textrm{m})^{-1}$,
where we have taken for concreteness $m_{e'} = f = 750$ GeV and $\widetilde \lambda_{b_R}/g_{\rho} = \lambda_{b_R}/g_{\rho} \simeq y_b/y_t \simeq 0.02$.
If this decay rate
were slightly smaller than this estimate, the displaced vertex of the $e'$ decay could be seen at LHCb, that has a sensitivity of about $40 \, \mu \textrm{m}$.

The scalar colour triplet $T$ can be either pair-produced or come from the decay chains of the fermionic comparts.
Being stable, it hadronizes into colour singlet bound states, either neutral ($T\bar{d} \equiv \mathcal{T}^0$) or singly charged ($T\bar{u} \equiv \mathcal{T}^-$)
and flies out of the detector. Their hadronic interactions are too small
to lead to detection \cite{CC}, although they could lead to transitions between $\mathcal{T}^0$ and $\mathcal{T}^-$ \cite{Farrar}. When the
$\triplet$-hadrons are neutral, they
will not be seen, but they will lead to events with large missing energy, while the charged ones behave as heavy muons undergoing ionization, an almost background-free signature.
The CDF collaboration sets a bound on the mass of a stop-like long-lived charged massive particle (CHAMP),
which very much resembles our scalar colour triplet,
$m_{\tilde t} \geqslant 250$ GeV at 95\% CL, which applies only if the CHAMP is charged in both the inner and outer trackers \cite{champ}.
At the LHC with
$100$ fb$^{-1}$ luminosity, the number of identifiable $\mathcal{T}^{\pm} \mathcal{T}^{\mp}$ events is estimated in $\sim 10000 - 300$ for
$m_{\mathcal{T}^{\pm}}= 0.5 - 1$ TeV \cite{CC}. The CMS collaboration is already putting bounds on CHAMPs \cite{cms}.
In particular, for a gluino-like particle, the bound already increased to 400 GeV.

We expect that the detection of $\mathcal{T}^{\pm}$ states, 
either from $T$ pair production or associated with tops and bottoms from the decay of heavier comparts,
will be the cleanest signature of this model.
When $\triplet$ is not detected, processes like $pp \rightarrow b^c \overline{b^c} \rightarrow t \bar{t} + \Lslash{E_T}$ or $pp \rightarrow t^c \overline{t^c} \rightarrow b \bar{b} + \Lslash{E_T}$,
\ie top or bottom pairs plus missing transverse energy,
could lead to the early discovery of the coloured fermionic comparts $b^c$ and $t^c$, while
the prospects for the colour singlets $\lx$, $e'$, $\nu'$ are more modest, due to the smaller production rates.
In general, the best search strategy for fermionic comparts at the LHC, in the presence of a lightest stable $T$ manifesting as missing energy, shall be similar to that
carried out by ATLAS and CMS for supersymmetry \cite{susysearch}, that is to look for events with jets and large missing transverse energy,
associated with a high $p_T$ lepton for those processes where the top is produced.
Consequently, bounds on the masses of the fermionic comparts could be extracted from these searches: squarks and gluinos as heavy as $\sim 0.5$ TeV are already probed by these kind of studies.

Since the scalar colour triplet $\triplet$ is stable thanks to the $B_3$ symmetry, one may wonder if its relic density may play a role as dark matter, or
more in general affect cosmology. A relic coloured particle is very constrained by the experimental bounds on its cross-section with nucleons,
even if it manifests in the form of neutral hadrons \cite{Starkman:1990nj}.
In any case,
one can argue that
the $\triplet$ relic density
will
be very suppressed since, besides annihilations through the SM gauge interactions,
$T$ being a composite state interacts strongly with the rest of the composite sector and thus it can annihilate efficiently to other lighter unstable composites,
such as longitudinal $W$ and $Z$ bosons \cite{Barbieri:2010mn}.
\\

Alternatively, consider the scenario where the lightest compart is either $\nu^c$ or $\nu'$.
This can be realized by taking $\lambda_{l^c}$ and/or $\lambda_{\nu'}$ sufficiently small (compared to $g_{\rho}$),
which may be compatible with EWSB and EWPTs.
Note that the third and fourth couplings of \eq{2co}  induce a mixing of $\nu^c$ and $\nu'$,
that plays a relevant role in their phenomenology. The lightest combination $\nu_l$ is
the stable compart in this scenario. The other comparts decay to SM$+\nu_l$ final states. The scalar triplet decays as
\beq
\triplet \rightarrow b \overline{\nu_{l,h}}~,~t \overline{e^c}
~,
\eeq
and the decay channels for the fermionic comparts are
\beq
e^c \rightarrow W^{(*)} \nu_{l,h} ~,~~~ t^c \rightarrow b b \overline{\nu_{l,h}}~,~~~ b^c  \rightarrow t b \overline{\nu_{l,h}}~, ~~~
\nu_h \rightarrow (Z,h) \nu_l~,~~~e' \rightarrow \nu_{l,h} b \bar{t},
\eeq
where $\nu_h$ is the orthogonal combination to $\nu_l$. Again, the decay channel $t^c \rightarrow W^{(*)} b^c$ or vice versa is present.
The signals at the LHC for this scenario are richer in top and bottom quarks than the previous one. When $t^c$, $b^c$, or $e'$ are doubly produced, they yield events with $b \bar{b} b \bar{b} + \Lslash{E_T}$ or $b \bar{b} t \bar{t} + \Lslash{E_T}$, which constitute an excellent opportunity for early LHC discovery. These final states are very
similar to supersymmetric scenarios where the third generation squarks are lighter than the other sfermions and
are produced mostly from gluino decays \cite{susy3}.

The stable $\nu_l$ could be a viable dark matter candidate.
In fact,  the fermions $l^c = (e^c, \nu^c)$ and $\nu'$ have the gauge quantum numbers
of a higgsino and a bino in supersymmetry (but $\nu'$ is a Dirac fermion). The analysis of dark matter
phenomenology will be similar, with a few significant differences: the higgsino-bino mixing provided
by the gauge couplings to the Higgs is replaced by $l^c - \nu'$ mixing provided
by the third and fourth Yukawa couplings in \eq{2co}; the $t$-channel annihilations into SM fermions
in supersymmetry proceed through sfermion exchange, while here $T$ is exchanged, via the first and last couplings
in \eq{2co}.
Note that
a purely $\nu^c$ dark matter would couple to the $Z$ boson and scatter on nuclei at the tree-level,
with a spin-independent cross section \cite{GW} a few orders of magnitude larger
than the upper bound from direct dark matter searches.
Therefore, $\nu_l$ should be mostly made of the $\SU(2)_L$ singlet state $\nu'$, in order to constitute a viable dark matter candidate.
Contrary to the case of the bino, the $\nu'$ annihilations into SM fermions are not chirality suppressed due to its Dirac nature.

As a matter of fact, a closely related
scenario has been studied in detail in a warped extra-dimension,
with $\SO(10)$ gauge symmetry in the bulk \cite{AS}.
There, states with the quantum numbers of our fermionic comparts are present and their phenomenology is discussed.
In particular, the lightest zero mode with $B_3\ne 0$ plays the role of dark matter.
Moreover, there are arguments that favour the analog of $\nu'$ to be the lightest, and a detailed
analysis of dark matter phenomenology is performed: while the relic density is mostly controlled by SM gauge interactions
for dark matter masses below $\sim 100$ GeV, for larger masses it becomes more relevant to consider
the interactions with the lightest Kaluza-Klein resonances of the $\SO(10)$ gauge bosons (we did not discuss
such multi-TeV vector resonances in this paper).

Finally, concerning flavour transitions in this scenario, $\triplet$ cannot mediate FCNCs at tree-level,
due to its non-zero $B_3$ charge. The only compart that couples linearly to the SM fermions is $b'$,
that can mediate flavour transitions at loop-level; however the bounds are milder than the one from $Zb\bar{b}$,
that already requires $m_{b'}$ to lie above one TeV.
Loop diagrams involving both $T$ and a lighter fermionic compart may provide some mild constraint.

\subsection{The model with $t_R \in \bf{\overline{16}}$ and $\U(1)_{B_I}$}

The phenomenology turns out to be completely different if the baryon number of the comparts is defined by the generator $(B-L)_{\SO(10)}$.
The absence of an external U(1) symmetry introduces extra couplings that allow any of the comparts to decay to SM fields and, in some cases,
to be singly produced.
With the help of Tables \ref{tabSM} and \ref{tabCP} it is easy to list all the allowed Yukawa interactions, as well as a few phenomenologically relevant
dim-5 operators, coupling the comparts to the SM fields:
\bea
{\rm 2~comparts:} &&
\laq \overline{q_L} l^c_R T~,~~~  \lambda_{b_R} b_R \nu'_R T^* ~,~~~\laq \overline{q_L}  q^c_R T^* ~,
\nonumber \\
&&\frac{\lambda_{b'} \lambda_{\qx}}{g_{\rho}} \qx_L b'_L H ~,~~~ \frac{\lambda_{\lx} \lambda_{e'}}{g_{\rho}} \lx_L e'_L H ~,~~~ \frac{\lambda_{\lx} \lambda_{\nu'}}{g_{\rho}} \lx_L \nu'_L H^* ~;
\nonumber \\
&&
\frac{\lambda_{e'}}{f} e'_L (\Sslash\partial T^*)  t_R  ~;
\label{2compBI} \\
{\rm 1~comparts:} &&
\laq \overline{q_L} b'_R H~,~~~\frac{\lambda_{q}^2}{g_\rho} q_{L} q_{L} T~, ~~~ \lambda_{b_R} b_R t_R T ~, ~~~\lambda_{b_R} b_R \qx_R H ~;
\nonumber \\
&& \frac{\lambda_{\qx}}{f} \qx_L (\Sslash\partial H^*) t_R ~.
\label{1compBI}
\eea
In identifying the allowed couplings above, we worked in the natural pNGB basis, where $H$ and $T$ have non-derivative interactions
only if they are ``contracted" with a coupling $\lambda_\psi$, that explicitly breaks $\SO(10)$;
e.g. a coupling $g_\rho q^c_R t_R H^*$ is not present, even if allowed by $\Gsym_{SM} \times \U(1)_{B_I}$.

Eq.~(\ref{1compBI}) shows that $\triplet$ can couple to two quarks (or two antiquarks), predominantly a bottom and a top.
The coupling to light generations are much smaller, since they are proportional to the corresponding Yukawas.
Therefore, $\triplet$ is a so-called (third-generation) diquark.
It can be singly produced at the LHC (or Tevatron), in association with a top
and possibly also a bottom, since its coupling to light quarks is comparatively negligible.
Still, we expect $\triplet$ pair production through gluon fusion to be the dominant production channel, unless its mass is unexpectedly large.
The triplet decays promptly to top-bottom final states,
\beq
\triplet \rightarrow \bar{t} \bar{b} ~,
\eeq
either left- or right-handed, although with different branching fractions. There exist several experimental studies on diquarks at Tevatron
\cite{TeVatrondiquark} and recently also at the LHC \cite{LHCdiquark}, although they focus on diquark resonant production and decay to dijets,
and therefore their bounds are easily evaded in our scenario.\footnote{
Nonetheless, the possibility is open that $T$ couples to light generations more strongly than naively expected (for instance to right-handed quarks, if their degree of compositeness is large). In that case the diquark could be copiously produced at the LHC, where it should be seen as a resonance in the channel $pp \rightarrow jj$. The exclusion limits from Tevatron and LHC collaborations are comparable, with model dependent mass bounds $m_T \gtrsim 1 \TeV$ \cite{TeVatrondiquark,LHCdiquark}.}
 Instead,
we expect the discovery of our third-generation diquarks to come from the study of events with $t\bar{t} b \bar{b}$ finals states,
when pair produced, or $t\bar{t} b$, from single production.
We also remark that this diquark does not mediate flavour-changing neutral currents at tree level, since it connects an up-type
quark to a down-type quark.

For what concerns the coloured fermion comparts, since $b'$ has the same couplings to the SM fields as in the case of $\U(1)_{B_E}$,
the same considerations for production and decay as above apply.
Now also the vector-like quark doublet $\qx$ couples to SM fields only, through the last two terms in \eq{1compBI}.
Therefore, $t^c$ and $b^c$ can be singly produced in association with a top or a bottom, just as $b'$.
Also, they predominantly decay to a bottom or a top and a longitudinal EW vector boson:
\beq
t^c \rightarrow th,~tZ_L,~~~b^c \rightarrow tW_L^-~.
\eeq
The lower bound on $m_{b'}$ from CDF discussed above also applies to $m_{b^c}$.
Similar searches exist for a toplike quark,
leading to the bound $m_{t^c} \gtrsim 260$ GeV \cite{CDFt}.
These limits could be strengthened studying correlations between $t^c$ and $b'$,$b^c$ decays \cite{FWTB}.

Finally, the colour singlet fermion comparts $\lx$, $e'$ and $\nu'$ carry a baryon number $\pm1$, and therefore they must
decay into three-quark final states. They actually do through $\triplet$, as apparent from the first two couplings
as well as the last one in \eq{2compBI}.
We find that the dominant decay channels are
\beq
\nu^c,~ \overline{\nu'} \rightarrow t b b~,~~~~~ e^c,~ \overline{e'} \rightarrow t b t~.
\eeq
Since these states can be significantly produced only in pairs,
 events with a high multiplicity of bottom and top quarks will be generated, \ie $b \bar{b} b \bar{b} t \bar{t}$ or $b \bar{b} t \bar{t} t \bar{t}$. Since $t \rightarrow b W$, and $W$ decays in turn either leptonically or hadronically, we expect events with
$\sim 6~b$'s,
plus (same-sign) lepton(s) and/or jets,  to be ideal channels for discovery.
However, the required luminosity is large, since the triquarks are produced through weak interactions only.
We believe that this model definitely deserves an accurate analysis of the third-generation diquark and triquark signatures.

\subsection{The model with  $t_R \in \textbf{10}$ and $\U(1)_{B_E}$}

\begin{table}[t]\begin{center}
\begin{tabular}{|c|ccc|c|}
\hline &&&& \\ [-2.5ex]
& $l^c$ & $l'$ & $t^c$ & $\triplet$ \\ \hline &&&& \\ [-2ex]
$\SU(3)_C$ & 1 & 1 & $\bar 3$  & 3 \\  [1ex]
$\SU(2)_L$ & 2 & 2 &  1  & 1\\  [1ex]
$\U(1)_Y$ & $\frac 1 2$ & $-\frac 1 2$ & $-\frac 2 3$  & $\frac 2 3$ \\  [1ex]\hline &&&& \\ [-2ex]
$\U(1)_{B_E}$ & $\frac 1 3$ &  $\frac 1 3$ & $\frac 1 3$  & 0\\ [1ex]
\hline
\end{tabular}
\caption{The gauge and baryon number assignments of the comparts in the case $(\emph{2})$, that is with $t_R \in {\bf 10}$.
The baryon number assignment external to $\SO(10)$ is adopted. \label{tabCP2}}
\end{center}\end{table}

Let us discuss the phenomenology in the case when the composite $t_R$ is embedded in the fundamental representation of $\K=\SO(10)$.
Since $\bf 10$ is a real representation, one needs to prevent a vector-like mass term, $m_\rho {\bf 10}\, {\bf 10}$, in order to keep $t_R$ massless
before EWSB. This is achieved
automatically by imposing baryon number conservation in the form of a symmetry $\U(1)_{B_E}$ external to $\SO(10)$.
On the contrary, $\U(1)_{B_I}$ would allow for such a mass term  and therefore it is not a viable option.
We display the quantum numbers of  the comparts
in Table \ref{tabCP2}, to promptly identify the allowed interactions between them and the SM particles in Table \ref{tabSM}.
Note that the baryon triality of the comparts
is non-zero, $B_3(l^c)=B_3(l')=1$ and $B_3(t^c)=B_3(T)=2$, so that they cannot decay into SM final states and the
lightest compart is stable.

There is in fact a unique allowed Yukawa interaction between comparts and SM particles,
\beq
\lambda_q \overline{q_L} l'_R T~.
\label{2yuk}\eeq
Therefore the phenomenology is controlled also by higher dimensional operators. Let us classify the allowed two-fermion operators
of dimension five:
\bea
{\rm 3~comparts:} &&
\frac{\lambda_{t^c}}{f} \overline{t^c_L} t_R T^* T^* ~,\nonumber \\
&&
\frac{\lambda_{l^c}}{f}\overline{l^c_L} t^c_R TH ~,~~~ \frac{\lambda_{t^c}}{f}\overline{t^c_L}l^c_R T^* H^* ~,~~~
\frac{\lambda_{l'}}{f}\overline{l'_L}t^c_R TH^* ~,~~~ \frac{\lambda_{t^c}}{f}\overline{t^c_L}l'_R T^* H  ~
; \label{3co2}\\
{\rm 2~comparts:}
&& \frac{\lambda_q\lambda_{l'}}{g_\rho}\frac{1}{m_\rho}\overline{q_L} (\Sslash{\partial}T) l'_L ~,~~~
\frac{\lambda_{\lx}}{f} \overline{\lx_L} t_R T^* H ~,~~~
\frac{\lambda_{l'}}{f} \overline{l'_L} t_R T^* H^* ~, \nonumber \\
&&
 \frac{\lambda_{l^c}}{f}\overline{l^c_L} l^c_R H^* H ~,~~~
\frac{\lambda_{l'}}{f}\overline{l'_L} l'_R H^* H ~,~~~
\frac{\lambda_{l'}}{f}\overline{l'_L}\lx_R H^* H^* ~,~~~ \frac{\lambda_{\lx}}{f}\overline{\lx_L} l'_R HH ~.
\label{2co2}
\eea
The last two terms of \eq{2co2} induce a mixing between $\nu^c$ and $\nu'$ of order $(\lambda_{\lx}/\lambda_{l'})^{\pm1} (v^2/f^2)$;
we call $\nu_l$ and $\nu_h$ the corresponding light and heavy mass eigenstates.

When $\triplet$ is lighter than the fermionic comparts, the former is stable, while the latter decay into SM$+\triplet$ final states,
through the coupling in \eq{2yuk} and those in the first line of \eq{3co2} and \eq{2co2}:
\beq
\nu_{l,h} \rightarrow t T^* ~, ~~~ e' \rightarrow b T^* ~, ~~~ e^c \rightarrow t W
 T^* ~,~~~
t^c \rightarrow t T^* T^*  ~.
\label{iipheno}
\eeq
The intermediate decay $e^c \rightarrow \nu_{l,h} W^{(*)}$ or vice versa is also possible, and similarly for $e'$.
When the $t^c$ decay channel in \eq{iipheno} is not open kinematically, $m_{t^c} < 2 m_{\triplet}$,
$t^c$ can decay to $ b b \bar{t} T$, through
a dim-7 operator, much in the same way as $e'$ in section \ref{phenomodel1}.
The phenomenology of the stable $T$ is basically the same as discussed for the model of section \ref{phenomodel1},
since its different charge appears difficult to measure.
It will appear as missing energy, when $\triplet$ hadronizes into a neutral bound state (e.g. $T\bar{u}$),
or it can also hadronize (with similar probability) to a charged bound state (e.g. $T\bar{d}$;
$T$ can be distinguished from $T^*$ because of the opposite charge of the hadron).
The processes in \eq{iipheno}
will resemble the supersymmetric ones for decays of neutralinos or charginos
with a stable stop in the final state, except for the decay of $t^c$ that has no counterpart in supersymmetry.

Alternatively, consider the possibility that the lightest compart is the neutral component of $l^c$ or $l'$,
more precisely the linear combination $\nu_l$. The decays of $T$ and of the
colour singlet comparts into SM$+\nu_l$ final states proceed through the channels
\beq
T \rightarrow t \overline{\nu_{l,h}}~, ~
b \overline{e'} ~,~~~
e' \rightarrow \nu_{l,h} W^{(*)}~,~~~ e^c \rightarrow \nu_{l,h} W^{(*)}~,~~~ \nu_h \rightarrow (Z,h) \nu_l ~.
\eeq
Finally, the only coloured fermion compart decays through the operators in the second line of \eq{3co2}, as
\beq
t^c \rightarrow \nu_{l,h} T^* ~,
\eeq
where $T^*$ and $\nu_h$ could either be real or virtual.

Concerning dark matter, this scenario resembles that of section \ref{phenomodel1}
with $\nu_l$ as lightest compart.
In the present case $\nu_l$ is a mixture of the $\SU(2)_L$ doublets components
$\nu'$ and $\nu^c$, that are both higgsino-like.
This makes it a problematic
dark matter candidate, for the same reason as $\nu^c$ in section \ref{phenomodel1}: such a Dirac fermion
scatters on nuclei at tree-level through $Z$ boson exchange, with a spin-independent cross section
by far larger than allowed by direct dark matter searches.
One can envisage two ways out:
either its number density today is much lower than the one required to explain dark matter (due to the rapid annihilation channel into $W$'s
and possible additional channels via the composite sector); or it is split into a pseudo-Dirac pair of fermions
with mass difference $\delta m \gtrsim 100$keV, e.g. due to a small mixing with an electroweak singlet Majorana fermion:
in this case the coupling to $Z$
of the lightest state is suppressed by $\delta m/m$ \cite{TSW} and it thus becomes a viable dark matter candidate.\\

Let us conclude with a comparison between the two possibilities we adopted to prevent baryon number violation.
The choice between $\U(1)_{B_E}$ and $\U(1)_{B_I}$
determines whether $\triplet$ is stable (or it decays to a lighter stable compart) or it behaves as a diquark, respectively.
We remind again that there are also alternative possibilities to suppress proton decay.
For example, if one adopts a $\U(1)_{3B + L}$ symmetry external to SO(10), $T$ would behave as a composite leptoquark mostly coupled
to the third generation, as studied in Ref.~\cite{Gripaios:2009dq}.
Clearly,  the phenomenology of the exotic fermions also depends on this choice.
In this paper we explored two viable possibilities, since both are consistent with gauge coupling unification and EWSB, with sensible
differences in terms of the underlying model-building, of constraints from EWPTs, and more importantly in terms of phenomenology at colliders.
What option is chosen by nature is an experimental question, at this stage.

\section{Conclusions \label{conclu}}

We performed a first comprehensive analysis
of composite-Higgs scenarios that account, at the same time,
for the stability of the EW scale and for gauge coupling unification.
We characterized in detail the structure of the composite sector,
required to achieve precision unification
while realizing EWSB in agreement with EWPTs.
These requirements determine the spectrum of light exotic particles below the compositeness scale,
to be observed in the first few years of the LHC operation.
The effective theories we presented constitute the low energy manifestation of fully natural theories valid up to the GUT scale,
since the EW-GUT hierarchy is stabilized by the strongly coupled dynamics of the composite sector.

The way precision unification is implemented can be summarized in two essential requirements: (\emph{i})
the global symmetry group $\mathcal G$ of the composite sector, that necessarily contains $\SU(3)_C \times \SU(2)_L \times \U(1)_Y$,
must be a simple group; (\emph{ii}) the SM right-handed top quark $t_R$ must be a composite state, as suggested by the large top mass.
Then, under the fairly generic assumptions discussed in section \ref{GCE}, it follows that the degree of unification at leading order is very good,
comparable to the one of the minimal supersymmetric SM,
with unification scale $M_{GUT} \sim 10^{15} \GeV$.

Given the additional requirement of custodial symmetry of the composite sector, that is demanded by EW precision data, we found that
the minimal implementation of composite unification entails a symmetry $\mathcal G=\SO(11)$, broken
spontaneously at a scale of the order of a few TeVs to $\mathcal K=\SO(10)$. It is amusing that the $\SO(10)$ technology, developed to study the GUT scale
sector in weakly coupled theories of unification, is brought down to earth in the composite GUT scenario, where it relates to the TeV scale particle content.
One remarkable example is the prediction of a colour triplet scalar $T$ associated with the Higgs doublet $H$:
together they fill the composite pNGB multiplet, transforming in the \textbf{10} representation of SO(10).

We find that the price to pay to make the composite GUT scenario realistic is high, but it is rewarded by the existence of
fully viable and relatively compact models, that address the hierarchy problem and the gauge unification as effectively
as low energy supersymmetry does,
and predict very distinctive signals in forthcoming collider searches.
Here is a summary of the main
difficulties to address in the construction of a composite GUT model,
and of the
solutions we proposed and investigated in our analysis:
\begin{itemize}

\item The compositeness of $t_R$
demands for  an ad hoc content of elementary chiral fermions: the SM ones, but without the top isosinglet, plus
a set of exotic ones
to pair with the SO(10) partners of $t_R$ and make them massive.
We have shown how this content can be simply understood, complementing the three SM generations with an
extra SO(10) multiplet of chiral fermions, and decoupling the top isosinglet with a mass $\sim M_{GUT}$.
This provides an anomaly-free set of elementary fermions which is exactly the one needed for the precise unification of gauge couplings.
The $t_R$ partners are vector-like fermions with masses below the compositeness scale
(as much as supersymmetric partners shall be found below the scale of supersymmetry breaking).
We studied the potentially dangerous contributions of these exotic fermions
to the EW precision parameters in two cases,  $t_R \in \overline{\textbf{16}}$ and  $t_R\in \textbf{10}$.
In the first case, the contributions from the exotic fermion $b'$
to $\widehat{T}$ and $\delta g_{b_L}$ constrain the allowed range for
$f$ and  we find the lower bound $f \gtrsim 500 \GeV$.
In the second case, the contributions to $\widehat{T}$ and $\delta g_{b_L}$ from the exotic fermions can be suppressed,
since they do not violate
$\SU(2)_L \times \SU(2)_R \times P_{LR}$, as long as they have equal masses.
The stronger constraints come from the usual correction to $\widehat{S}$ of composite-Higgs models,
and from the negative contribution to $\widehat{T}$ from $m_h$. We find a lower
bound $f \gtrsim 600 \GeV$, that could be relaxed
taking a large custodial violating mass difference between the exotic fermions.

\item Too fast proton decay (and too large neutrino masses) must be forbidden
imposing that the composite sector and its couplings to the elementary sector respect baryon (and lepton) number;
this requirement compares with
$R$-parity in the case of supersymmetry.
We identified two consistent ways to assign baryon number to the composite states,
adopting a symmetry $\U(1)_B$ that is
either external or internal to the SO(10) global symmetry.
In particular, the lightest composite particles, that is to say $T$ and the partners of $t_R$,
have different baryon numbers in the two cases.
Baryon number violation is thus forbidden in spite of a scalar sector that contains both a Higgs doublet and a colour triplet.
However, the $\U(1)_B$ symmetry
seems to require that the chiral elementary fermions do not fill complete GUT multiplets: it remains
an open challenge to realize the needed splitting at the GUT scale, in analogy with the doublet-triplet splitting problem of supersymmetric GUTs.

\item The minimization of the pNGB potential $V_{eff}(H,T)$ depends on the interplay between
the large SM couplings, $g_s$ and $y_t$, and two free parameters, the exotic fermion coupling $\lambda_x$ and the inter-composite coupling $g_\rho$.
We demonstrated that $V_{eff}$ is compatible with EWSB with no colour breaking in a sizable portion of parameter space.
This region is reduced significantly by EWPTs, introducing the so-called little hierarchy problem.
The fine-tuning is customarily estimated
by the parameter $\xi = v^2/f^2$.
As discussed above, we gratifyingly find that our composite GUT models are fully compatible with $\xi \simeq 0.1$ (10\% tuning),
comparable to minimal composite-Higgs models.
However, a sufficiently large exotic coupling $\lambda_x$ is necessary to achieve EWSB, and its value is strongly correlated with $\xi$ so that,
to comply with EWPTs, $\lambda_x$ is restricted to a narrow range that depends on $g_\rho$.

\end{itemize}

The analysis of EWSB and EWPTs allowed us to identify the favoured range for the parameters $f$, $g_{\rho}$ and $\lambda_x$.
This in turn determines the masses of the light composite states: the Higgs, the triplet $T$, and the exotic fermions $x$'s.
In the case with $t_R \in \overline{\textbf{16}}$, we find that the exotic fermions are expected to lie in the range $m_x = \lambda_x f \sim 1-3 \TeV$
(assuming $\xi \lesssim 0.1$, to bar unnecessary fine-tuning),
although some $x$'s are allowed to have a significantly lighter mass.
The mass of the Higgs rapidly grows with $\lambda_x$, so that heavier $x$'s correspond to a heavier Higgs. We find $m_h$ values
typically larger than 400 GeV, and up to 800 GeV.
The mass of the triplet is dominated by gluon loops, yielding $m_T \gtrsim 1 \TeV$, increasing with $m_{\rho}=g_\rho f$.
In the case with $t_R \in \overline{\textbf{10}}$ lighter states are expected,
with masses as light as the experimental lower bounds.
This is due to the lower value of $\lambda_x$ needed to realize EWSB
and to the smaller number of exotic fermions, with respect to the previous case.

Having delimited the territory of viable composite GUT models, we performed a first survey of the phenomenology at colliders.
The distinctive features of this scenario is that the Higgs doublet and the right-handed top are composite, and they are
both accompanied by their $\SO(10)$ partners (comparts).
Were these comparts identified experimentally,
one would obtain a suggestive indication in favour of grand unification.
The phenomenology of comparts strongly depends on their baryon number, since this determines, along with the SM gauge symmetries, their couplings
to the SM particles. We identify two main scenarios, characterized by broadly different signatures:
\begin{itemize}
\item[(\emph{i})] When $\U(1)_B$ is chosen to be external to SO(10),
the lightest compart with non-zero baryon triality is stable. This will manifest at the LHC either as a charged track associated to a heavy particle, or as large missing transverse energy. In both cases, top and/or bottom quarks are expected in the final state. The stable compart, if colourless and neutral,
may account for the dark matter energy density.
\item[(\emph{ii})] When $\U(1)_B$ is internal to SO(10), there is no stable compart. The triplet $T$ behaves as a third generation diquark,
while the colour singlet fermion comparts are third generation triquarks: they decay respectively to two or three quarks, predominantly tops and bottoms,
what yields striking events such as $pp \rightarrow b \bar{b} b \bar{b} t \bar{t}$ or $b \bar{b} t \bar{t} t \bar{t}$.
\end{itemize}
Since the coloured comparts have the largest production cross sections, they could be copiously produced in the initial LHC running,
so they are perfect targets for early discovery.
The signals have a significant intersection with those for third generation squarks and gluinos in the supersymmetric scenario.
 Still there are a few distinguishing features, as we detailed in section \ref{pheno}:
 some comparts resemble fourth generation quarks,
and there is the possibility of no stable particles with, correspondingly, events with higher multiplicity of tops and bottoms.

Finally, let us remark that it will be more difficult to demonstrate the composite nature of the new particles at the LHC.
Previous analyses to characterize the compositeness  of the Higgs and of
the top indicate the need of precision measurements of their couplings, that require large integrated luminosity.
We expect that the compositeness of the comparts could be deciphered conducting similar investigations.
This would ultimately distinguish them from elementary states with identical quantum numbers
(e.g. a stable colour triplet $T$ shares many similarities with a stable stop).
These compositeness measurements will probe new physics associated to the scale $f$,
which we expect to be close to the EW scale by naturalness arguments,
and they may be the first hint for a strongly interacting sector, even when the resonances with mass $\sim m_\rho$
are too heavy to be directly observed.

We thus suggest that the first new physics signals at the LHC may come from (coloured) particles
associated with a strongly interacting sector, that must accompany the
Higgs and the top quark when the global symmetry of this sector unifies the EW and the colour interactions.
These new states must be lighter than the compositeness scale
and their discovery would point to a more accurate
gauge coupling unification, with respect to the SM field content.
In the absence of these signals, the idea of composite GUTs with a natural EW scale may be ruled out within a few years.
If instead some of its specific predictions were confirmed,  this scenario could receive substantial support.

\section*{Acknowledgements}

We thank Roberto Contino, Arthur Hebecker, Slava Rychkov and especially Alex Pomarol for the many enlightening discussions.
This work was supported in part by the Spanish Research Project CICYT-FEDER-FPA-2008-01430 and
Consolider-Ingenio 2010 Programme CPAN (CSD2007-00042).
The work of MF was also supported
by the Marie-Curie Reintegration Grant PERG06-GA-2009-256374 within the European Community FP7.
The work of JS was also supported by the Spanish MEC FPU grant No. AP2006-03102.




\end{document}